\documentclass[preprint,10pt,3p]{elsarticle}

\usepackage{amssymb}
\usepackage{tikz}
\usepackage{mathtools}
\usepackage{multirow}
\usepackage{nomencl}
\usepackage{color}
\usepackage{framed}
\usepackage{soul}
\usepackage{url}
\usepackage{hyperref}
\usepackage{eurosym}
\usepackage{natbib}
\usepackage{graphicx}
\usepackage{caption}
\usepackage{xcolor}
\usepackage{amsmath}
\usepackage{subcaption}
\usepackage{booktabs}
\usepackage{colortbl}
\usepackage{bm}
\usepackage[linesnumbered,ruled]{algorithm2e}
\usepackage{tikz}

\usetikzlibrary{shapes.geometric, arrows}
\tikzstyle{startstop} = [ellipse, minimum width=3cm, minimum height=1cm,text centered, draw=black]
\tikzstyle{io} = [trapezium, trapezium left angle=70, trapezium right angle=110, minimum width=3cm, minimum height=1cm, text centered, text width=8cm, draw=black]
\tikzstyle{process} = [rectangle, minimum width=3cm, minimum height=1cm, text centered, text width=5cm, draw=black]
\tikzstyle{decision} = [diamond, aspect=2, minimum width=2.5cm, minimum height=1cm, text centered, text width=3cm, draw=black]
\tikzstyle{arrow} = [thick,->,>=stealth]
\usepackage[edges]{forest}
\makenomenclature

\begin{document}
\begin{frontmatter}


\title{A multi-dimensional unsupervised machine learning framework for clustering residential heat load profiles}

\author[label1]{Vasilis Michalakopoulos\corref{cor1}}
\ead{vmichalakopoulos@epu.ntua.gr}
\cortext[cor1]{Corresponding author}
\author[label1]{Elissaios Sarmas}
\author[label2]{Viktor Daropoulos}
\author[label2]{Giannis Kazdaridis}
\author[label2]{Stratos Keranidis}
\author[label1]{Vangelis Marinakis}
\author[label1]{Dimitris Askounis}
\address[label1]{Decision Support Systems Laboratory, School of Electrical \& Computer Engineering, National Technical University of Athens, Greece}
\address[label2]{domx IoT Technologies, Greece}

\begin{abstract}


Decarbonizing the heating sector is central to achieving the energy transition, as heating systems are provide essential space heating and hot water in residential and industrial environments. A major challenge lies in effectively profiling large clusters of buildings to improve demand estimation and enable efficient Demand Response (DR) schemes. This paper addresses this challenge by introducing an unsupervised machine learning framework for clustering residential heating load profiles, focusing on natural gas space heating and hot water preparation boilers. The profiles are analyzed across five dimensions: boiler usage, heating demand, weather conditions, building characteristics, and user behavior. We apply three distance metrics—Euclidean Distance (ED), Dynamic Time Warping (DTW), and Derivative Dynamic Time Warping (DDTW)—and evaluate their performance using established clustering indices. The proposed method is assessed considering 29 residential buildings in Greece equipped with smart meters throughout a calendar heating season (i.e. 210 days). Results indicate that DTW is the most suitable metric, uncovering strong correlations between boiler usage, heat demand, and temperature, while ED highlights broader interrelations across dimensions and DDTW proves less effective, resulting in weaker clusters. These findings offer key insights into heating load behavior, establishing a solid foundation for developing more targeted and effective DR programs.

\end{abstract}

\end{frontmatter}

\clearpage

\begin{table*}[!t]
\scriptsize
\begin{framed}
\nomenclature{ED}{Euclidean Distance}
\nomenclature{DTW}{Dynamic Time Wrapping}
\nomenclature{DDTW}{Derivative Dynamic Time Wrapping}
\nomenclature{CHI}{Calinski-Harabasz Index}
\nomenclature{DBI}{Davies-Bouldin Index}
\nomenclature{SIL}{Silhouette Score}
\nomenclature{GMM}{Gaussian Mixture Model}
\nomenclature{PCA}{Principal Component Analysis}
\nomenclature{DR}{Demand Response}
\nomenclature{ML}{Machine Learning}
\nomenclature{ToM}{Time of Maximum}
\nomenclature{PCs}{Profile Classes}
\nomenclature{LTDH}{Low Temperature District Heating}
\nomenclature{DHW}{Domestic Hot Water}
\nomenclature{PAM}{Partitioning Around Medoids}
\nomenclature{HAC}{Hierarchical Agglomerative Clustering}
\nomenclature{HLP}{Heating Load Profiles}
 \printnomenclature

\end{framed}
\end{table*}
\section{Introduction}
\label{sec:intro}

Boilers are pressure vessels used in buildings and industrial facilities for heating water or producing steam. They are primarily used for providing space heating for buildings in temperate climates as well as for producing hot water and steam required by users such as laundries and kitchens. For space heating, boilers function like chillers in central air-conditioning systems and provide steam or hot water to different parts of the building for heating \cite{jayamaha2007energy}. Water boilers, categorized as either hot water boilers or steam boilers, have the capability to combust fossil fuels such as oil, gas, and coal (some operate using electric current).  Advancements in science and escalating demands compelled these systems to undergo intricate evolution over time. Now, each manufacturer follows its own philosophy, and this results in several types of boilers having different characteristics that perform the same task, with small differences in performance, savings and wear\cite{TSOUMALIS2021117256}. 

Needless to say, boilers still serve the purpose of delivering space heating warmth and producing hot water under various settings. Therefore, an efficient boiler has also a significant influence on heating-related energy savings \cite{ganapathy2002industrial}. We are coming to a conclusion that a substantial amount of energy can be saved by adopting energy saving measures and by improving the overall boiler efficiency \cite{BARMA2017970}. Adding to the previous, it is important to implement real-time monitoring systems for energy efficiency and also to detect the performance degradation of boilers \cite{baldi2017real}. There are many applications in literature, by connecting computers with multiple industrial or residential boilers with predictive and optimal control laws \cite{9292251}. Analyzing the values of various parameters provided by the digital interfaces of modern boilers is essential for drawing conclusions about their functioning and by extent optimize their operation.

Presently, there is a growing inclination towards energy efficiency, with a concurrent transition to greener technologies already underway. An update regarding the shift from conventional energy sources, such as fossil fuels, to more environmentally friendly forms of energy and setting the goal for reducing net greenhouse gas emissions by at least 55\% by 2030, compared to 1990 levels, was issued by the European Union (EU) in 2019 \cite{greendeal}. As stated, buildings in the EU are responsible for 40\% of the overall energy consumption and 36\% of greenhouse gas emissions \cite{Energyefficiencyinbuildings}. Moreover, the Clean Energy for All Europeans package \cite{foralleuropeanspackage}, serves as a comprehensive guide for energy efficiency in buildings, introducing numerous initiatives in this domain. Conserving energy and emissions involves implementing diverse strategies, including the reduction of natural gas usage for heating residential structures. Achieving this goal involves optimizing both the efficiency and functionality of domestic boilers. To manage energy use and maintain comfort effectively, it's essential to integrate advanced controls, smart sensors, weather data to ML pipelines and algorithms. Accurate demand forecasting involves analyzing various factors such as environmental conditions, building characteristics and specifications and user habits and behavior. As the volume of household data increases, segmenting households into manageable groups becomes critical for optimizing energy use and identifying opportunities for savings and/or Demand Response (DR) participation.

The key contributions of our work can be summarized as follows:
\begin{itemize}
    \item To the best of out knowledge, the current study is the first that incorporates a multi-dimensional heating load profile clustering methodology, including heat demand, temperature, building characteristics, boiler specifications, and user preferences. 

    \item We evaluate two 
 unsupervised Machine Learning (ML) clustering algorithms, under three well-known clustering indices, utilizing three different distance metrics and unique features to determine their effectiveness and applicability.

     \item Through extensive experiments the correlations between different dimensions are uncovered, opening new avenues for researchers aimed at developing well-suited DR programs.
    
    \item The focus is on identifying key dimensions for accurately clustering households and identifying energy-saving potential and DR prospects, tested on a real-world dataset comprised of 30 households located in Greece.
\end{itemize}


The subsequent sections of this paper unfold as follows. Section \ref{sec:litrev} delves into the problem setting and provides an extensive review of relevant literature. The intricacies of the devised methodology are expounded upon in Section \ref{sec:method}. Section \ref{sec:casestudy} dives into a practical implementation, offering insights derived from an empirical application grounded in a real-world case study. In Section \ref{sec:discusion} follows the discussion on the cross-dimensional correlations and the limitations and practical implications. Finally, the paper concludes with Section \ref{sec:conclusions}, with a summary of key findings and concluding remarks.

\section{Problem Setting \& Literature Review}
\label{sec:litrev}

\subsection{Problem Setting}

In the past, many homes utilized outdated technology boilers, characterized by a simple operational mode, operating under either full load or being idle. This basic on/off functionality poses a challenge to achieving optimal temperature regulation within the house. Nowadays, the majority of  heating systems provide modulated control, thus enabling the tuning of the system's response under multiple levels.
However, without incorporating proper control principles and techniques for tuning the operation of the employed heating system, it is not possible to jointly improve the achieved energy efficiency while meeting the user specified thermal comfort limits. In contemporary times, the integration of advanced controls with the state-of-the-art heating systems has rendered the aforementioned task achievable. Consequently, researchers delved into innovative solutions, exploring the integration of smart sensors, weather data, ML algorithms, and adaptive controls. Forecasting, whether short-term or long-term, plays a crucial role in improving the efficiency of heating systems and optimizing their energy consumption.

Several critical variables must be considered and analyzed to accomplish this task successfully. Firstly, monitoring the prevailing outdoor weather conditions, especially temperature variations, is crucial due to the fact that it directly impacts the performance of heating systems along with the preferences of building occupants \cite{tsolkas2023dynamic}. In addition, the ability to monitor heating system parameters and energy consumption in real-time is essential for implementing efficient resource management and providing valuable insights into the dynamic behavior of the heating system. Adding to the previous, understanding user behavior and building specializations is a key aspect, allowing for the customization of the system to align with individual user needs \cite{testasecca2023recent}. These factors collectively shape a strategy for tackling the intricacies involved in forecasting the energy demand of residential space heating systems. 

In this context, the ability to deliver accurate demand predictions rely heavily on considering all the key aforementioned parameters, such as weather, building, user and heating system profiles. However, the significance amplifies when it comes to the optimal management of large portfolios of residential heating systems, for achieving energy savings or implementing DR actions towards environmental sustainability \cite{skaloumpakas2024reshaping}.
As the population of participating households and the amount of available historical data continue to grow, the application of dedicated forecasting models at household level becomes challenging and practically impossible \cite{michalakopoulosdatadriven}. This fact emphasizes the need for implementing innovative approaches 
for segmenting participating households into more homogeneous and manageable groups, which can reduce the variability and uncertainty of demand and ultimately simplify the demand profiling and forecasting procedures \cite{MICHALAKOPOULOS20242244}.
Indicative application scenarios may include the ability to pinpoint residences with the highest potential for energy savings and/or for DR participation. As a result, heating load profiles grounded in diverse facets of the problem, which enable a better understanding of consumption patterns and behaviors, are the solution to the identified problem and at the same time the motivation for this study.

\subsection{Related Work}

A plurality of studies have been conducted on electrical load profiling and segmentation that drove the research in the field that is currently being explored \cite{michalakopoulos2024machine, CZETANY2021111376}. More specifically, {\citeauthor{mcloughlin2015clustering} \cite{mcloughlin2015clustering} evaluated K-means, K-medoids and Self-Organising Maps (SOM) in order to cluster households in Ireland based on their electricity use, and derived a series of ten Profile Classes (PC's) that characterize the electricity use within the home. A similar approach was followed by \citeauthor{okereke2023k} \cite{okereke2023k} using only the K-means algorithm. An alternative strategy to identify building electricity usage profiles was followed by \citeauthor{li2018identification} \cite{li2018identification} who combined Gaussian Mixture model based clustering for intra-building clustering and hierarchical clustering for inter-building clustering respectively. An advantage of this approach compared to single step methods such as PAM or hierarchical clustering was the significant reduced computational cost. Furthermore, \citeauthor{ESKANDARNIA2022103618} \cite{ESKANDARNIA2022103618} utilized autoencoders with dimensionality reduction techniques to address the problem at hand. Nevertheless, the related work and corresponding literature that are going to be assessed are ought to be targeted towards creating heating load profiles based on the earlier stated axes of the problem, namely temperature, building, user, heating load and boiler characteristics. The choice of the heating method is versatile, as the creation of these profiles should be detached from the particular heat source, enabling independence from the source itself and directing attention towards other relevant considerations. This adaptability extends beyond boiler systems to encompass various alternatives, including district heating and heat pumps. The central objective is to separate the profile development process from the specifics of the heating technology, allowing for a broader focus on relevant factors.

A notable study by \citeauthor{GMM} \cite{GMM} introduces a Gaussian Mixture Model (GMM) methodology, considering parameters such as heat demand, temperature, and user preferences. The work emphasizes the application of district heating systems to address heating requirements and enhance the accuracy of prediction models. \citeauthor{WANG2019113373} \cite{WANG2019113373} conducted research on clustering district heating users based on consumption patterns, proposing novel approaches for extracting features from consumption data and revealing typical daily patterns even after removing the influence of ambient temperature. The GMM algorithm as an unsupervised learning technique was again utilized for the purposes of the study, much like the approach of \citeauthor{MA2014464} \cite{MA2014464} which delves into the quantitative assessment of energy consumption patterns impacts on heat consumption by considering time and building types as key factors. The study demonstrates that classifying buildings based on their functions, such as commercial, apartment, and office, proves to be an effective approach. Further enriching the understanding of versatile heating options, \citeauthor{LUMBRERAS2023105732} \cite{LUMBRERAS2023105732} determines the relevant number of typical profiles connected to district heating, using the K-means algorithm, followed by the use of Classification and Regression Trees (CART) to predict profile types based on external variables like calendar and climatic factors. A distinct approach is taken by \citeauthor{DOCARMO2016171} \cite{DOCARMO2016171}, again using the K-means algorithm, examining daily load profiles of energy demand for heating in dwellings with heat pumps, utilizing cluster and regression analyses to discern patterns linked to socio-economic and technical attributes. \citeauthor{JESPER2021100085} \cite{JESPER2021100085}, introduces a method for predicting such profiles for industry and commerce, analyzing natural gas load profiles whilst employing the K-means algorithm for clustering based on dependency on mean daily ambient temperature. {\citeauthor{gianniou2018clustering} \cite{gianniou2018clustering} used the K-means algorithm to cluster the daily load profiles of district heating single family households in Aarhus, Denmark and also studied the correlation among heat consumption, building and occupant characteristics using logistic regression. In the work of \citeauthor{calikus2019data} \cite{calikus2019data} the K-shape clustering algorithm was employed to group customers from Sweden with similar heat load profiles, who were connected to two district heating networks. The case study was conducted on buildings from six different categories, such as school buildings, commercial buildings etc. A different approach was adopted by \citeauthor{ma2017variation} \cite{ma2017variation} who utilised the Partitioning Around Medoids (PAM) clustering algorithm to identify the daily heating energy usage profiles of higher education buildings in Norway relying on the Pearson Correlation Coefficient as a dissimilarity measure. \citeauthor{8633397} \cite{8633397} contributes to the broader field by proposing an electric water heater control strategy based on dynamic programming and K-means, addressing the challenges of minimizing peak consumption during critical periods. Unlike previous studies, \citeauthor{LI2021120194} \cite{LI2021120194} proposes a FCM clustering algorithm based on particle swarm optimization (PSO), based on a dataset of natural gas consumption bills, to identify distinct gas consumption patterns, considering both increasing block tariffs (IBTs) and temperature factors. 

The analysis of local low temperature district heating (LTDH) systems requires a comprehensive understanding of factors influencing overall energy demand, including both supply system properties and building characteristics. The study from \citeauthor{KALLERT2018122} \cite{KALLERT2018122} introduces a novel approach utilizing a VBA tool to create stochastic user profiles, forming the foundation for generating heating load profiles (HLP) that account for randomized electricity and domestic hot water (DHW) demand as well as body heat profiles based on inhabitants' behavior, thus offering a versatile tool for analyzing and designing complex energy systems like LTDH supply schemes. As a novel contribution, the methodology proposed by \citeauthor{FERRARI2023104306} \cite{FERRARI2023104306} study conducts a comprehensive comparison of energy implications stemming from different building internal heat load profiles derived from current standards.

\begin{table}[t!]
    \centering
    \begin{tabular}{cccccc|c}
        \hline
        & \multirow{2}{*}{Heat} & \multirow{2}{*}{Temperature} & \multirow{2}{*}{Building} & \multirow{2}{*}{Boiler} & \multirow{2}{*}{User} & \multirow{2}{*}{Heating Type} \\
        & Demand & & & & \\
        \hline
        \cite{GMM} & \checkmark & \checkmark & & & \checkmark & District Heating \\
        \cite{WANG2019113373} & \checkmark  & & & & & District Heating \\
        \cite{LI2021120194} & \checkmark & \checkmark & & & & Natural Gas \\
        \cite{DOCARMO2016171} & \checkmark & & \checkmark &  & \checkmark & Heat Pumps \\
        \cite{JESPER2021100085} & \checkmark & \checkmark & &  & & Natural Gas\\
        \cite{KALLERT2018122} & \checkmark & \checkmark & &  & \checkmark & District Heating\\
        \cite{FERRARI2023104306} & \checkmark &  & \checkmark &  & \checkmark & Internal Heating Standards\\
        \cite{MA2014464} & \checkmark &  & \checkmark &  &  & District Heating\\
        \cite{LUMBRERAS2023105732} & \checkmark & \checkmark & \checkmark &  &  & District Heating\\
        \cite{LUMBRERAS2023105732} & \checkmark &  &  &  &  \checkmark & Heat Pumps\\
        \cite{8633397} & \checkmark &  &  &  &  & Electrical boilers\\
        \cite{gianniou2018clustering} & \checkmark &  & \checkmark &  & \checkmark & District Heating\\
        \cite{calikus2019data} & \checkmark &  &  &  &  & District Heating\\
        \cite{ma2017variation} & \checkmark &  &  &  &  & District Heating \\
        \hline
        \multirow{2}{*}{This paper} & \multirow{2}{*}{\checkmark} & \multirow{2}{*}{\checkmark} & \multirow{2}{*}{\checkmark} & \multirow{2}{*}{\checkmark} &  \multirow{2}{*}{\checkmark} & \multirow{2}{*}{Gas Boilers} \\
        & & & & & & \\
    \end{tabular}
    \caption{Overview of the related literature review}
    \label{tab:lit_rev}
\end{table}

A comprehensive overview of the objectives and fundamental characteristics of the various research endeavors discussed earlier is presented in Table \ref{tab:lit_rev}. The papers delineated in the Table are centered on addressing the challenge of clustering heating load profiles under different scenarios. These profiles are categorized based on distinct characteristics, including user-related aspects, building features, weather influences, and heating load attributes. Our study uniquely explores and comprehensively addresses all of these dimensions, making it the only research of its kind to delve into such a broad and in-depth analysis. The overarching goal is to elevate the efficacy of the segmentation process, supplementing energy efficiency within the heating domain by strategically orchestrating these diverse elements aiming to pinpoint homes with the most promising energy savings potential, thereby optimizing resource allocation and fostering sustainable heating practices and DR programs.

\section{Methodology}
\label{sec:method}

The methodology outlined in Figure \ref{fig:Solution} proceeds as follows: Initial data collection involves gathering boiler and sensor data from both indoor and outdoor spaces within the building, which is then processed to detect outliers and anomalies. Subsequently, key dimensions of the problem are identified, followed by a tailored feature engineering process designed to suit the characteristics of each dimension for use in the Euclidean distance (ED) approach. Meanwhile, plain time-series data are utilized for the Dynamic Time Warping (DTW) and Derivative Dynamic Time Warping (DDTW) approaches, respectively. Clustering algorithms are then applied to the input data created across a range of cluster numbers, from two to nine, with evaluation based on established metrics of similarity and dissimilarity. The optimal number of clusters is determined for each dimension, distance measure, and the problem as a whole. Finally, the outcomes of each scenario are assessed to identify the most effective approach for each dimension and their interrelations within the problem context.

\begin{figure*}[ht!] 
\centering
 \makebox[\textwidth]{\includegraphics[width=.8\paperwidth]{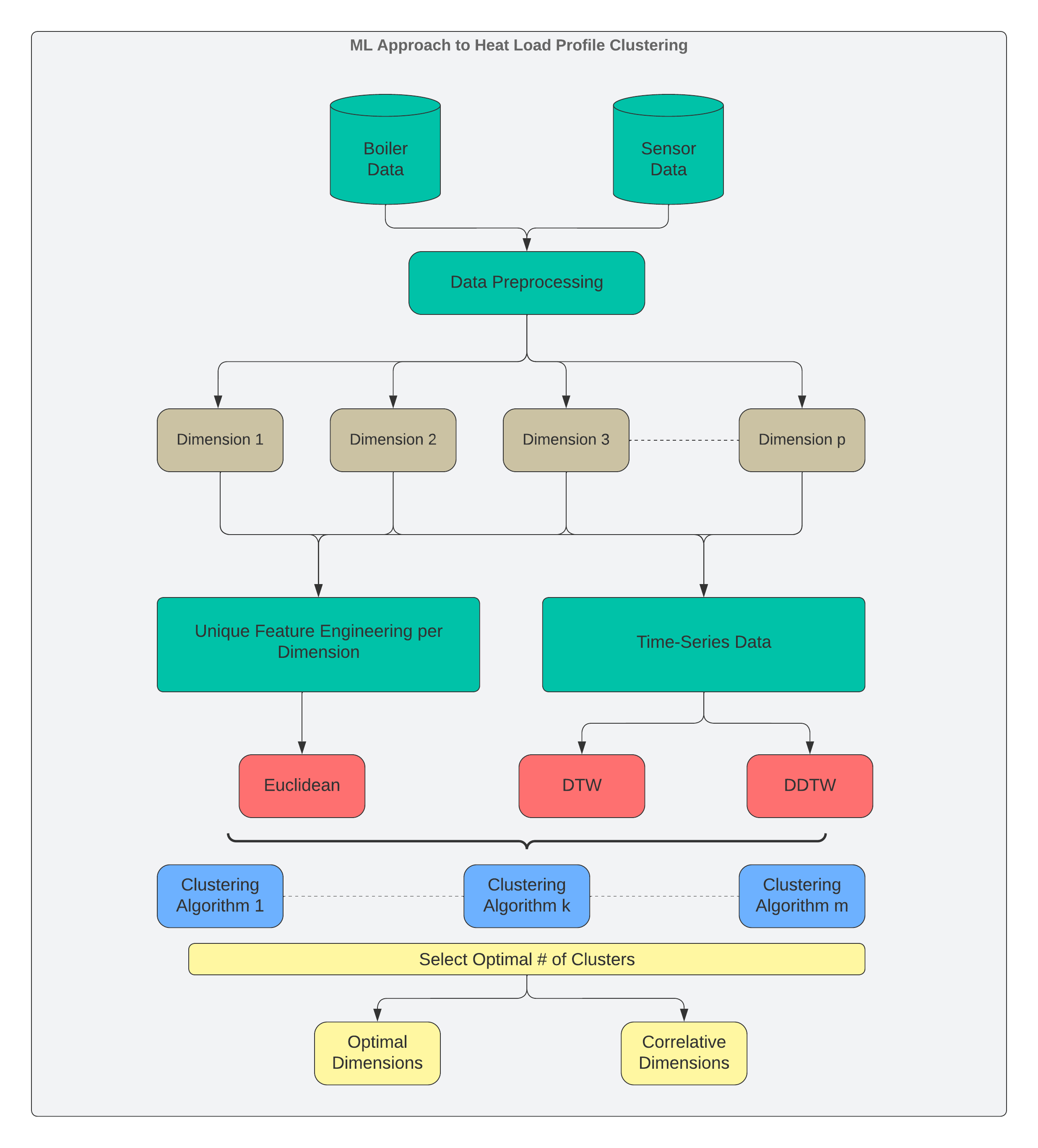}}
\caption{Proposed Methodology.}
\label{fig:Solution}
\end{figure*}

\subsection{Heating load profiles}

The most critical dimensions of the problem, pertaining to the heating load profiles of gas heating boilers, have been discerned to include heat demand, specific buildings, individual boilers, user behavior, and outdoor temperature variations. These dimensions collectively form the core components that influence and shape the heating load profiles under examination. By clustering based on these dimensions, it becomes possible to group similar heating load profiles together, providing insights into potential areas for improvement and tailored solutions to enhance overall performance and user satisfaction.

As previously indicated, a distinct feature engineering strategy has been devised for each identified dimension in order to enhance the results coming from the ED approach, which will be thoroughly analyzed. The mean profiles for each dimension for each minute of the day have been deprived from the original data, and thus the following analysis will be based on them.

\subsubsection{Heat demand dimension}
\label{subsec:heatingload}
To capture the intricacies of heat demand within the dataset, we applied feature engineering techniques. Two primary methods were employed:

\begin{itemize}
    \item Time-of-Maximum Analysis (ToM):
        \begin{itemize}
            \item Identified peak periods of heat demand by determining the time of maximum load for each profile.
            \item Categorized data into distinct time ranges, namely early morning, morning, noon, evening, night, and late night and created binary columns indicating whether the time of maximum load fell within each specified range.
            \item Analytically the time intervals:
                \begin{itemize}
                    \item Early Morning (5:00 - 7:59) 
                    \item Morning (8:00 - 9:59)
                    \item Noon (10:00 - 16:59)
                    \item Evening (17:00 - 21:59)
                    \item Night (20:00 - 23:59)
                    \item Late Night (0:00 - 4:59)
                \end{itemize}
            
        \end{itemize}
        
    \item Statistical Analysis:
        \begin{itemize}
            \item The statistical measures chosen for this analysis are as follows: 
                \begin{itemize}
                    \item Mean
                    \item Standard Deviation
                    \item Maximum
                    \item Minimum 
                    \item 25th percentile
                    \item 50th percentile
                    \item 75th percentile
                \end{itemize}
        \end{itemize}
\end{itemize}

Following feature engineering, the dataset underwent transformation to integrate the newly derived features. This involved expanding the original dataset to incorporate columns representing peak periods and statistical metrics computed for each heating load profile. The enhanced dataset formed the foundation for subsequent clustering analysis.

\subsubsection{Temperature dimension}

In addressing the dimension concerning temperature, it became evident that the data presented a considerable degree of complexity due to the susceptibility of sensors to outdoor conditions. Consequently, the recorded measurements occasionally deviated from the expected ones, necessitating an extensive outlier detection and data cleaning process to ensure the dataset's suitability for analysis. Despite these challenges, the feature engineering efforts primarily focused on statistical analysis, mirroring the methodologies mentioned in Subsection \ref{subsec:heatingload}, namely mean, standard deviation, maximum, minimum and the three percentiles (25th, 50th and 75th). Notably, the Time-of-Maximum (ToM) analysis, which proved effective in heat demand, was deemed inadequate for temperature analysis primarily because peak temperature hours are typically consistent, and thus was omitted from consideration.

\subsubsection{Building dimension}
\label{subsec:building}
In addressing the building dimension of the problem, our focus shifted towards selecting characteristics that reflect the behavior of heating load profiles based on distinct building attributes. Particularly, we delved into discerning the disparity between indoor and outdoor temperatures during periods when the boiler remains inactive. This approach aimed to showcase the influence of the building envelope and size alone on the convergence of these two temperature time series. By isolating and analyzing this discrepancy, we sought to glean insights into the inherent thermal dynamics governed by the structural composition and insulation efficacy of buildings.

Again the ToM analysis was excluded and the statistical one remained. The difference lies into the statistical measures used to assess this temperature time series difference. More specifically, the mean and standard deviation values remained and the rest were replaced with more appropriate ones for this specific case. More specifically: 

The ToM analysis was excluded for the building dimension, while the statistical measures remain. However, the statistical metrics utilized to assess the temperature time series difference were tailored to this specific case. Specifically, the mean and standard deviation values were retained, while the remaining metrics were substituted with more suitable ones:

\begin{itemize}

    \item Rate of Change: Represents the average rate of change of the time series values over time, offering insights into the trend or pattern of evolution. A higher rate of change may suggest rapid fluctuations in temperature, potentially indicating inefficiencies in the building's envelope.
    
    \item Kurtosis: Measures the peakedness or flatness of the distribution of feature values. High kurtosis values may indicate a peaked distribution, suggesting that temperature differences are concentrated around a central value with limited variability.
    
    \item Skewness: Calculates the asymmetry of the distribution of the time series, indicating whether the data is skewed towards higher or lower values. Positive skewness values suggest predominance of instances where indoor temperatures exceed outdoor temperatures.

    \item Variance: Represents the average squared deviation of the time series values from their mean, providing a measure of the dispersion or spread of the data. Higher variance values suggest greater variability in temperature differences, indicating fluctuations in thermal conditions within the building. 
\end{itemize}

\subsubsection{Boiler dimension}

In assessing heating load profiles for boilers, we focus on the difference between the temperature inside the boiler and the temperature of the water when it comes back from the radiators. This is important because boilers are crucial for controlling indoor temperature and how well heating systems work. By looking at this temperature difference, we can understand how efficiently the boiler is working and how well it's heating the building. However, because of the very high temperatures involved, the measurements recorded often showed a significant difference from what was expected. Consequently, meticulous data cleaning procedures were imperative to ensure the reliability and accuracy of the dataset within this dimension.

The decision to employ the same statistical analysis as elucidated in Subsection \ref{subsec:building} was grounded in the nature of the dimension, which revolves around analyzing temperature difference time series. 


\subsubsection{User dimension}

Within the user dimension, our focus lies on analyzing the target room temperature set by users, which serves as a crucial parameter guiding the operation of the heating system. The user-defined target temperature represents the desired level of comfort and plays a pivotal role in determining the heating load profiles. To gain deeper insights into user preferences and system performance, we employ the same statistical analysis techniques utilized in assessing Subsection \ref{subsec:heatingload}. Specifically, we compute mean, standard deviation, maximum, minimum and the three percentiles (25th, 50th and 75th) of the target room temperature time series. Given that the target room temperature is typically set to remain constant throughout the day and night, the convergence of time series data is not as critical as in other dimensions where bigger fluctuations of the temperatures were present. Therefore, our focus shifts to uncovering the intricacies within the aforementioned time series. This entails identifying subtle variations or patterns in user preferences and system behavior, which may not be readily apparent but are nonetheless essential for optimizing heating system performance and enhancing user comfort.

\subsection{Distance Measures}

This subsection provides information about the distance measures utilized by the clustering algorithms in this paper. 

\subsubsection{Euclidean distance (ED)}

ED is the most common distance metric. For two time series, $X = (x_1, x_2, ..., x_n)$ and $Y = (y_1, y_2, ..., y_n)$ in an \(n\)-dimensional space, can be defined as the arithmetic square root of the sum of the squares of the differences of the values on all dimensions, which can be calculated by Equation \ref{eq:ed} \cite{deza2009encyclopedia}.

\begin{equation}
\label{eq:ed}
d_{(x, y)} = \sqrt{\sum_{i=1}^{n} (x_i - y_i)^2}
\end{equation}

\subsubsection{Dynamic time warping (DTW) distance}

DTW algorithm has earned its popularity by being extremely efficient as the time-series similarity measure which minimizes the effects of shifting and distortion in time by allowing “elastic” transformation of time series in order to detect similar shapes with different phases \cite{senin2008dynamic}. Let \(X\) and \(Y\) represent two time series. Suppose $X = (x_1, x_2, ..., x_m)$ has length \(m\), and $Y = (y_1, y_2, ..., y_n)$ has length \(n\), DTW yields optimal solution in the \(O(MN)\) time \cite{muller2007dynamic}. Specifically, a warping path is denoted as \(W = (w_1, w_2, \ldots, w_k)\), where each \(w_l = |x_i - y_j|\), and the overall weight of the optimal warping path is used as the DTW distance as shown in 
Equation \ref{eq:dtw} \cite{9382996}. 

\begin{equation}
\label{eq:dtw}
\text{DTW}(X, Y) = \min_{W} \sum_{l=1}^{n} w_l, \quad w_l \in W
\end{equation}

\subsubsection{Derivative dynamic time warping distance (DDTW)}

While DTW has found success across various domains, it may yield pathological results. The main problem  with  DTW  is  that  the  algorithm  may  fail  to  find  obvious, natural alignments in two sequences simply because a feature (i.e peak, valley, inflection point,  plateau  etc.)  in  one  sequence  is  slightly  higher  or  lower  than  its  corresponding feature in the other sequence \cite{Keogh2001DerivativeDT}. To address this issue,  a modification of DTW that focuses on the higher-level feature of "shape" rather than considering the Y-values of the data points was proposed by \citeauthor{Keogh2001DerivativeDT} and it was named Derivative Dynamic Time Warping (DDTW) due to the fact that it obtains information about the shape from the first derivative of the sequences. DDTW’s time complexity is \(O(MN)\), which is the same as  DTW and can be defined as seen in Equations \ref{eq:first_ddtw} - \ref{eq:third_ddtw}:

\begin{equation}
\label{eq:first_ddtw}
D_q[i] = \frac{{(q_i - q_{i-1}) + \left(\frac{{(q_{i+1} - q_{i-1})}}{2}\right)}}{2}, \quad 1 < i < m
\end{equation}

\begin{equation}
\label{eq:second_ddtw}
D_q[1] = D_q[2]
\end{equation}

\begin{equation}
\label{eq:third_ddtw}
D_q[m] = D_q[m - 1]
\end{equation}

Where \(q_i\) represents the sequence of data points and \(c_j\) denotes the centroids, the distance measure \(d(q_i, c_j)\) in DDTW is not Euclidean; instead, it is based on the squared difference of the estimated derivatives of \(q_i\) and \(c_j\) \cite{Keogh2001DerivativeDT}.

\subsection{Evaluation Metrics}

This subsection delves into the insights of the evaluation metrics utilized to assess the results of the clustering algorithms in this paper. Prior to the explanation of the metrics, it is noteworthy that while the optimal values for SIL and CHI are attained by selecting the clustering configuration that maximizes them, the situation is reversed for DBI where the ideal cluster separation is determined by minimizing its value.

\subsubsection{Silhouette Score (SIL)}
SIL stands out as one of the most widely used measures for evaluating cluster validity. It assesses each sample based on the mean intra-cluster distance (\( a \)) and the mean nearest-cluster distance (\( b \)) \cite{rousseeuw1987silhouettes}. SIL for an one distinct sample is calculated using the formula:

\begin{equation}
S_i = \frac{b_i - a_i}{\max(a_i, b_i)}
\end{equation}

In this context, \( b_i \) denotes the distance from a sample to the nearest cluster it does not belong to. It's essential to note that SIL is applicable only when the number of labels satisfies the condition \( 2 \leq n_{\text{labels}} \leq n_{\text{samples}} - 1 \).

\subsubsection{Davies-Bouldin Index (DBI)}

The Davies-Bouldin Index (DBI) \cite{davies1979cluster} serves as a crucial evaluation metric for determining the best number of clusters, showed by \(K\). The index is defined as follows:
\begin{equation}
DBI = \frac{1}{K} \sum_{i=1}^{K} \max_{j \neq i} \left( \frac{S_i + S_j}{d_{ij}} \right)
\end{equation}
where:
\begin{itemize}
    \item \(S_i\) represents the average distance between each data point within cluster \(i\) and its centroid.
    \item \(d_{ij}\) is the distance between the centroids of clusters \(i\) and \(j\).
\end{itemize}

\subsubsection{Calinski-Harabasz Index (CHI)}

The Calinski-Harabasz Index \cite{calinski1974dendrite}, proposed by \citeauthor{calinski1974dendrite} in \citeyear{calinski1974dendrite}, serves as a key evaluation metric for identifying the optimal number of clusters, denoted by \(K\), in the clustering process. It is defined as follows:
\begin{equation}
\label{eqn:fifth}
    CHI = \frac{B(K)}{W(K)} \times \frac{N - K}{K-1}
\end{equation}
where:
\begin{itemize}
    \item \(B(K)\) stands for the between-cluster sum of squares, indicating the dispersion between the cluster centroids.
    \item \(W(K)\) indicates the within-cluster sum of squares, measuring the compactness of data points within clusters.
    \item \(N\) is the total number of points.
\end{itemize}

\subsection{Clustering Algorithms}

This subsection provides information about the unsupervised ML-based clustering algorithms utilized in this study.

\subsubsection{K-means}

One of the most commonly utilized unsupervised learning techniques in ML that is frequently employed for clustering analysis is the K-means algorithm \cite{kmeans}. This method involves dividing a given dataset into k clusters. The main aim of the K-means algorithm is to minimize an objective function $J$, as shown in \ref{eq:kmeans} , which is determined by the sum of squared distances between each data point and the cluster center it's assigned to.

\begin{equation}
\label{eq:kmeans}
    J = \sum_{i=1}^{k} \sum_{j=1}^{n} ||x_j - c_i||^2
\end{equation}

In this context, $k$ represents the number of clusters, $c_i$ signifies the centroid of cluster $i$, $x_j$ stands for a data point, and $\|x_j - c_i\|$ represents the Euclidean distance between data point $x_j$ and cluster centroid $c_i$.

The essence of the K-means algorithm lies in its iterative process of assigning each data point to its closest centroid and adjusting the centroid of each cluster according to the mean of all data points within it. This iterative refinement persists until convergence is attained. Renowned for its simplicity of implementation and computational efficiency, the algorithm is adept at handling large datasets \cite{kmeans_compute}. Over time, numerous enhancements and adaptations to the K-means algorithm have surfaced, encompassing diverse approaches such as alternative distance metrics, varied initialization techniques, and refined convergence criteria.

\subsubsection{Hierarchical Agglomerative Clustering (HAC)}

Hierarchical agglomerative clustering (HAC) is a clustering technique that starts with each data point as its own cluster and progressively merges clusters based on their similarity. The process begins by treating each data point as a single cluster. Then, at each iteration, it merges the two closest clusters into a single cluster until only one remains, forming a hierarchical tree or a so called dendrogram. HAC can be characterized as greedy, in the algorithmic sense. A sequence of irreversible algorithm steps is used to construct the desired data structure \cite{hac_1}.

One advantage of HAC clustering is its ability to reveal the underlying structure of the data through the hierarchical dendrogram. This allows for the exploration of different levels of granularity in the clustering results. Conversely, HAC can pose computational challenges, particularly with large datasets, due to the need for computing and storing pairwise distances between all data points.

HAC can be described as an algorithm in the following manner:

\begin{itemize}
  \item Let $D = \{x_1, x_2, \ldots, x_n\}$ be the dataset consisting of $n$ data points.
  
  \item At each iteration, the algorithm computes the pairwise distance between all clusters and merges the two closest clusters based on a chosen linkage criterion. $d(C_i, C_j)$ represent the distance between clusters $C_i$ and $C_j$.
\end{itemize}

Equation \ref{eq:agglomerative} defines the process of agglomerative clustering, where $C_{\text{new}}$ represents the newly formed cluster resulting from merging clusters $C_i$ and $C_j$ with the minimum distance $d(C_i, C_j)$. This formulation captures the iterative nature of HAC, which progressively combines clusters based on their pairwise distances until a single cluster remains. Equivalently, HAC can be thought of as a tree structure such that leaves correspond to individual data points and internal nodes represent the cluster of their descendant leaves \cite{hac}.

\begin{equation}
\label{eq:agglomerative}
C_{\text{new}} = \arg\min_{C_i, C_j} d(C_i, C_j)
\end{equation}

\section{Case study}
\label{sec:casestudy}

This section provides details on the case study conducted to evaluate the clustering performance of the proposed methodology for heat demand data of natural gas heating boilers. The dataset description is included as well as the experimental design of the clustering experiments. The results are presented as different scenarios for each of the dimension utilized in this study.

\subsection{Data Set}
\label{subsec:dataset}
The data used for this analysis was captured by IoT devices installed at multiple residential buildings and comprises a number of features related to the indoor and outdoor conditions of each pilot house as well as to operational data of the corresponding gas boilers installed at each house. An overview of the collected features is presented in Table \ref{table:dataset}.

\begin{table}[ht]
\centering
\begin{tabular}{|l|l|l|l|}
\hline
\textbf{Feature} & \textbf{Type} & \textbf{Description} \\ \hline
blr\_mod\_lvl & Float & The percentage of maximum boiler modulation capacity  \\ \hline
blr\_t & Float & Current temperature of the water inside the boiler  \\ \hline
heat & Boolean & Flag indicating whether boiler's circulator is active  \\ \hline
flame & Boolean & Flag indicating if combustion is happening or not  \\ \hline
water & Boolean & Flag indicating if water for domestic usage is requested  \\ \hline
t\_out & Float & Current outside temperature  \\ \hline
t\_ret & Float & Water temperature entering the boiler after being circulated \\ \hline
t\_r & Float & Current room temperature  \\ \hline
t\_r\_set & Float & The target room temperature the heating system is instructed to reach \\ \hline
t\_set & Float & The boiler's target temperature\\ \hline
nodata & Boolean & Indicates invalid or missing data  \\ \hline
time & DateTime & Time of measurement  \\ \hline
\end{tabular}
\caption{Indoor, Outdoor and Gas Boiler Features}
\label{table:dataset}
\end{table}

The raw data is captured using various sampling frequencies by the sensors and may also contain noise, thus a series of simple pre-processing steps are necessary to prepare the data for further processing. Specifically, after identifying temporal windows with missing data (e.g., when network is down), we concluded that resampling all measurements to 1-minute intervals is sufficient in order to capture informative variance in the temporal domain whilst eliminating some of the high frequency-noise content. For the purposes of this case study, data captured from 29 households during the heating season from October 2022 to April 2023 are utilized. Furthermore, a visual representation of a typical residential household e.g. a 24-hour period of the indoor and the boiler temperature, as well as the modulation level are as illustrated in \ref{fig:dataset}.

\begin{figure}[ht!]
    \centering
    \begin{subfigure}[b]{1\linewidth}
        \centering
        \includegraphics[width=.8\linewidth]{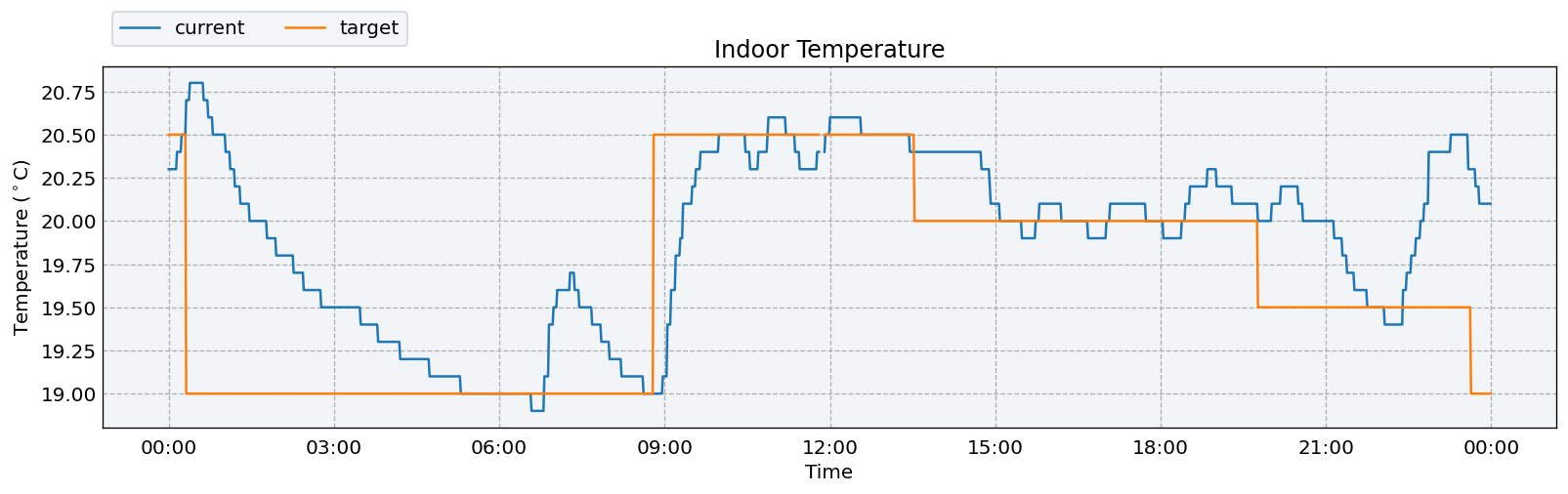}
        \caption{Comparison between indoor and target indoor temperature}
        \label{fig:dataset_indoor}
    \end{subfigure}
    \\
    \begin{subfigure}[b]{1\linewidth}
        \centering
        \includegraphics[width=.8\linewidth]{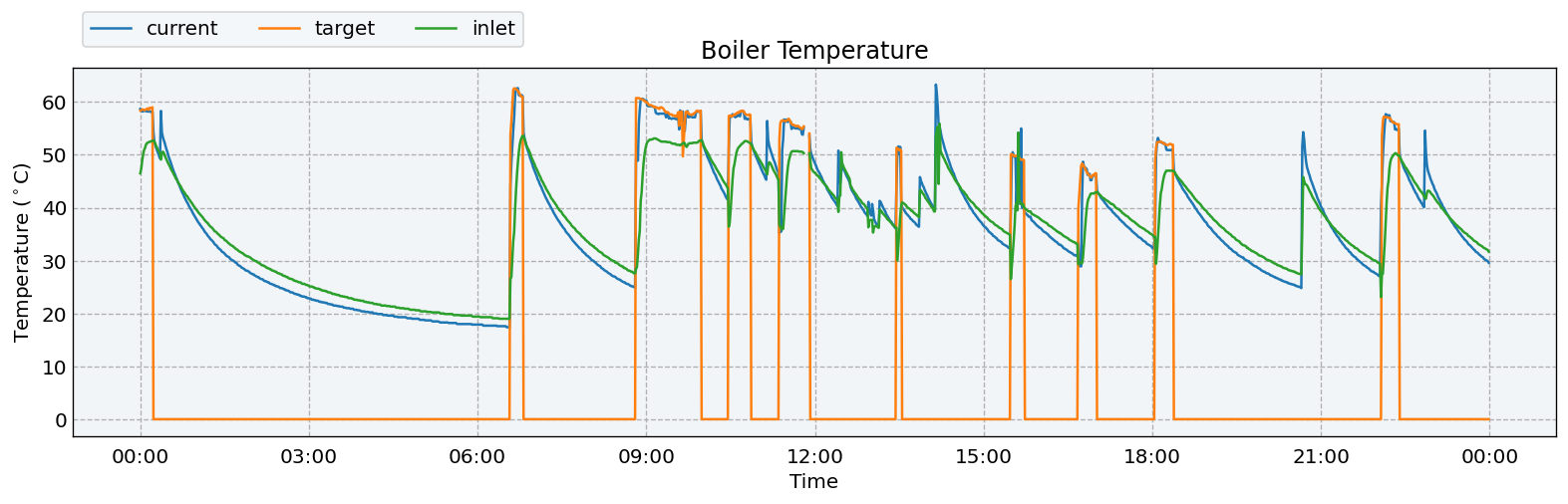}
        \caption{Comparison between boiler, target boiler and returning water temperature}
        \label{fig:dataset_boiler}
    \end{subfigure}
    \\
    \begin{subfigure}[b]{1\linewidth}
        \centering
        \includegraphics[width=.8\linewidth]{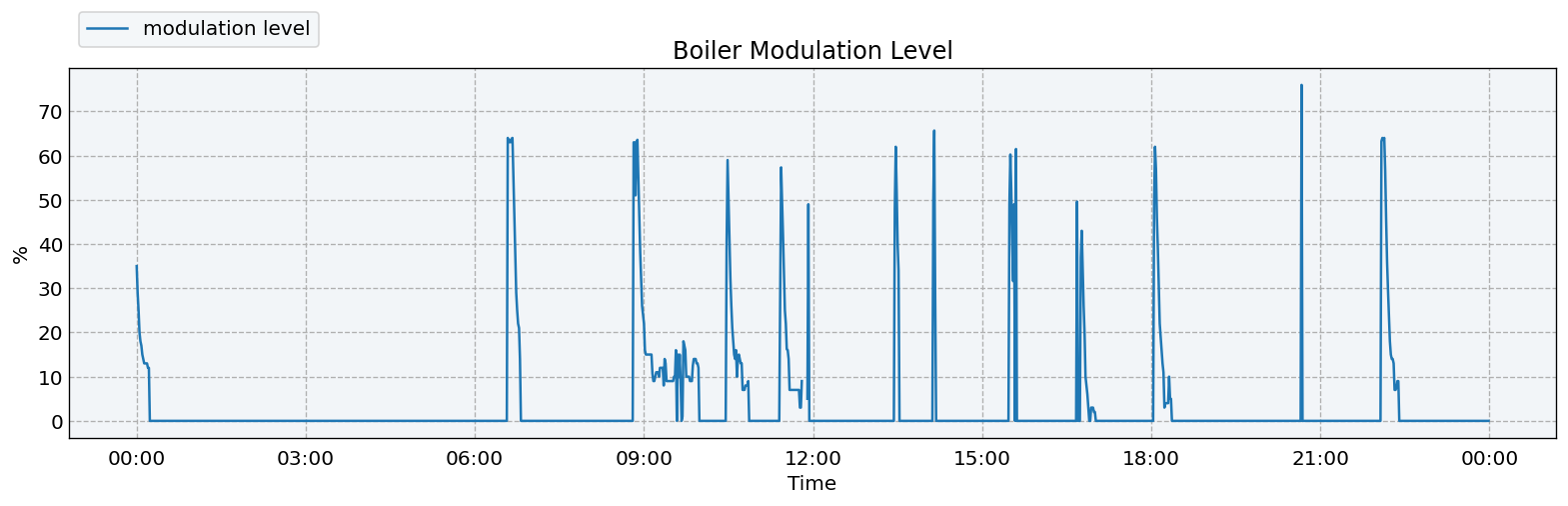}
        \caption{Boiler modulation level}
        \label{fig:dataset_modulation}
    \end{subfigure}
    \caption{Data from a typical residential heating system during a day}
    \label{fig:dataset}
\end{figure}

Although pre-processing steps were initially conducted to organize the dataset after receiving raw data from the IoT sensors, additional preparation was required for our analysis. Delving deeper into the process, we implemented specific filters to fine-tune the values and eliminate outliers likely stemming from sensor malfunction or miscalculation. Concerning outdoor temperature, any values exceeding 25C$^{\circ}$ were omitted, as such temperatures are not typical during winter months. Additionally, for the target indoor temperature set by users, instances where the boiler was inactive resulted in inaccurate measurements with excessively low values. These were consequently filtered out and substituted with the lowest temperature recorded while the boiler was operational.

\subsection{Clustering results}
\label{subsec:clustering_results}

We tested two heterogeneous clustering algorithms, the K-means algorithm and the HAC algorithm employing three distance metrics: ED, DTW, and DDTW which allowed the experiments to capture sensitivity to sequence shape, temporal alignment, and robustness to data variability. The evaluation was based on three well-known clustering indices SIL, CHI, and DBI to comprehensively assess cluster quality and separation. The algorithms have been assessed in scenarios that span between $2 \leq k \leq 10 $, where k is the number of clusters. It is worth noting that after careful examination of the clustering results obtained for all three distance metrics DTW, DDTW, and ED, it becomes evident that the optimal number of clusters is three. Across multiple iterations and dimensions, the most common scenario was that the three-cluster configuration consistently yielded the highest metric scores, demonstrating robustness and reliability. Notably, while the two-cluster configuration occasionally exhibited superior scores, further analysis revealed inherent drawbacks. More specifically, in the two-cluster scenario, one cluster often emerged significantly smaller or even remain totally empty. This deficiency persisted across all distance calculation methods, including DTW, DDTW, and ED. Such occurrences hindered our ability to discern subtle intricacies within the dataset, highlighting the importance of selecting a cluster configuration that not only optimizes scoring metrics but also ensures meaningful and trustworthy cluster formations. Therefore, the final range of cluster numbers to be presented is $3 \leq k \leq 10$.

An additional noteworthy aspect is that HAC results will not be reported for the DDTW metric. This decision was based on the observation that even with an increase in the number of clusters, the resulting clusters remained minimal or empty across all dimensions. This phenomenon suggests that the inherent characteristics of DDTW, likely related to its distance calculation methodology, hinder the effective formation of substantial clusters through HAC. Consequently, only the K-means algorithm was explored to accommodate the unique properties of DDTW. The clustering results for each facet of the problem -namely boiler, heat demand, temperature, building and user dimension - will be presented separately as different scenarios.

\subsubsection{Scenario I : Boiler dimension}

The first scenario involves the assessment of the variation in boiler and return water temperature differences for each house over the course of a day, as illustrated in Figure \ref{fig:appendix1}. The hours of the day are represented as the first minute of each hour, showing that the data are given in such a frequency. It provides insights into the boiler profiles of the heating systems in different households, indicating fluctuations in the thermal gradients of the boilers throughout the day. This analysis shows the difference between the different systems located in each household and the base that the algorithms are going to build their clustering approach on.

In Figure \ref{fig:boiler_sil} the SIL score for every number of clusters and algorithm for the three distance metrics is depicted. As can be seen from the plots, the number three is best number of clusters for this particular case achieving high SIL scores. For DTW, the number four is quite close to the previous but not exceeding it. Table \ref{tab:cluster_metrics} displays the SIL, DBI, and CHI scores for each scenario, with the cluster number set to three based on the fact that this is the optimal number of clusters identified by all metrics as stated in \ref{subsec:clustering_results}. It's important to highlight that the highest score is desirable for both SIL and CHI, indicating better clustering quality. However, for DBI, lower scores are preferred, signifying better cluster separation. Notably, in the provided table, the values for each evaluation metric are optimal for DTW and they are the same for both algorithms, showcasing that in this case both algorithms produces the exact same labels. 

The clustering algorithms produced identical labels for both ED and DTW metrics. This suggests that they achieved the same clustering results regardless of the algorithm used. However, the evaluation metric scores were notably higher for DTW. This indicates that the plain time-series data exhibited clear and distinct patterns, requiring no additional features for separation. While the results from the ED approach were promising with the inclusion of extra features, it's important to recognize that these outcomes might not be as accurate without those additional features. The effectiveness of the ED approach could vary depending on the complexity of the dataset and the nature of the underlying patterns.

\begin{figure*}[h!] 
\centering
 \makebox[\textwidth]{\includegraphics[width=.7\paperwidth]{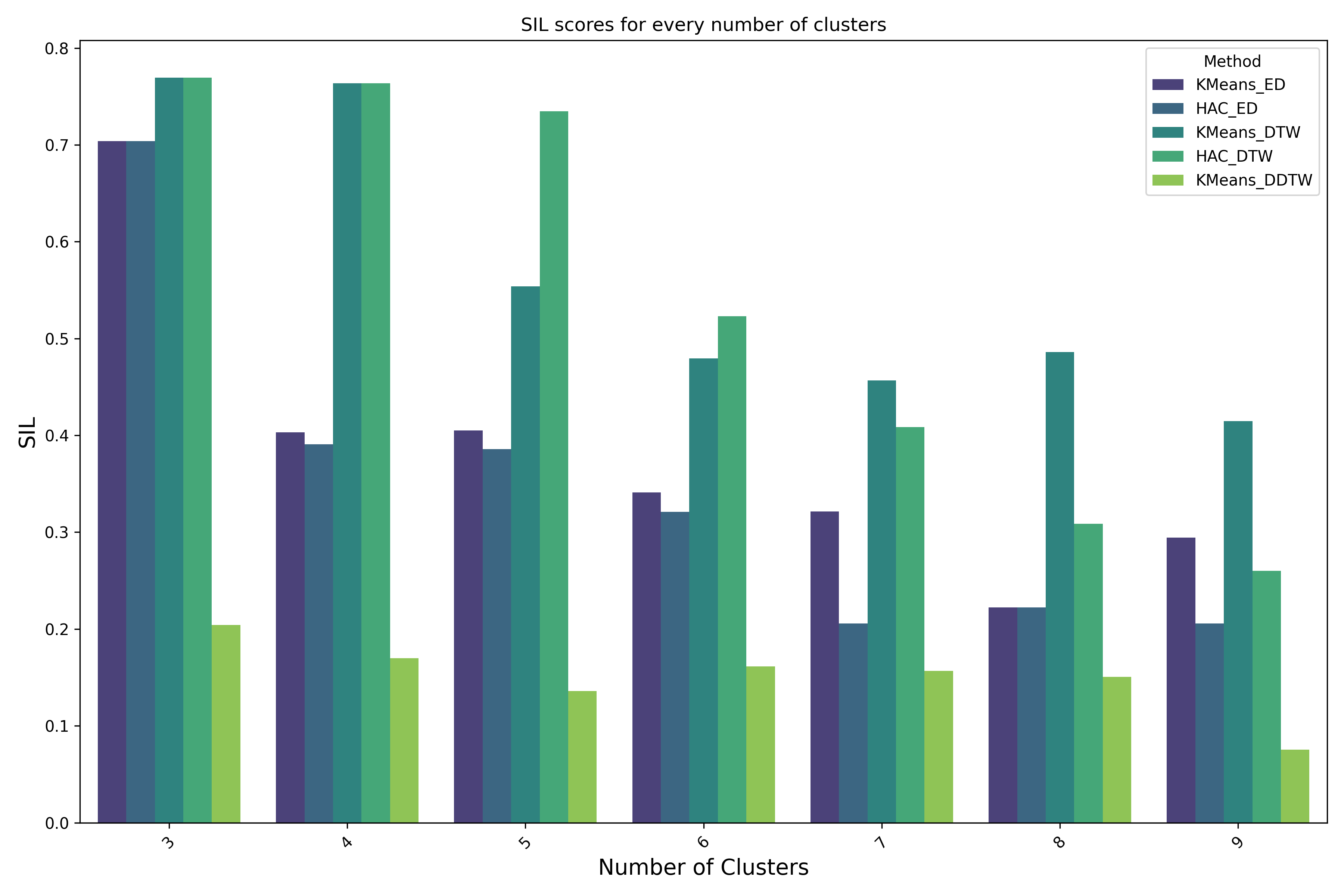}}
\caption{IL score for every number of clusters and algorithm for the three distance metrics in the boiler dimension}
\label{fig:boiler_sil}
\end{figure*}

\begin{table}[h!]
\centering
\caption{Cluster evaluation metrics for the boiler dimension}
\label{tab:cluster_metrics}
\begin{tabular}{|c|c|c|c|}
\hline
Scenario & Silhouette & DBI & CHI \\ \hline
K-Means with ED & 0.703640 & 0.284858 & 146.516525 \\ \hline
HAC with ED & 0.703640 & 0.284858 & 146.516525 \\ \hline
\textbf{K-Means with DTW} & \textbf{0.769452} & \textbf{0.178791} & \textbf{197.323287} \\ \hline
\textbf{HAC with DTW} & \textbf{0.769452} & \textbf{0.178791} & \textbf{197.323287} \\ \hline
K-Means with DDTW & 0.204362 & 1.565751 & 11.503813 \\ \hline
\end{tabular}
\end{table}

As previously noted, the labels between ED and DTW are exactly the same, highlighting their high metric scores. Table \ref{tab:boiler_euclidean_ddtw_matrix} illustrates the differences in labels between those selected from ED and DTW, compared with the corresponding ones chosen by DDTW. Upon examination, it's evident that the labels agree only for cluster 1. Conversely, for clusters 0 and 3, the DDTW approach results in a larger separation of houses between them compared to DTW and ED. Despite the clear indications provided by the evaluation metric results regarding the effectiveness of the two sets of labels, we integrate an additional visual analysis step to assess the clustering outcomes. In Figure \ref{fig:boiler_pca}, we present the results of Principal Component Analysis (PCA) \cite{abdi2010principal} applied to the boiler dataset. Each subfigure represents the PCA analysis for a specific distance metric: DTW in Figure \ref{fig:boiler_dtw_pca} and DDTW in Figure \ref{fig:boiler_ddtw_pca}. In both subfigures, the data points are represented in a two-dimensional space, where the axes denote the principal components obtained from PCA. The x-axis represents the first principal component (PC1), capturing the direction of maximum variance in the data. Conversely, the y-axis represents the second principal component (PC2), capturing the second most significant source of variance orthogonal to PC1 \cite{paul2013methodological}. Each data point is colored according to its assigned cluster label, allowing us to visualize the distribution of clusters in the reduced feature space.

\begin{table}[h!]
\centering
\caption{ED-DTW vs. DDTW}
\label{tab:boiler_euclidean_ddtw_matrix}
\begin{tabular}{c|ccc}
\hline
&0 & 1 & 2 \\ \hline
0 & 8 & 1 & 14 \\
1 & 0 & 1 & 0 \\
2 & 1 & 0 & 3 \\ \hline
\end{tabular}
\end{table}

\begin{figure}[htbp]
  \centering
  \begin{subfigure}[b]{0.4\textwidth}
    \includegraphics[width=\textwidth]{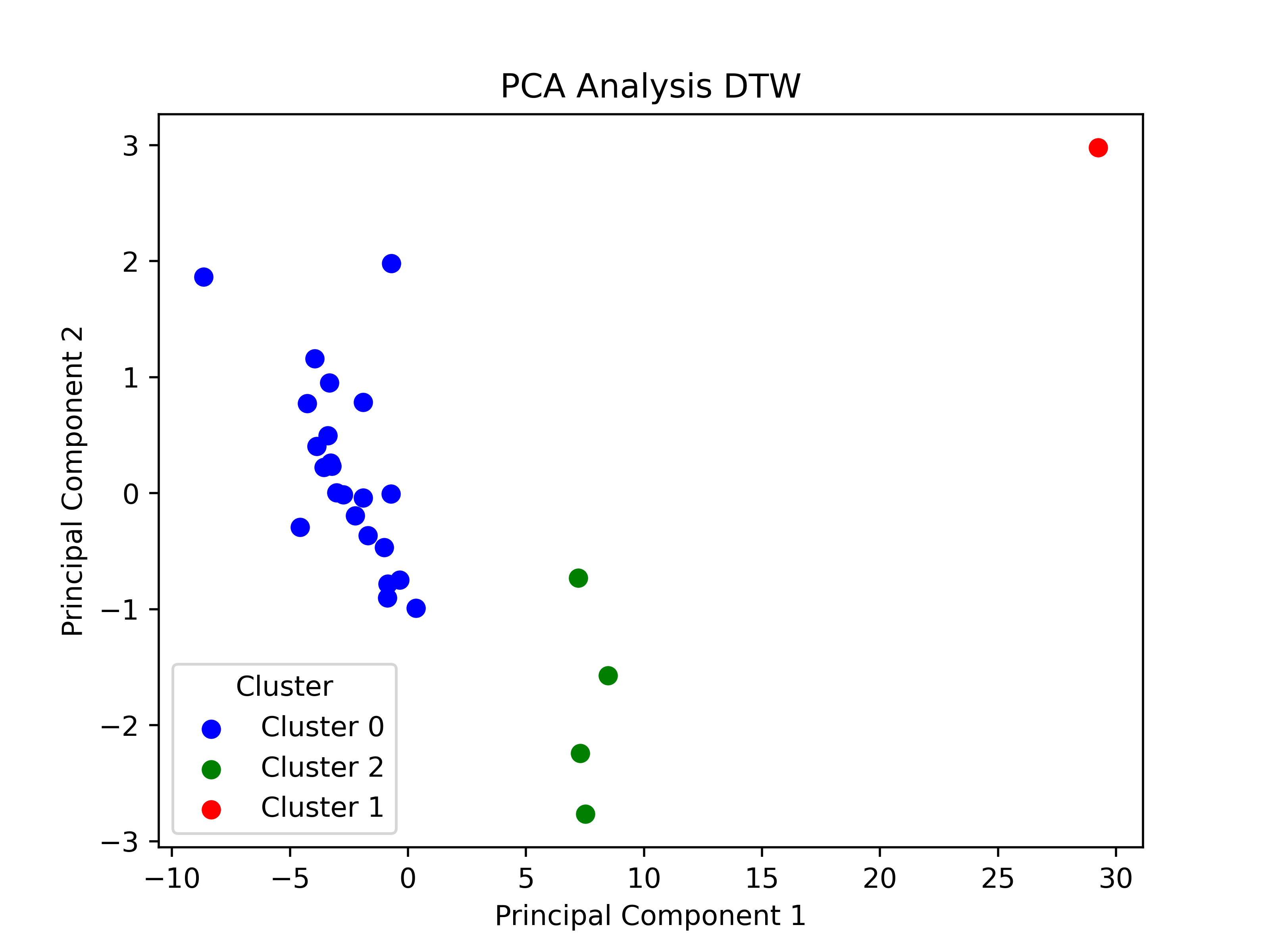}
    \caption{PCA Analysis - DTW}
    \label{fig:boiler_dtw_pca}
  \end{subfigure}
  \hfill
  \begin{subfigure}[b]{0.4\textwidth}
    \includegraphics[width=\textwidth]{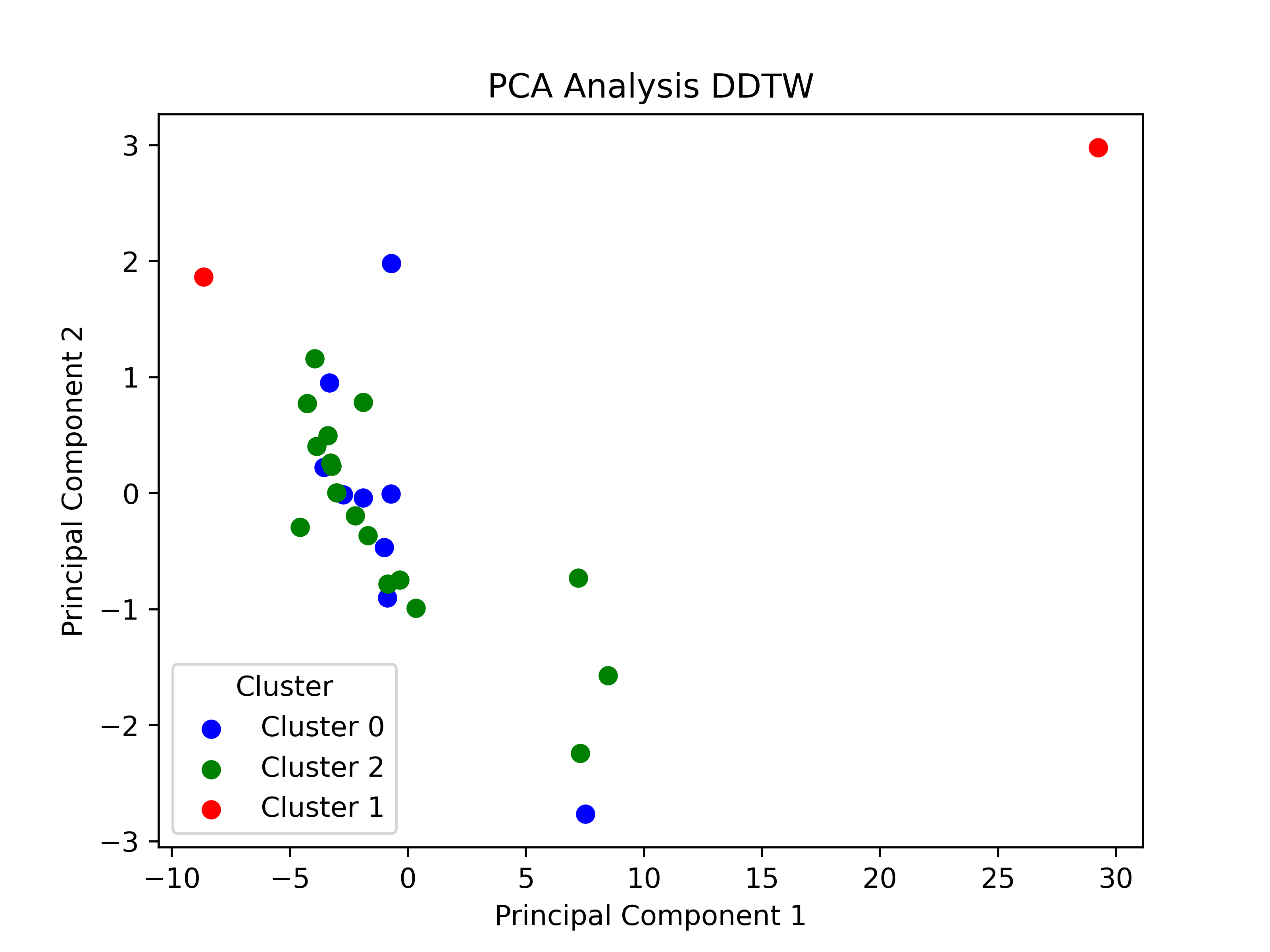}
    \caption{PCA Analysis - DDTW}
    \label{fig:boiler_ddtw_pca}
  \end{subfigure}
  \caption{PCA Analysis in the boiler dimension}
  \label{fig:boiler_pca}
\end{figure}

Concluding the analysis for the boiler dimension, according to the results from Table \ref{tab:cluster_metrics} and the PCA plots in Figure \ref{fig:boiler_pca} it is clear that the labels from produced from ED and DTW are better defined in terms of cluster similarity and dissimilarity. Thus, in Figure \ref{fig:cluster_diagrams} a visualization that showcases all clusters individually in separate sub-figures, accompanied by vital statistical metrics such as mean, and median is depicted. This unified representation provides a holistic perspective of the clustering outcomes, facilitating a deeper comprehension of each cluster's characteristics. The mean serves as a measure of central tendency, representing the average behavior of data points within a cluster, while the median provides insights into the central point of the distribution, serving as the basis for center-based clustering algorithms. 

\begin{figure}[h!]
  \centering
  \begin{subfigure}[b]{0.3\textwidth}
    \includegraphics[width=\textwidth]{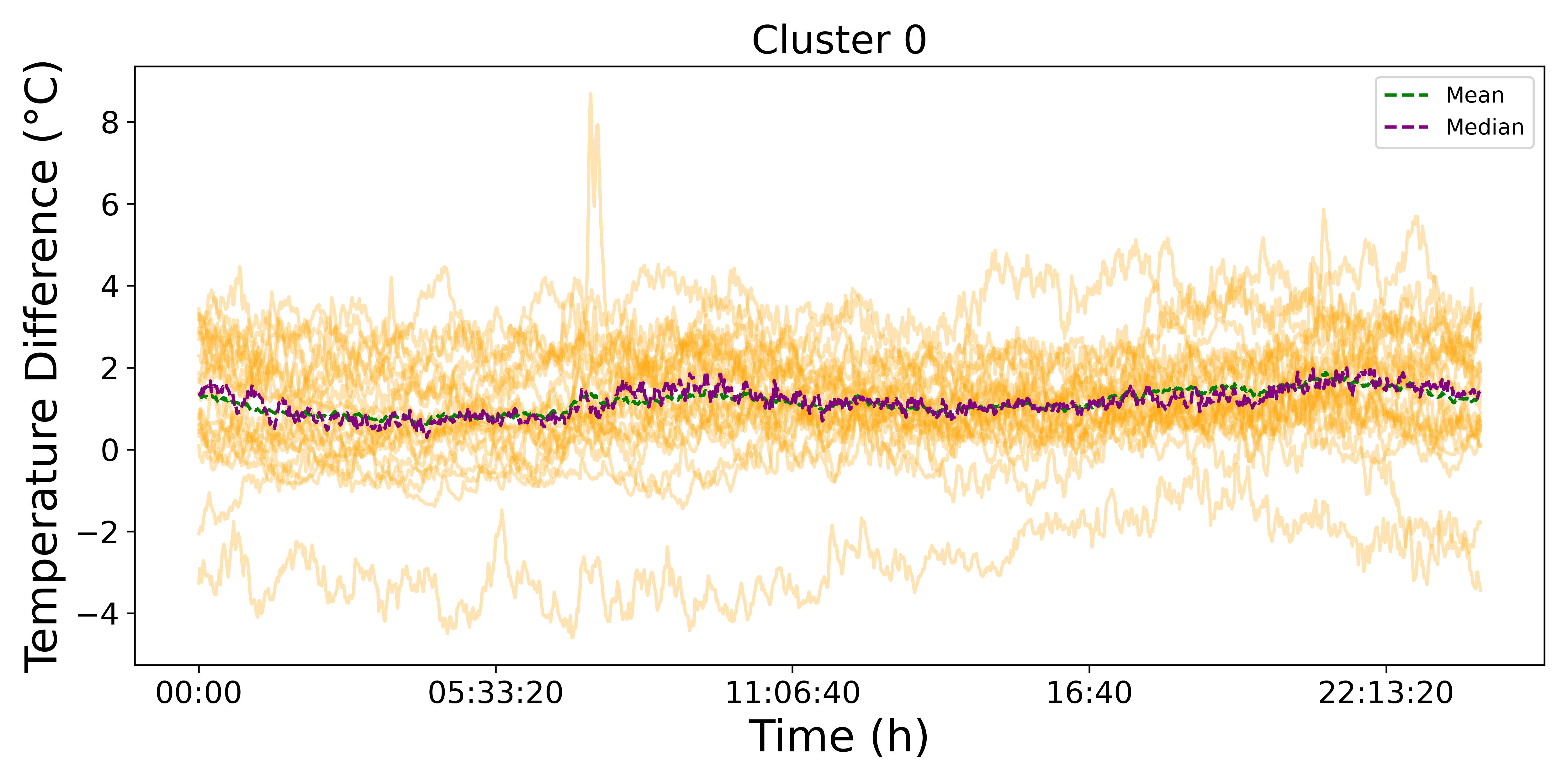}
    \caption{Cluster 0}
    \label{fig:cluster_0}
  \end{subfigure}
  \begin{subfigure}[b]{0.3\textwidth}
    \includegraphics[width=\textwidth]{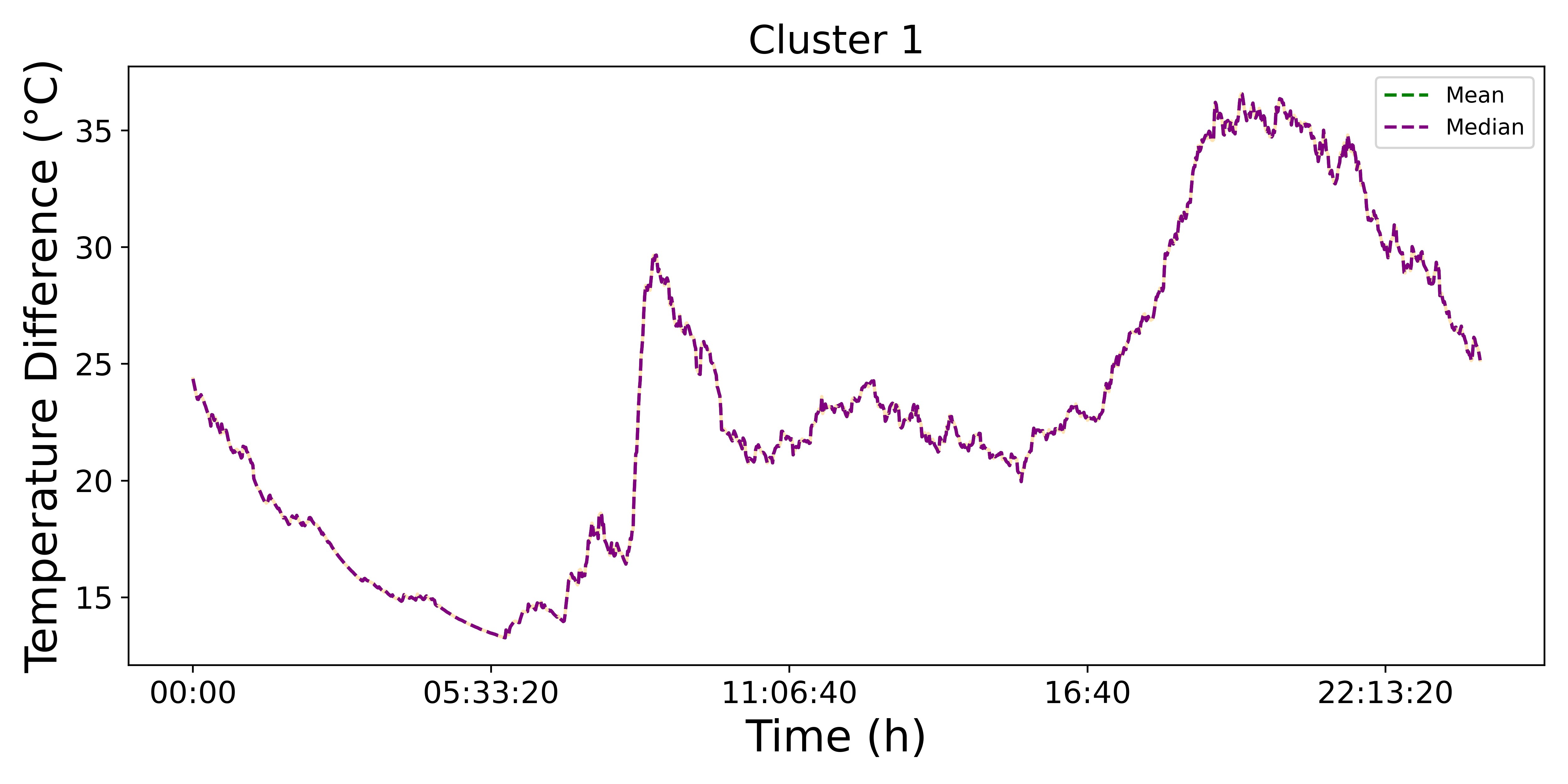}
    \caption{Cluster 1}
    \label{fig:cluster_1}
  \end{subfigure}
  \begin{subfigure}[b]{0.3\textwidth}
    \includegraphics[width=\textwidth]{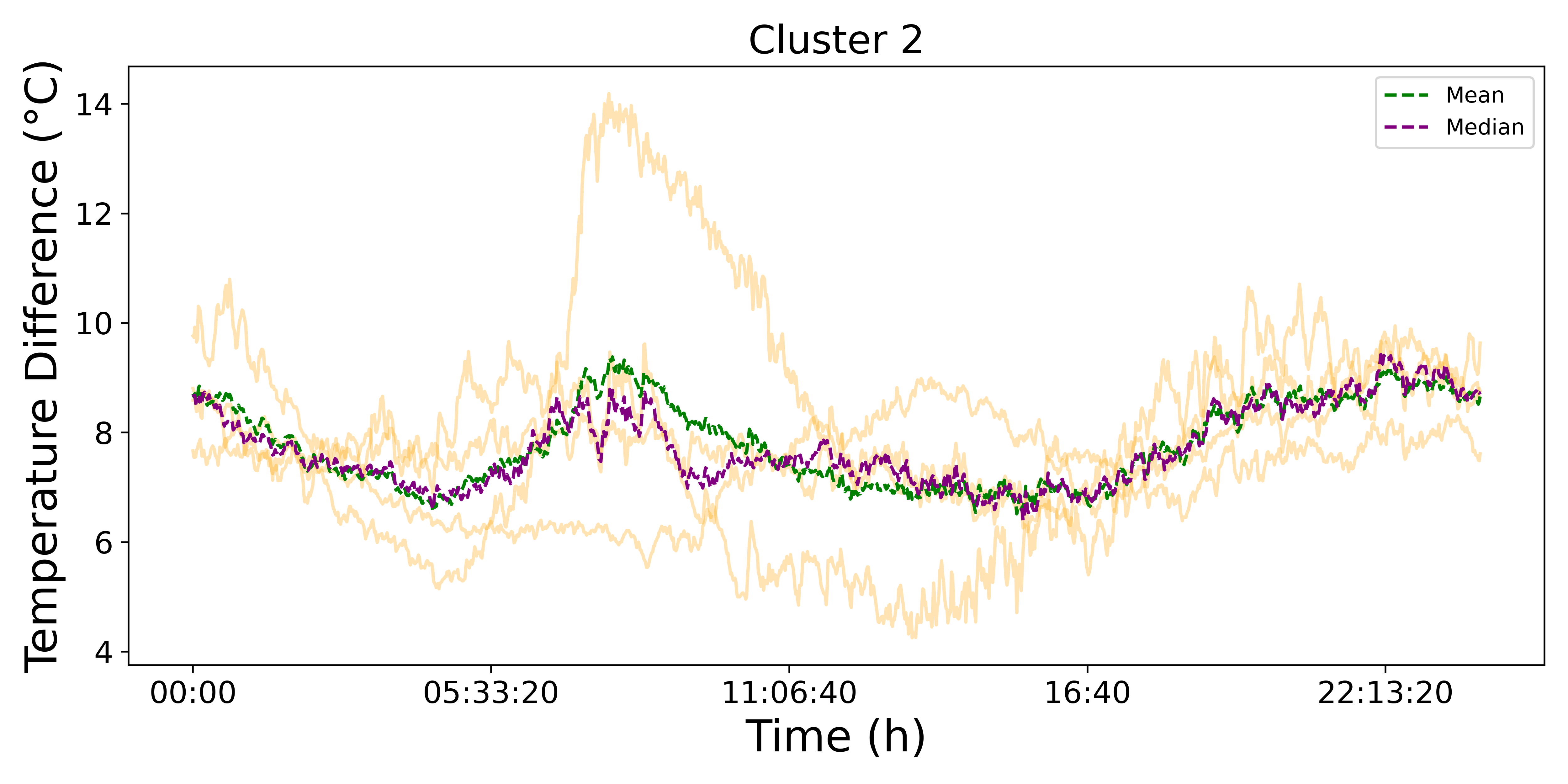}
    \caption{Cluster 2}
    \label{fig:cluster_2}
  \end{subfigure}
  \caption{Clusters based on K-means and DTW for the boiler dimension}
  \label{fig:cluster_diagrams}
\end{figure}

\subsubsection{Scenario II: Heat demand dimension}

The plot showcased in Figure \ref{fig:modulation_level} demonstrates the variability in boiler modulation levels across different households throughout the day, represented as the start of each hour. This visualization offers crucial insights into the modulation behavior of the specific heating system installed in each house, revealing swings in heat demand expressed as percentages of the modulation usage.

The SIL score for every number of clusters and algorithm utilized for the three distances metrics is shown in Figure \ref{fig:heating_sil}. As we have mentioned earlier the HAC algorithm is excluded from further analysis for DDTW, but in this case DTW also leads to the formation of empty clusters with HAC and thus it has been excluded as well from the Figure. In all the presented bars, except for the one depicting K-means with ED, the ideal number of clusters is three. Given the similarity in results between three and four clusters in this particular case, we opt for three clusters for the sake of analysis. Additionally, the choice of three clusters is preferred over four because the fourth cluster comprises only one house. Table \ref{tab:cluster_metrics_modulation}, showcases the evaluation metrics scores for each scenario chosen, with the cluster number designated as three. Remarkably, the values for each evaluation metric in the provided table are optimal for DTW. The evaluation metrics between K-means and HAC for ED are different due to the fact that two labels were different between the two algorithms. HAC labels are selected for further analysis due to their superior performance across all evaluation metrics.

\begin{figure*}[h!] 
\centering
 \makebox[\textwidth]{\includegraphics[width=.8\paperwidth]{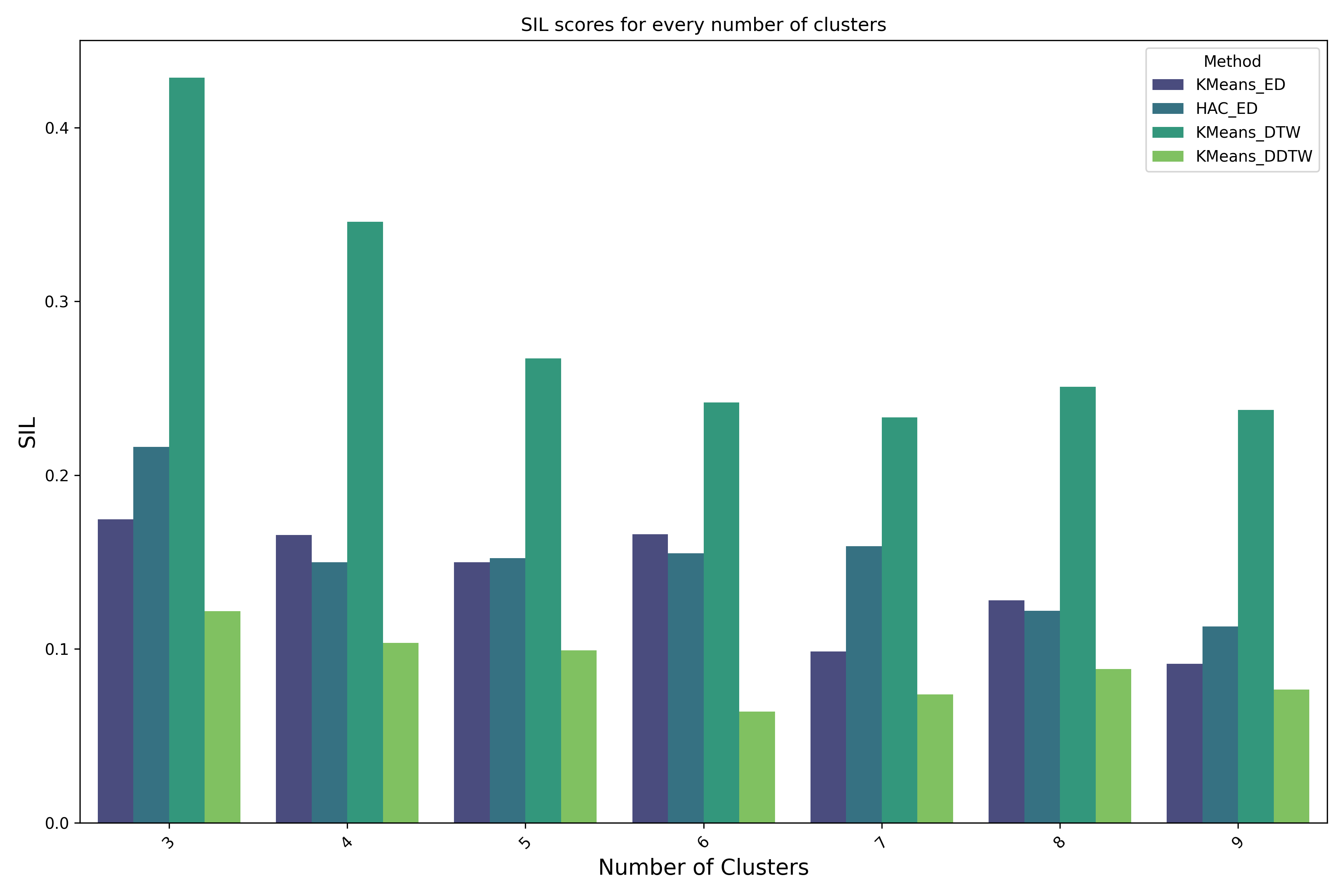}}
\caption{IL score for every number of clusters and algorithm for the three distance metrics in the heating dimension}
\label{fig:heating_sil}
\end{figure*}

\begin{table}[h!]
\centering
\caption{Cluster evaluation metrics for the heat demand dimension}
\label{tab:cluster_metrics_modulation}
\begin{tabular}{|l|c|c|c|}
\hline
Algorithm & Silhouette & DBI & CHI \\ \hline
K-Means with ED & 0.174537 & 1.536718 & 9.738654 \\ \hline
HAC with ED & 0.216338 & 1.497139 & 9.249262 \\ \hline
\textbf{K-Means with DTW} & \textbf{0.428794} & \textbf{0.759205} & \textbf{34.586499} \\ \hline
K-Means with DDTW & 0.121640 & 1.693275 & 4.394497 \\ \hline
\end{tabular}
\end{table}
 
Since the labels between the different distance metrics were different, Table \ref{tab:comparison_matrices_modulation} indicates the similarity matrices between them. This label analysis shows that the corresponding labels selected from ED, DTW and DDTW reveal notable differences. In \ref{subtab:edvsdtw} a substantial similarity in clustering labels between ED and DTW is observed, evidenced by the non-zero elements along the diagonal and upper triangle. However, discrepancies arise, notably in clusters 0 and 2, where the clustering labels diverge. Similarly, \ref{subtab:edvsddtw} reveals a comparison between trends, with close to zero elements concentrated along the diagonal and upper triangle, again, clusters 0 and 2 exhibit differences in labeling between the two metrics. Transitioning to the \ref{subtab:dtwvsddtw} significant disparities in clustering labels between DTW and DDTW are noted, particularly in cluster 1, highlighting distinct clustering patterns captured by each approach.

Figure \ref{fig:heat_demand_pca} illustrates the PCA analysis applied to the heat demand dataset. As depicted in Figure \ref{fig:pca_ddtw_modulation}, the PCA analysis aligns with previous findings, indicating that DDTW does not correspond well to clustering. Turning our attention to Figures \ref{fig:pca_ed_modulation} and \ref{fig:pca_dtw_modulation}, representing ED and DTW respectively, we observe distinct cluster separations based on different criteria for each method. However, DTW exhibits a clearer separation in the first principal component, suggesting superior performance in cluster delineation. In summary, considering the results presented in Table \ref{tab:cluster_metrics_modulation} alongside these figures, the DTW approach emerges as the optimal solution for this problem. In the visualizations of each cluster in Figure \ref{fig:clusters_modulation}, we gain a detailed insight into how the data points are grouped together based on their similarity in heat demand patterns, employing both K-means clustering and the DTW similarity measure.

\begin{table}[h!]
\centering
\caption{Similarity matrices for the heat demand dimension}
\label{tab:comparison_matrices_modulation}
\begin{subtable}{0.3\linewidth}
\centering
\caption{ED vs. DTW Matrix}
\label{subtab:edvsdtw}
\begin{tabular}{c|ccc}
\hline
& 0 & 1 & 2 \\ \hline
0 & 7 & 0 & 3 \\
1 & 6 & 9 & 0 \\
2 & 1 & 0 & 2 \\
\hline
\end{tabular}
\end{subtable}
\begin{subtable}{0.3\linewidth}
\centering
\caption{ED vs. DDTW Matrix}
\label{subtab:edvsddtw}
\begin{tabular}{c|ccc}
\hline
& 0 & 1 & 2 \\ \hline
0 & 3 & 0 & 7 \\
1 & 11 & 1 & 3 \\
2 & 1 & 0 & 2 \\
\hline
\end{tabular}
\end{subtable}
\begin{subtable}{0.3\linewidth}
\centering
\caption{DTW vs. DDTW Matrix}
\label{subtab:dtwvsddtw}
\begin{tabular}{c|ccc}
\hline
& 0 & 1 & 2 \\ \hline
0 & 7 & 1 & 6 \\
1 & 8 & 0 & 1 \\
2 & 0 & 0 & 5 \\
\hline
\end{tabular}
\end{subtable}
\end{table}

\begin{figure}[ht!]
  \centering
  \begin{subfigure}[b]{0.3\textwidth}
    \includegraphics[width=\textwidth]{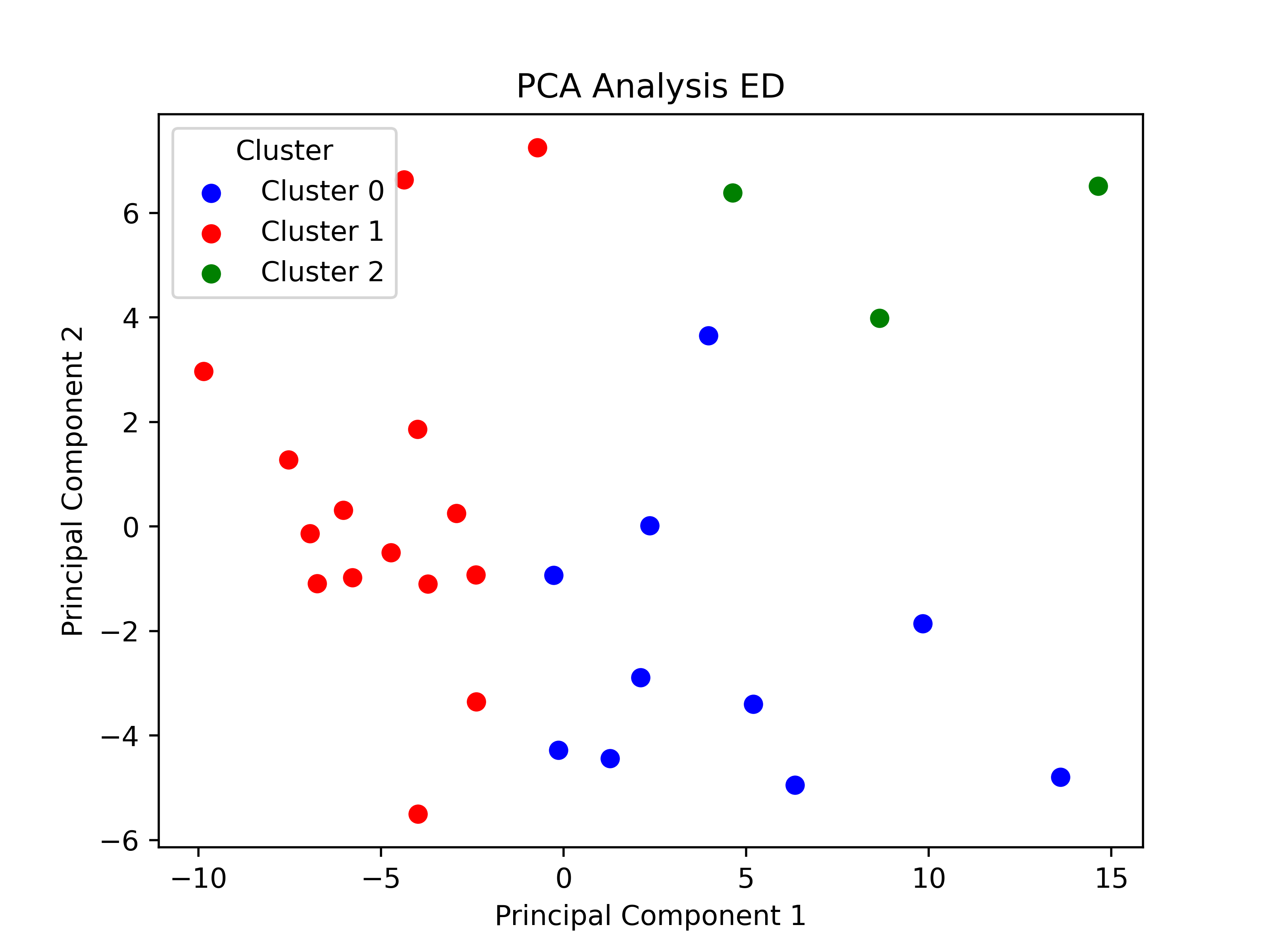}
    \caption{PCA Analysis - ED}
    \label{fig:pca_ed_modulation}
  \end{subfigure}
  \begin{subfigure}[b]{0.3\textwidth}
    \includegraphics[width=\textwidth]{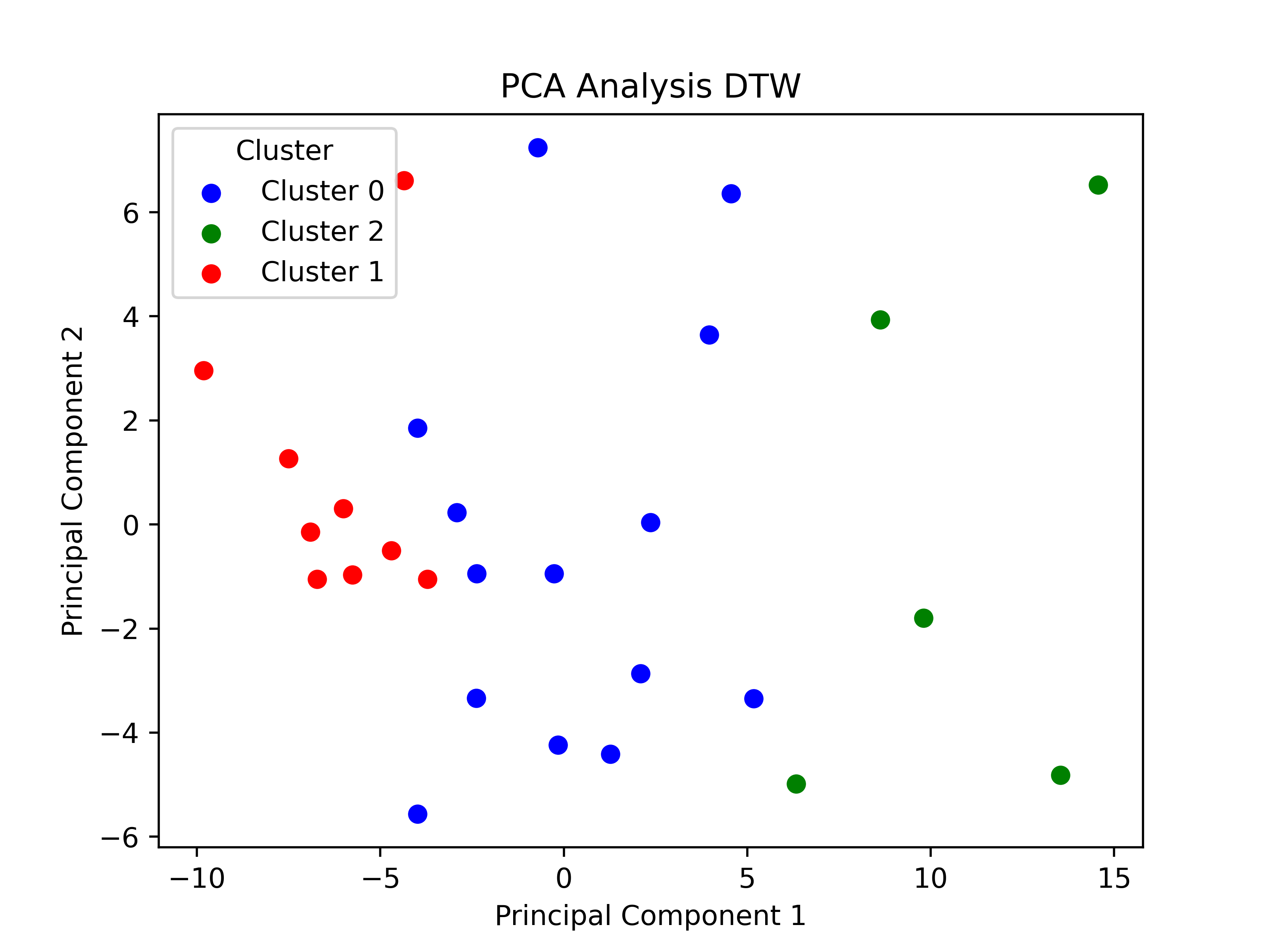}
    \caption{PCA Analysis - DTW}
    \label{fig:pca_dtw_modulation}
  \end{subfigure}  
  \begin{subfigure}[b]{0.3\textwidth}
    \includegraphics[width=\textwidth]{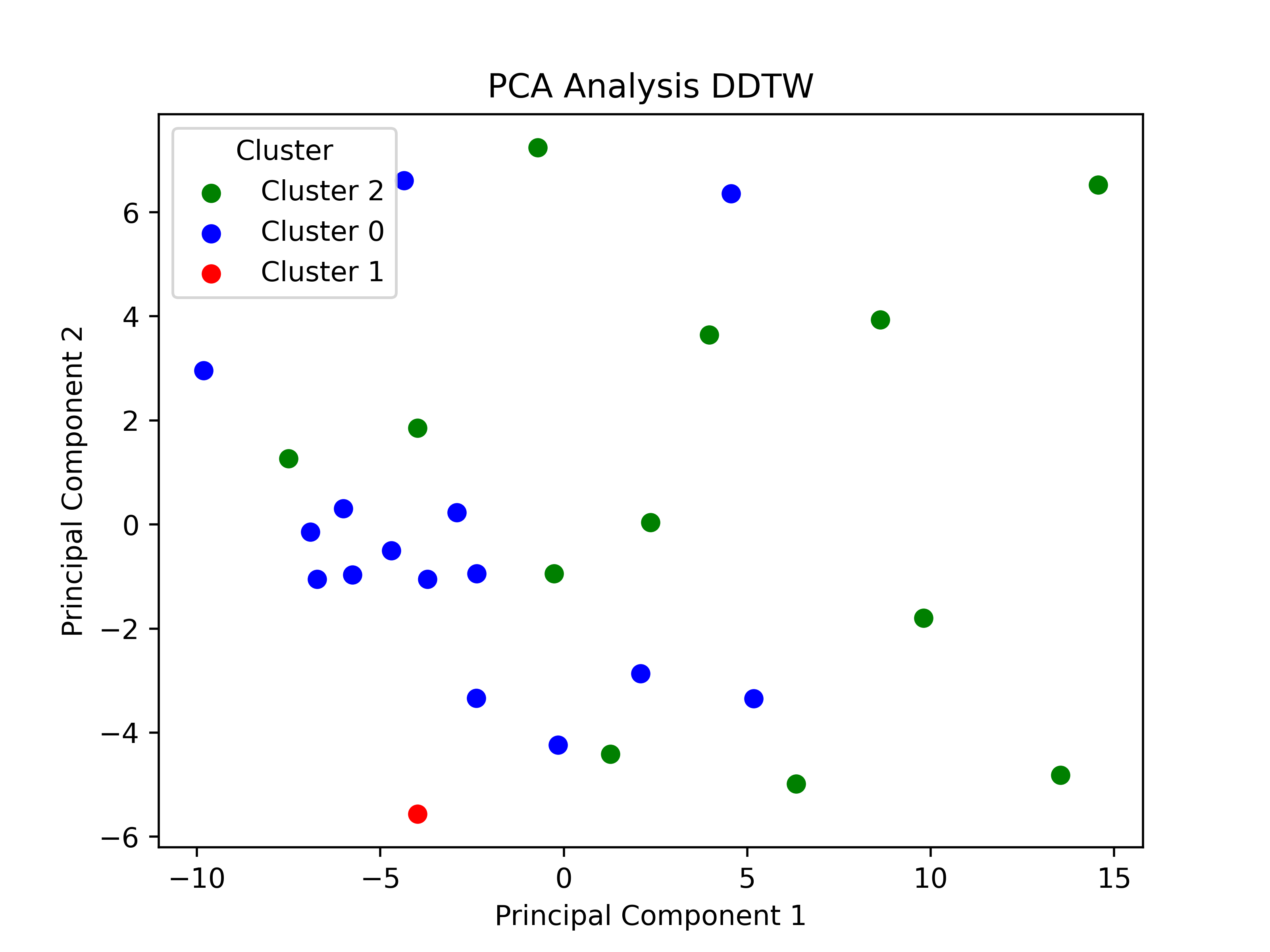}
    \caption{PCA Analysis - DDTW}
    \label{fig:pca_ddtw_modulation}
  \end{subfigure}
  \caption{PCA analysis in the heat demand dimension}
  \label{fig:heat_demand_pca}
\end{figure}

\begin{figure}[ht!]
  \centering
  \begin{subfigure}[b]{0.3\textwidth}
    \includegraphics[width=\textwidth]{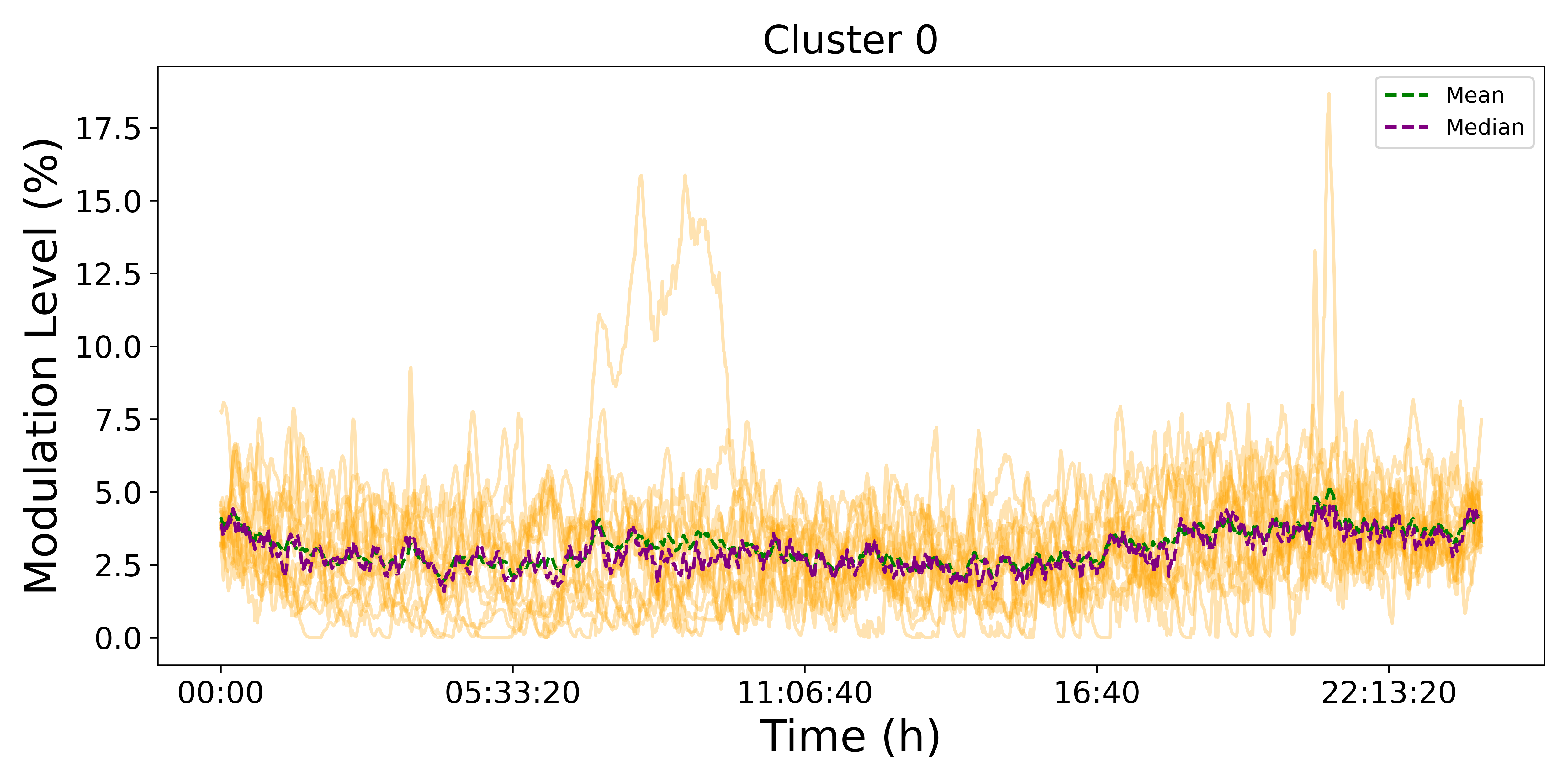}
    \caption{Cluster 0}
    \label{fig:cluster_0_modulation}
  \end{subfigure}
  \begin{subfigure}[b]{0.3\textwidth}
    \includegraphics[width=\textwidth]{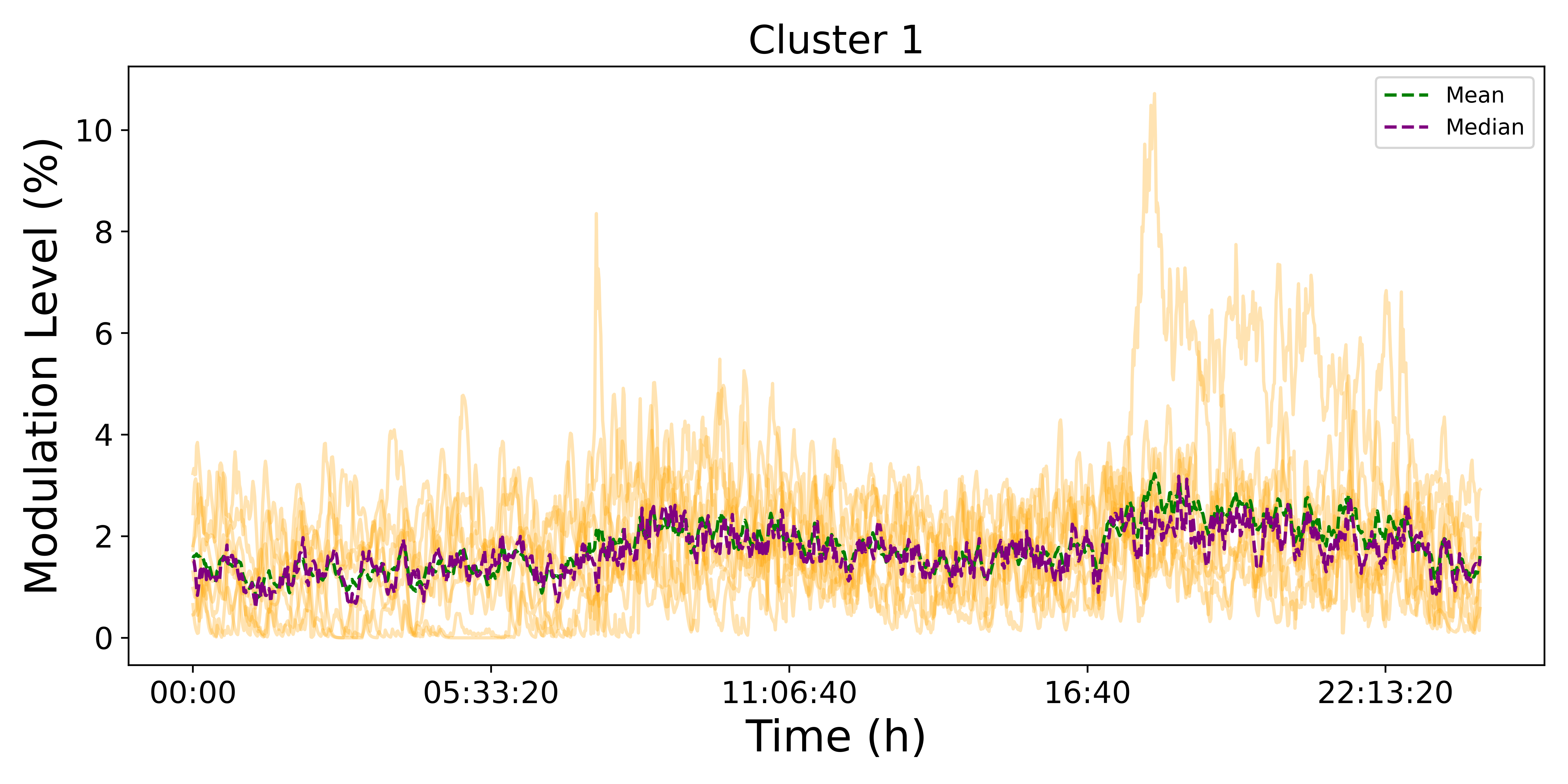}
    \caption{Cluster 1}
    \label{fig:cluster_1_modulation}
  \end{subfigure}
  \begin{subfigure}[b]{0.3\textwidth}
    \includegraphics[width=\textwidth]{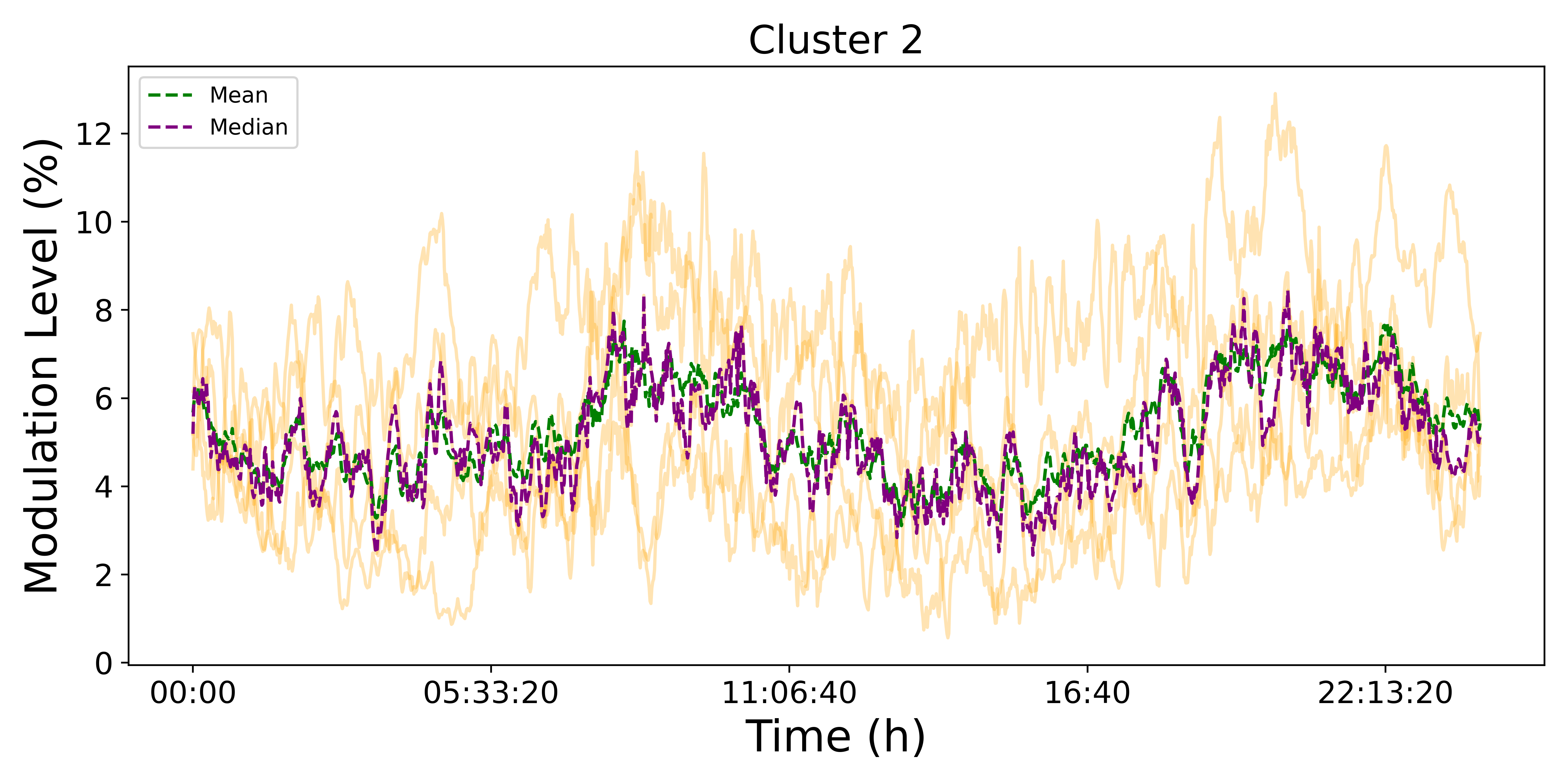}
    \caption{Cluster 2}
    \label{fig:cluster_2_modulation}
  \end{subfigure}
  \caption{Clusters based on K-means and DTW for the heat demand dimension}
  \label{fig:clusters_modulation}
\end{figure}

\subsubsection{Scenario III: Temperature dimension}

Figure \ref{fig:outdoor_temperature} illustrates the fluctuation in outdoor temperatures recorded by sensors across various households over the course of the day, with each data point corresponding to the beginning of every hour. It highlights notable patterns, such as temperature peaks occurring around midday and fluctuations throughout the day, all depicted in degrees Celsius.

In this case as well the resulting labels from K-means and HAC for both ED and DTW were exactly the same, so no further analysis between algorithms is necessary at this point. Figure \ref{fig:temperature_sil} displays the SIL scores for various clustering algorithms and distance metrics. Across all plots, the optimal number of clusters is three. Table \ref{tab:cluster_metrics_temperature} presents the evaluation metric scores for each scenario, with three clusters being the selected number. Interestingly, DTW consistently yields optimal values across all evaluation metrics for both K-means and HAC.

\begin{figure*}[h!] 
\centering
 \makebox[\textwidth]{\includegraphics[width=.8\paperwidth]{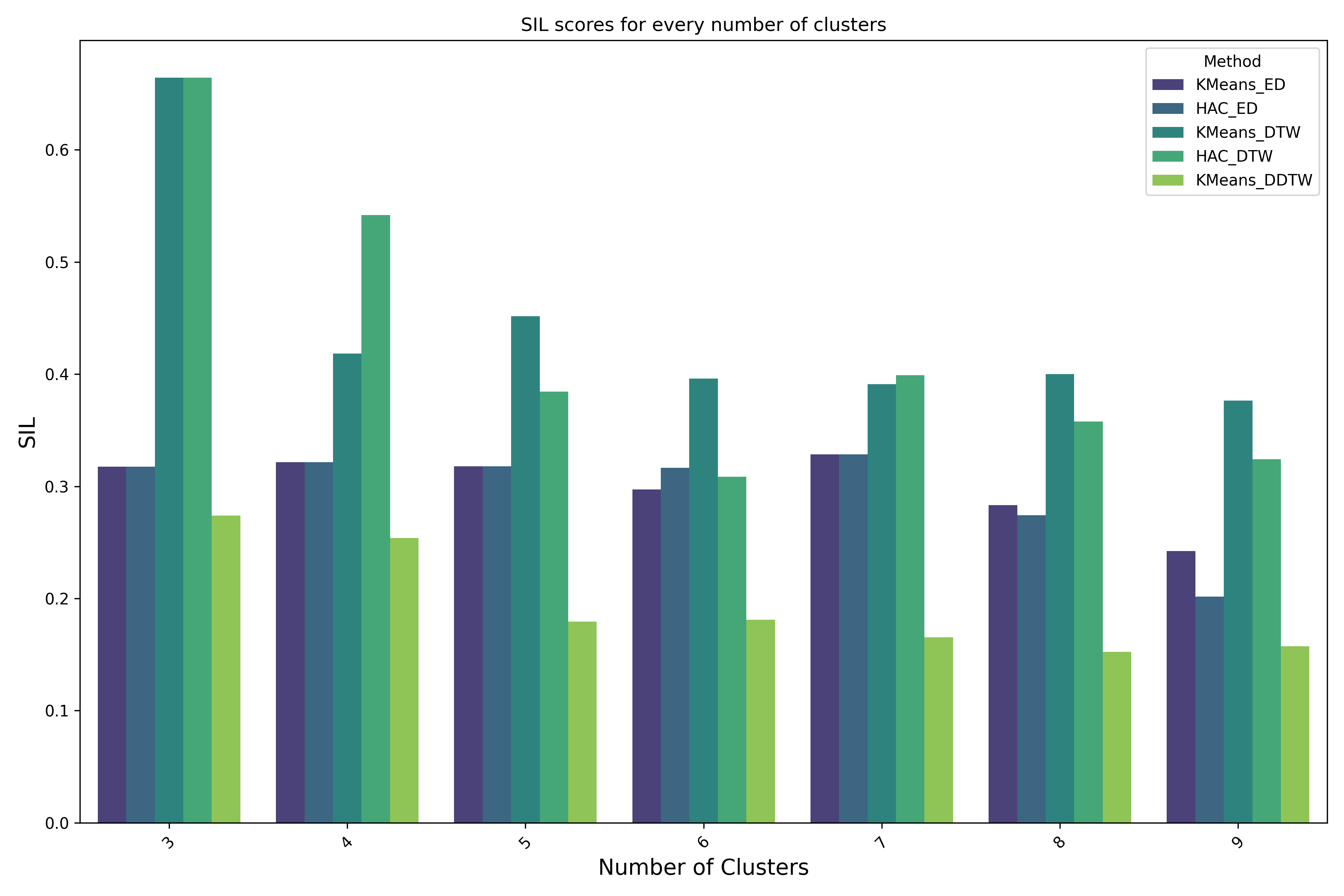}}
\caption{IL score for every number of clusters and algorithm for the three distance metrics in the temperature dimension}
\label{fig:temperature_sil}
\end{figure*}

\begin{table}[h!]
\centering
\caption{Cluster evaluation metrics for the temperature dimension}
\label{tab:cluster_metrics_temperature}
\begin{tabular}{|l|c|c|c|}
\hline
Algorithm & Silhouette & DBI & CHI \\ \hline
K-means with ED & 0.358195 & 0.818539 & 25.022211 \\ \hline
HAC with ED & 0.358195 & 0.818539 & 25.022211 \\ \hline
\textbf{K-means with DTW} & \textbf{0.674398} & \textbf{0.430472} & \textbf{33.784205} \\ \hline
\textbf{HAC with DTW} & \textbf{0.674398} & \textbf{0.430472} & \textbf{33.784205} \\ \hline
K-means with DDTW & 0.325206 & 1.168875 & 20.332338 \\ \hline
\end{tabular}
\end{table}

Table \ref{tab:similarity_matrices_temperature} provides insight into the differences in clustering labels across various distance metrics. Here as well, the comparison between labels from ED, DTW, and DDTW reveal substantial discrepancies. Subtable \ref{subtab:ed_vs_dtw_temp} highlights that there is a big chunk of houses that with ED were split into two different clusters in contrast with DTW that merged them into one single cluster, leaving cluster 0 with only one house. Similarly, Subtable \ref{subtab:dtw_vs_ddtw_temp} showcases minimal similarities between DTW and DDTW, with nearly zero elements along the diagonal and upper triangle, indicating divergent labeling, particularly in clusters 0 and 2. Transitioning to Subtable \ref{subtab:ed_vs_ddtw_temp}, significant disparities between ED and DDTW clustering labels are observed.

Figure \ref{fig:temperature_pca} presents the PCA analysis applied to the outdoor temperature dataset. As depicted in Figure \ref{fig:pca_ddtw_temperature}, the PCA analysis aligns with previous findings, indicating poor correspondence of DDTW with clustering. Examining Figures \ref{fig:pca_ed_temperature} and \ref{fig:pca_dtw_temperature}, representing ED and DTW respectively, distinct cluster separations based on different criteria for each method are observed. However, DTW exhibits clearer separation in the first principal component. The only concern raised is that in the case of ED the separation seems quite accurate while also assigning more than one value to cluster 0.

\begin{table}[h!]
\centering
\caption{Similarity matrices for the temperature dimension}
\label{tab:similarity_matrices_temperature}
\begin{subtable}{0.3\linewidth}
\centering
\caption{Euclidean vs. DTW Matrix}
\label{subtab:ed_vs_dtw_temp}
\begin{tabular}{c|ccc}
\hline
& 0 & 1 & 2 \\ \hline
0 & 0 & 12 & 1 \\
1 & 1 & 12 & 0 \\
2 & 0 & 0 & 2 \\
\hline
\end{tabular}
\end{subtable}
\begin{subtable}{0.3\linewidth}
\centering
\caption{Euclidean vs. DDTW Matrix}
\label{subtab:ed_vs_ddtw_temp}
\begin{tabular}{c|ccc}
\hline
& 0 & 1 & 2 \\ \hline
0 & 2 & 4 & 7 \\
1 & 0 & 7 & 6 \\
2 & 0 & 2 & 0 \\
\hline
\end{tabular}
\end{subtable}
\begin{subtable}{0.3\linewidth}
\centering
\caption{DTW vs. DDTW Matrix}
\label{subtab:dtw_vs_ddtw_temp}
\begin{tabular}{c|ccc}
\hline
& 0 & 1 & 2 \\ \hline
0 & 0 & 1 & 0 \\
1 & 2 & 9 & 13 \\
2 & 0 & 3 & 0 \\
\hline
\end{tabular}
\end{subtable}
\end{table}

\begin{figure}[ht!]
  \centering
  \begin{subfigure}[b]{0.3\textwidth}
    \includegraphics[width=\textwidth]{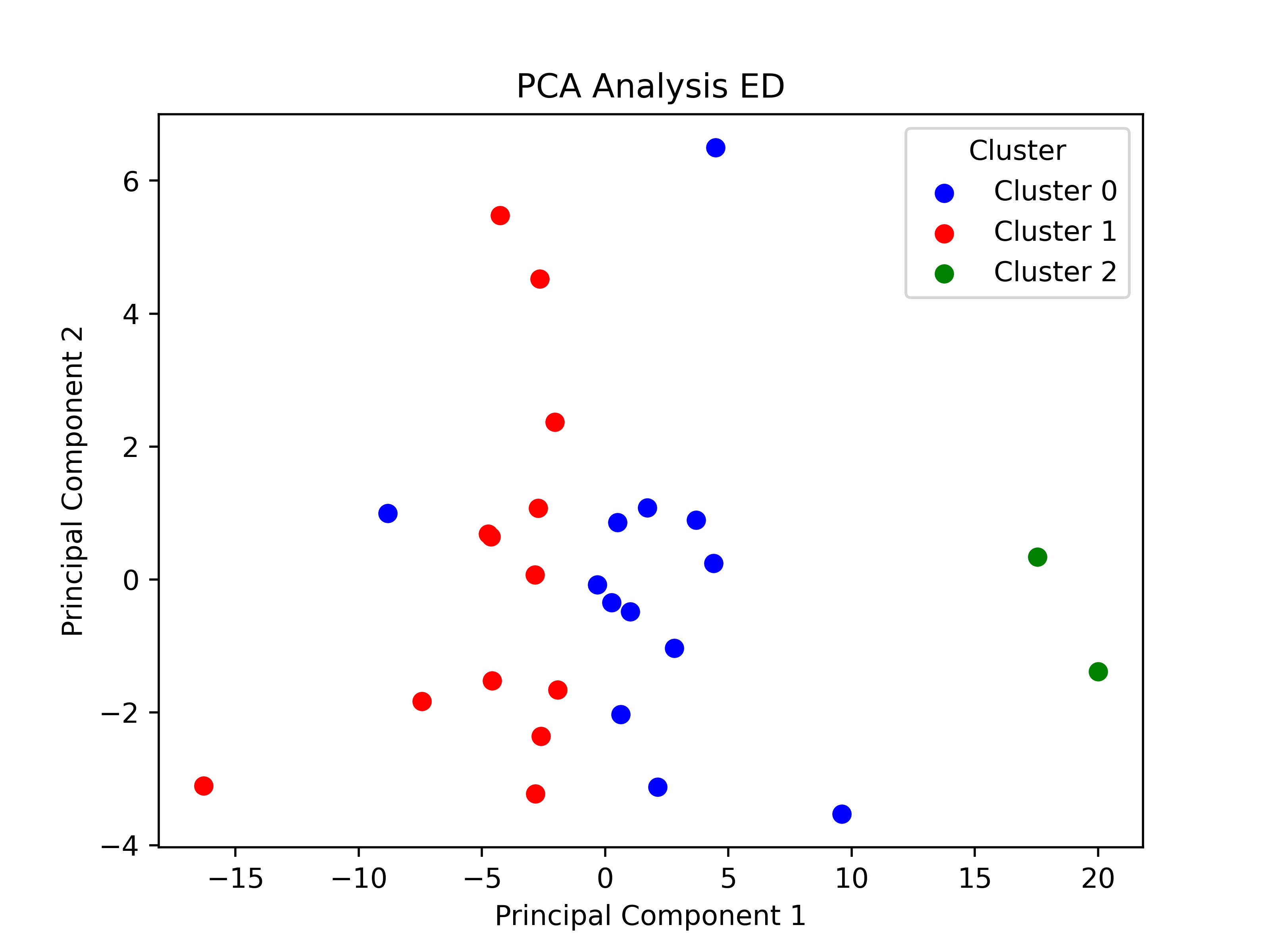}
    \caption{PCA Analysis - ED}
    \label{fig:pca_ed_temperature}
  \end{subfigure}
  \begin{subfigure}[b]{0.3\textwidth}
    \includegraphics[width=\textwidth]{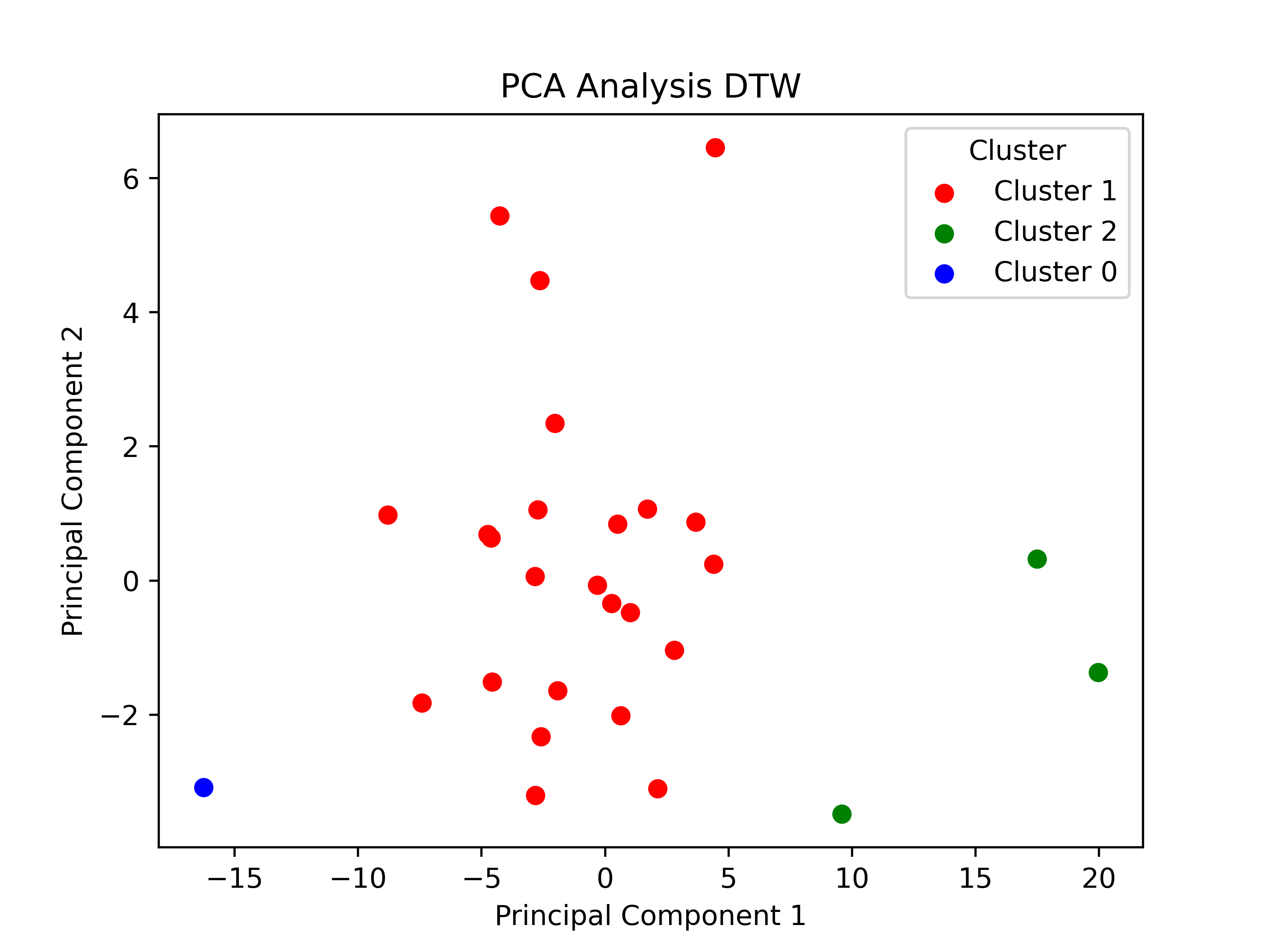}
    \caption{PCA Analysis - DTW}
    \label{fig:pca_dtw_temperature}
  \end{subfigure}
  \begin{subfigure}[b]{0.3\textwidth}
    \includegraphics[width=\textwidth]{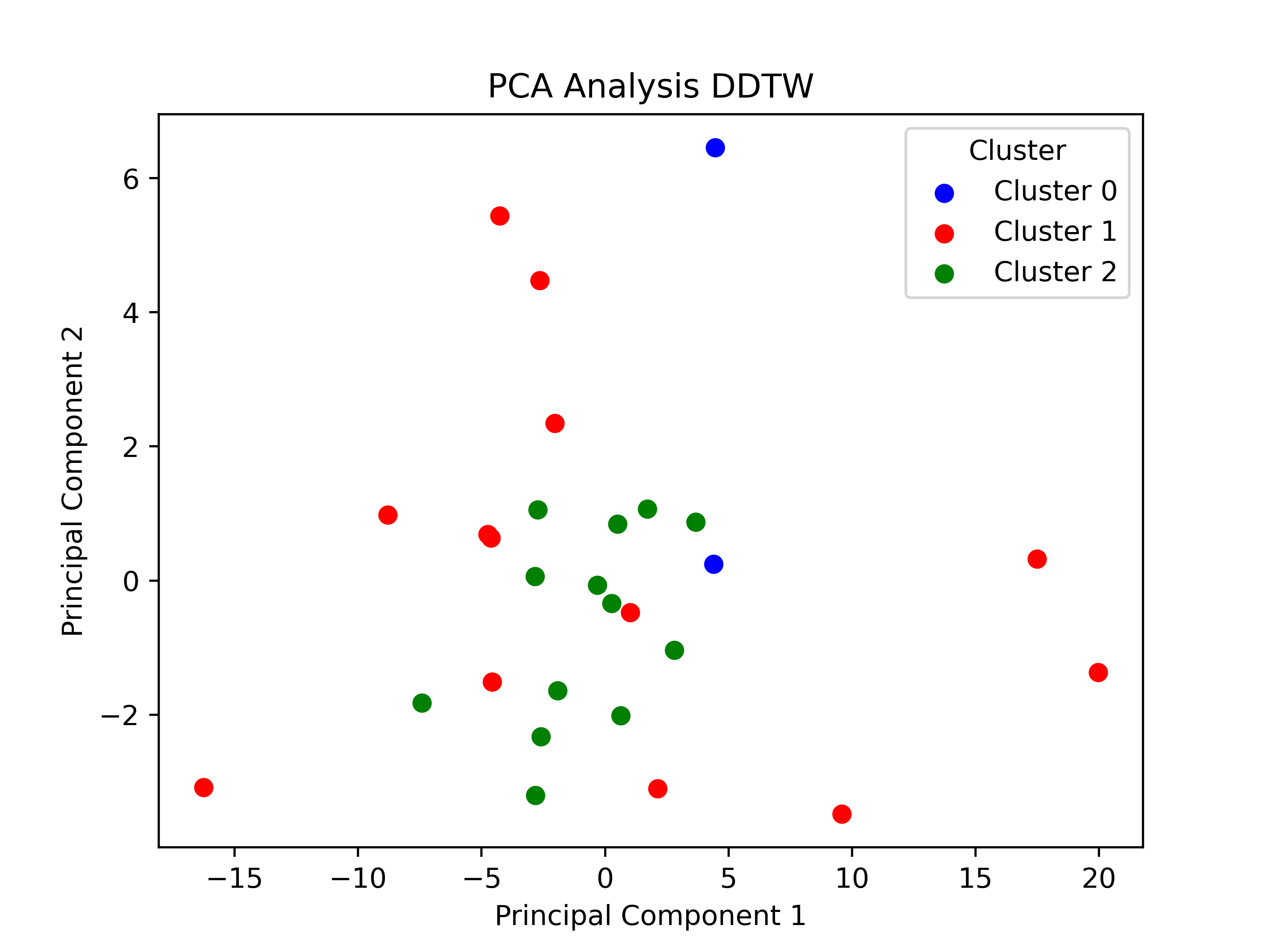}
    \caption{PCA Analysis - DDTW}
    \label{fig:pca_ddtw_temperature}
  \end{subfigure}
  \caption{PCA analysis in the temperature dimension}
  \label{fig:temperature_pca}
\end{figure}

In summary, considering the results presented in Table \ref{tab:cluster_metrics_temperature} alongside these figures, the DTW approach emerges as the optimal solution for this problem. Visualizations of each cluster in Figure \ref{fig:clusters_temperature} offer detailed insights into how data points are grouped based on their similarity in heat demand patterns, utilizing both K-means clustering and the DTW similarity measure. As can be witnessed again, the clusters are based on the amplitude of the values but also the kind of fluctuations that are present in each time series.

\begin{figure}[ht!]
  \centering
  \begin{subfigure}[b]{0.3\textwidth}
    \includegraphics[width=\textwidth]{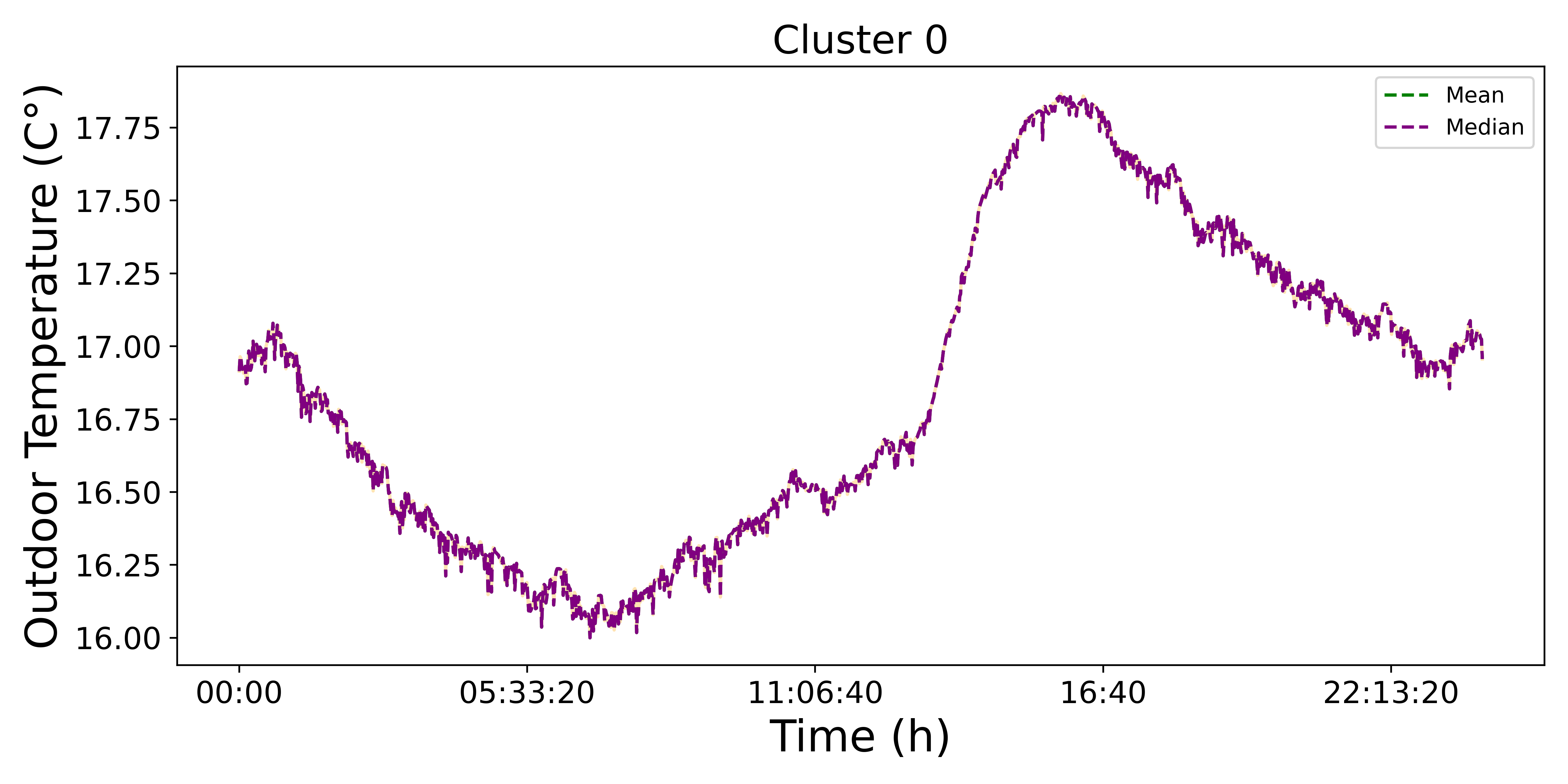}
    \caption{Cluster 0}
    \label{fig:cluster_0_temperature}
  \end{subfigure}
  \begin{subfigure}[b]{0.3\textwidth}
    \includegraphics[width=\textwidth]{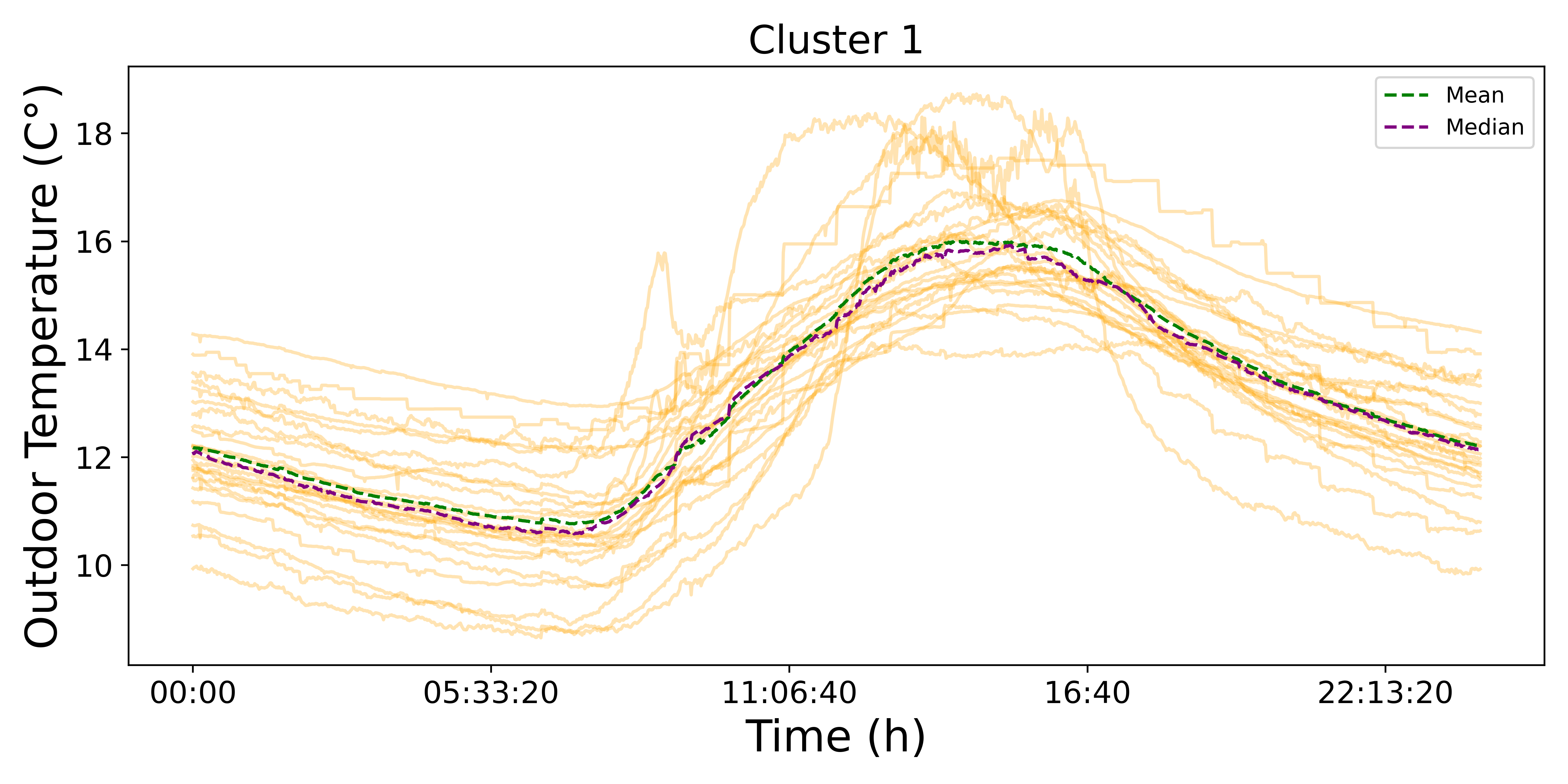}
    \caption{Cluster 1}
    \label{fig:cluster_1_temperature}
  \end{subfigure}
  \begin{subfigure}[b]{0.3\textwidth}
    \includegraphics[width=\textwidth]{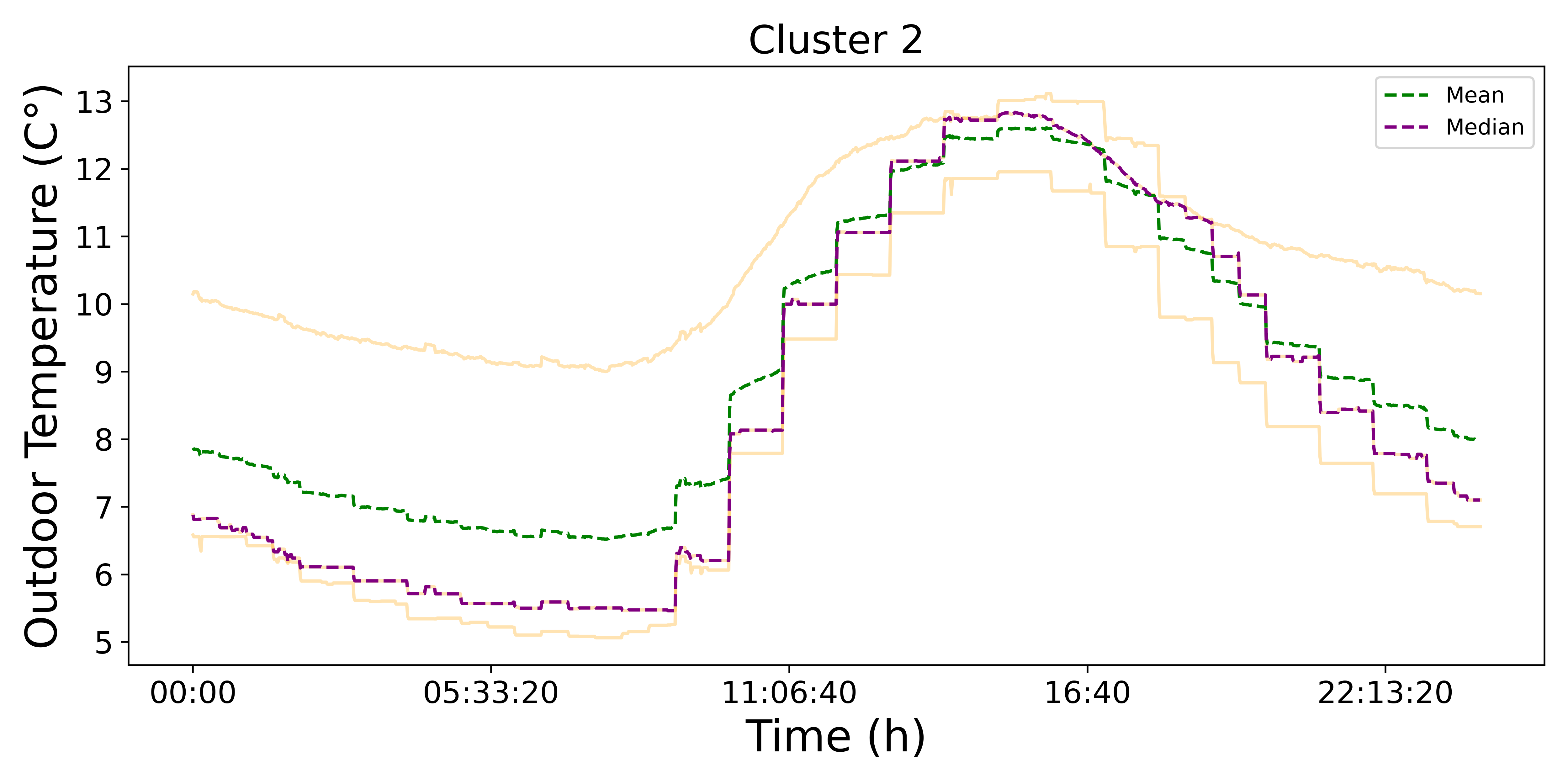}
    \caption{Cluster 2}
    \label{fig:cluster_2_temperature}
  \end{subfigure}
  \caption{Clusters based on K-means and DTW for the temperature dimension}
  \label{fig:clusters_temperature}
\end{figure}

\subsubsection{Scenario IV: Building dimension}

For the building dimension specifically, our investigation focused on differentiating indoor and outdoor temperatures during periods of boiler inactivity. This method aimed to highlight the impact of the building envelope alone on the alignment of these two temperature time series. Figure \ref{fig:building} showcases the diversity in the disparity between indoor and outdoor temperatures across various households over the course of the day, using the beginning of each hour as the point of reference.

\begin{figure*}[h!] 
\centering
 \makebox[\textwidth]{\includegraphics[width=.8\paperwidth]{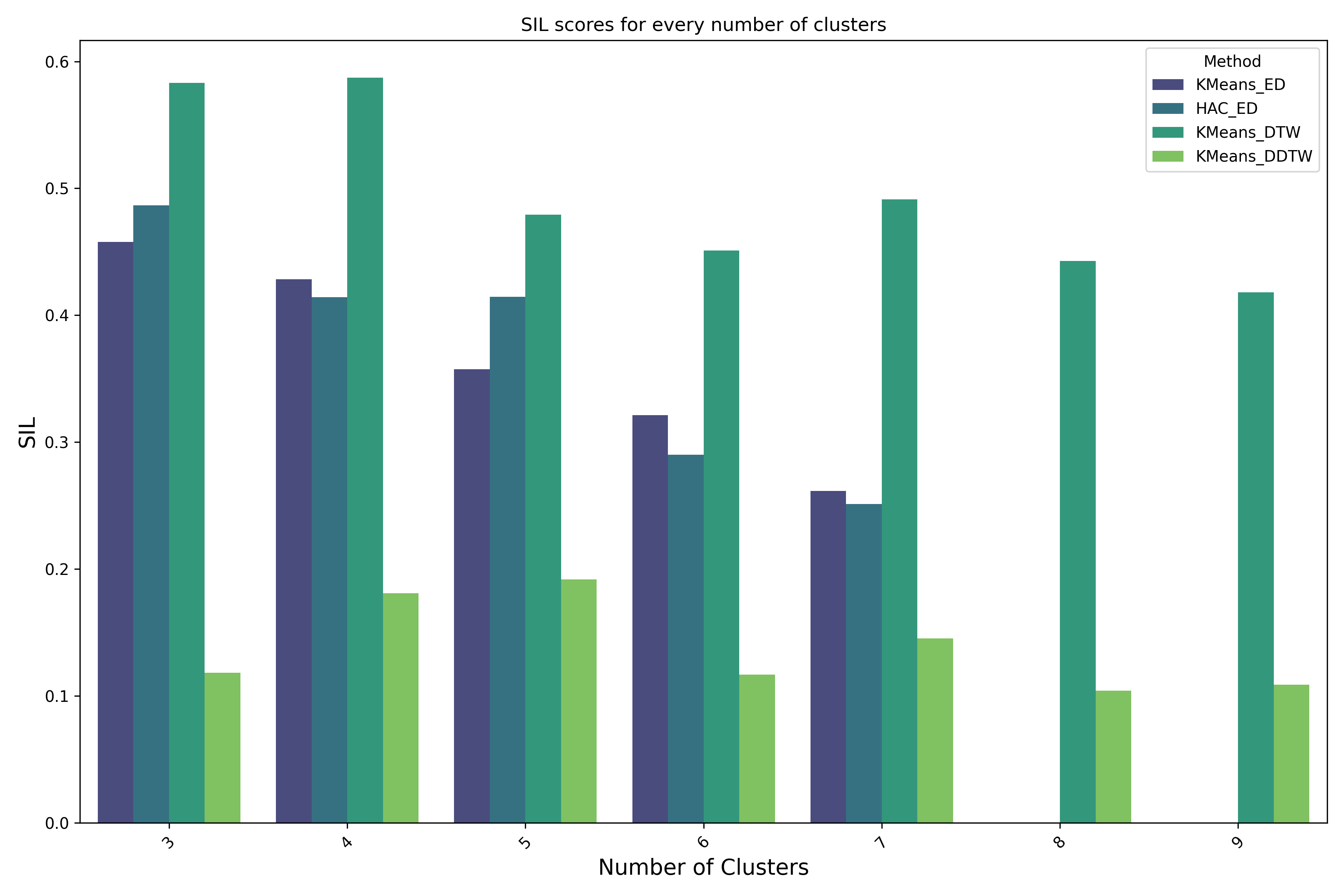}}
\caption{IL score for every number of clusters and algorithm for the three distance metrics in the building dimension}
\label{fig:house_sil}
\end{figure*}

For the current scenario, the labels for HAC and K-means were exactly the same for ED and DTW. In Figure \ref{fig:house_sil}, SIL scores are depicted for all different clustering algorithms and distance metrics, except for HAC and DTW as empty clusters were created so the case was deemed for no further analysis. As can be witnessed in the plot the best SIL for K-means with HAC is four. Since the variance lies in the third digit after the decimal point, we will persist with our analysis, selecting three as the preferred option. A comprehensive overview of the evaluation metric scores for each scenario is provided in Table \ref{tab:cluster_metrics_building}, specifically with three clusters being the chosen count. Notably, DTW produced the best metric scores but being extremely close with ED specifically for CHI.

\begin{table}[h!]
\centering
\caption{Cluster evaluation metrics for the building dimension}
\label{tab:cluster_metrics_building}
\begin{tabular}{|l|c|c|c|}
\hline
Algorithm & Silhouette & DBI & CHI \\ \hline
K-Means with ED & 0.457836 & 0.686432 & 33.754995 \\ \hline
HAC with ED & 0.457836 & 0.686432 & 33.754995 \\ \hline
\textbf{K-Means with DTW} & \textbf{0.583171} & \textbf{0.538681} & \textbf{34.150596} \\ \hline
K-Means with DDTW & 0.118132 & 1.802669 & 14.653911 \\ \hline

\end{tabular}
\end{table}

At this point, in Table \ref{tab:similarity_matrices_builing} a comparison of labels from the ED, DTW, and DDTW methods reveals notable inconsistencies. Subtable \ref{subtab:ed_vs_dtw_building} illustrates that ED and DTW produced almost the exact same labels, with only one house making the difference. That fact showcases the similarity between the two methods for this dimension. On the other hand, Subtables \ref{subtab:dtw_vs_ddtw_building} and \ref{subtab:ed_vs_ddtw_building} depicts substantial differences between the clustering labels of ED and DTTW, as well as between DTW and DDTW, with minimal overlap observed.

\begin{table}[h!]
\centering
\caption{Similarity matrices for the building dimension}
\label{tab:similarity_matrices_builing}
\begin{subtable}{0.3\linewidth}
\centering
\caption{Euclidean vs. DTW Matrix}
\label{subtab:ed_vs_dtw_building}
\begin{tabular}{c|ccc}
\hline
& 0 & 1 & 2 \\ \hline
0 & 14 & 0 & 0 \\
1 & 0 & 11 & 0 \\
2 & 1 & 0 & 2 \\
\hline
\end{tabular}
\end{subtable}
\begin{subtable}{0.3\linewidth}
\centering
\caption{Euclidean vs. DDTW Matrix}
\label{subtab:ed_vs_ddtw_building}
\begin{tabular}{c|ccc}
\hline
& 0 & 1 & 2 \\ \hline
0 & 3 & 2 & 9 \\
1 & 7 & 1 & 3 \\
2 & 1 & 0 & 2 \\
\hline
\end{tabular}
\end{subtable}
\begin{subtable}{0.3\linewidth}
\centering
\caption{DTW vs. DDTW Matrix}
\label{subtab:dtw_vs_ddtw_building}
\begin{tabular}{c|ccc}
\hline
& 0 & 1 & 2 \\ \hline
0 & 3 & 2 & 10 \\
1 & 7 & 1 & 3 \\
2 & 1 & 0 & 1 \\
\hline
\end{tabular}
\end{subtable}
\end{table}

Moving on, in Figure \ref{fig:building_pca} the PCA analysis for each list of labels produces by the clustering algorithms for each distance metric for the building dimension is illustrated. More specifically, for ED and DTW the results indicate great promise as the clusters seem to be separated in order, showcasing strong similarity. 

\begin{figure}[ht!]
  \centering
  \begin{subfigure}[b]{0.3\textwidth}
    \includegraphics[width=\textwidth]{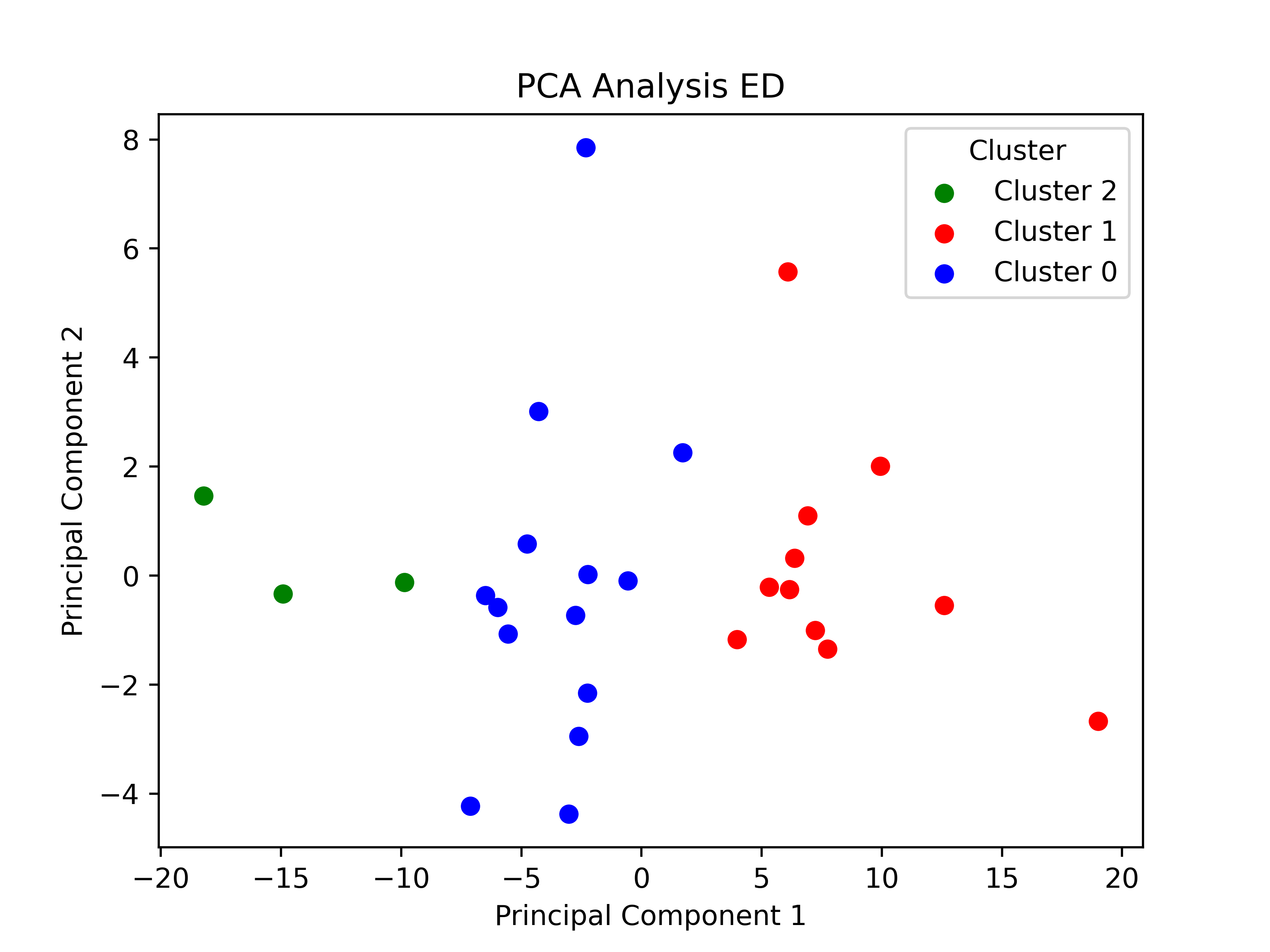}
    \caption{PCA Analysis - ED}
    \label{fig:pca_ed_building}
  \end{subfigure}
  \begin{subfigure}[b]{0.3\textwidth}
    \includegraphics[width=\textwidth]{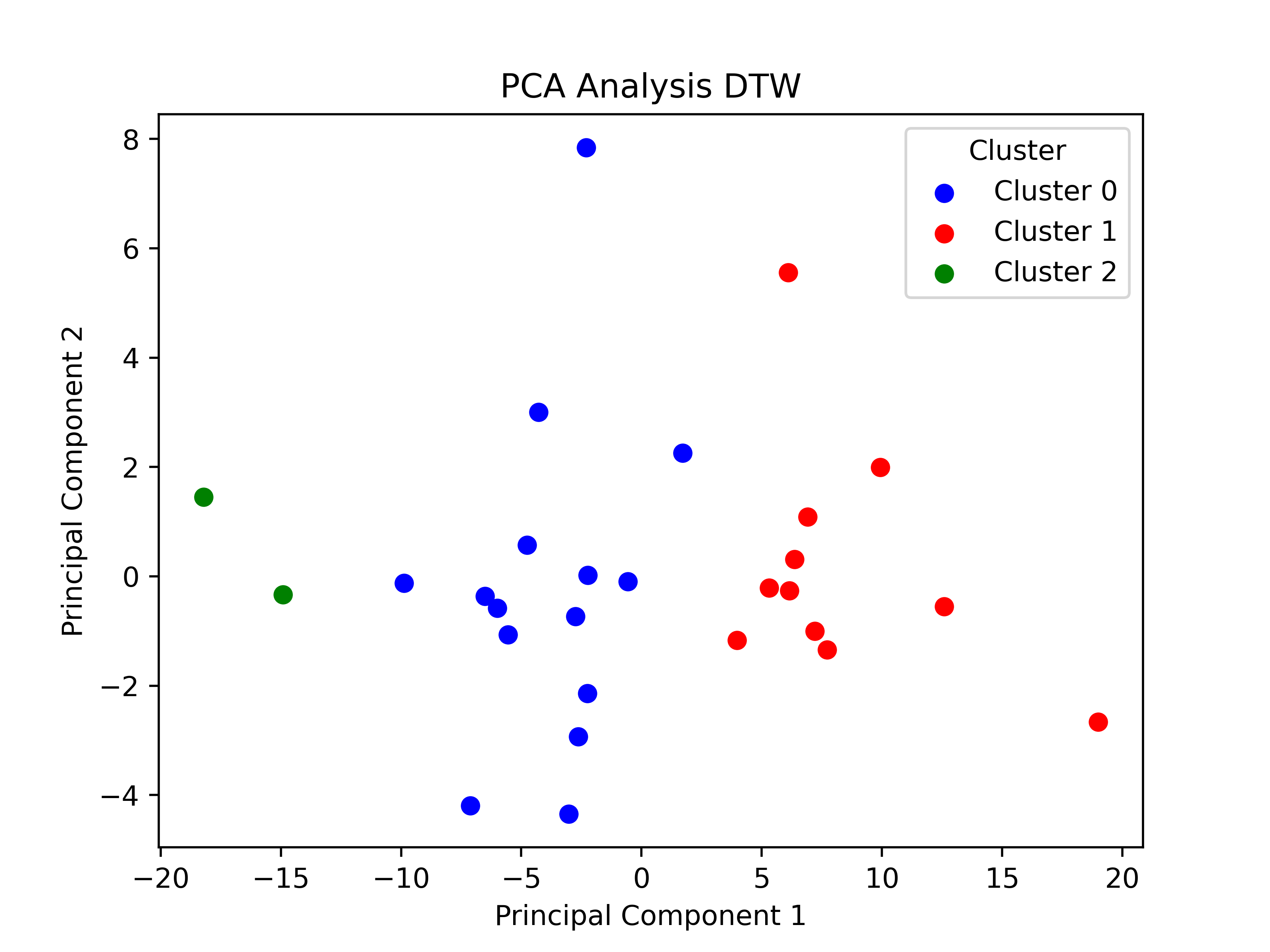}
    \caption{PCA Analysis - DTW}
    \label{fig:pca_dtw_building}
  \end{subfigure}
  \begin{subfigure}[b]{0.3\textwidth}
    \includegraphics[width=\textwidth]{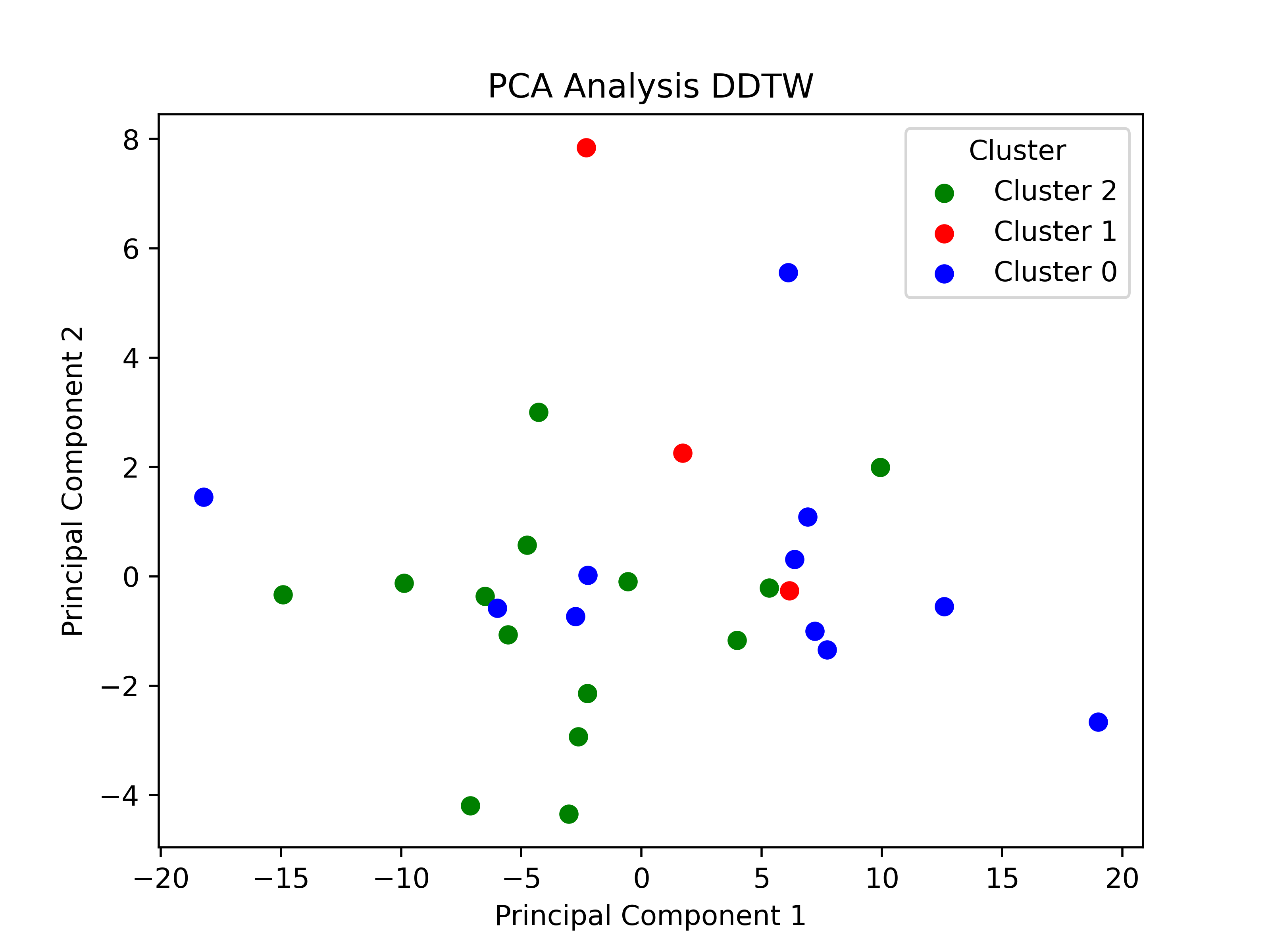}
    \caption{PCA Analysis - DDTW}
    \label{fig:pca_ddtw_building}
  \end{subfigure}
  \caption{PCA analysis in the building dimension}
  \label{fig:building_pca}
\end{figure}

Concluding, when evaluating the results provided in Table \ref{tab:cluster_metrics_building} alongside the accompanying figures, it becomes evident that the DTW approach stands out as the most effective solution for addressing this problem with ED being one house short. The visual representations of each cluster in Figure \ref{fig:clusters_building} provide comprehensive insights into how data points are organized according to their similarity in building characteristics, leveraging both K-means clustering and the DTW similarity measure. Once more, it is apparent that the clustering is not solely based on the magnitude of the values but also on the specific fluctuations present within each time series.

\begin{figure}[ht!]
  \centering
  \begin{subfigure}[b]{0.3\textwidth}
    \includegraphics[width=\textwidth]{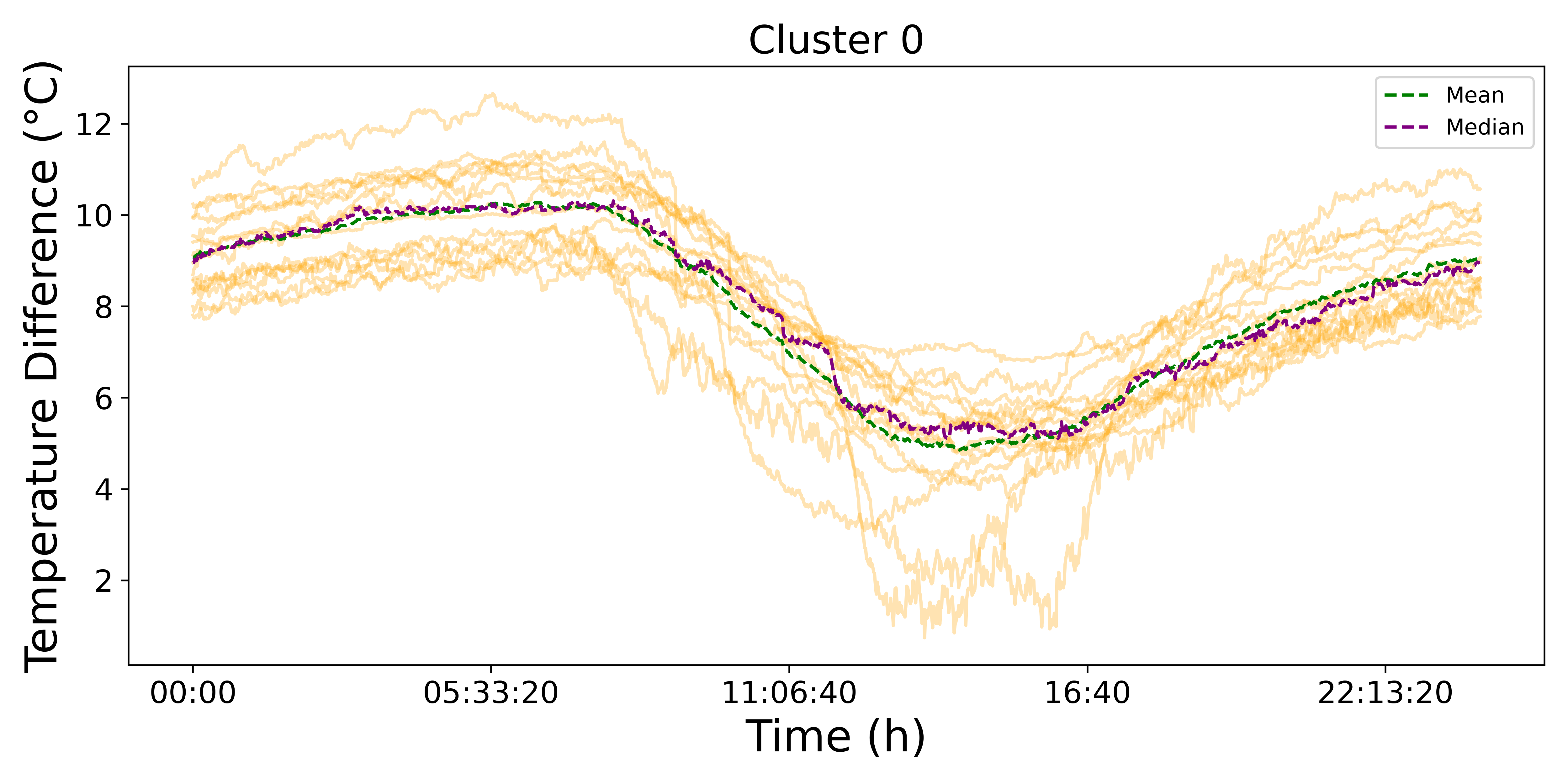}
    \caption{Cluster 0}
    \label{fig:cluster_0_building}
  \end{subfigure}
  \begin{subfigure}[b]{0.3\textwidth}
    \includegraphics[width=\textwidth]{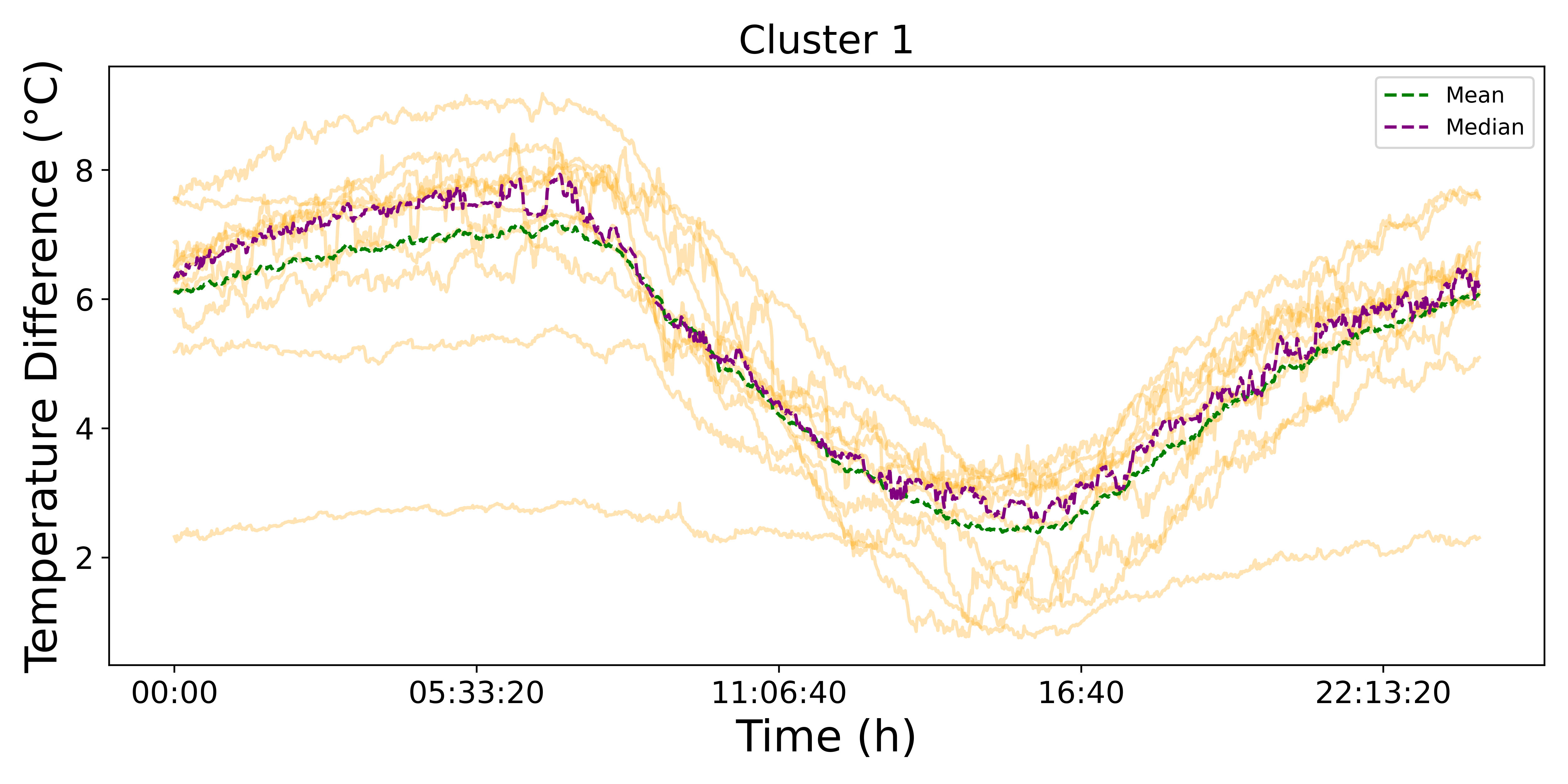}
    \caption{Cluster 1}
    \label{fig:cluster_1_building}
  \end{subfigure}
  \begin{subfigure}[b]{0.3\textwidth}
    \includegraphics[width=\textwidth]{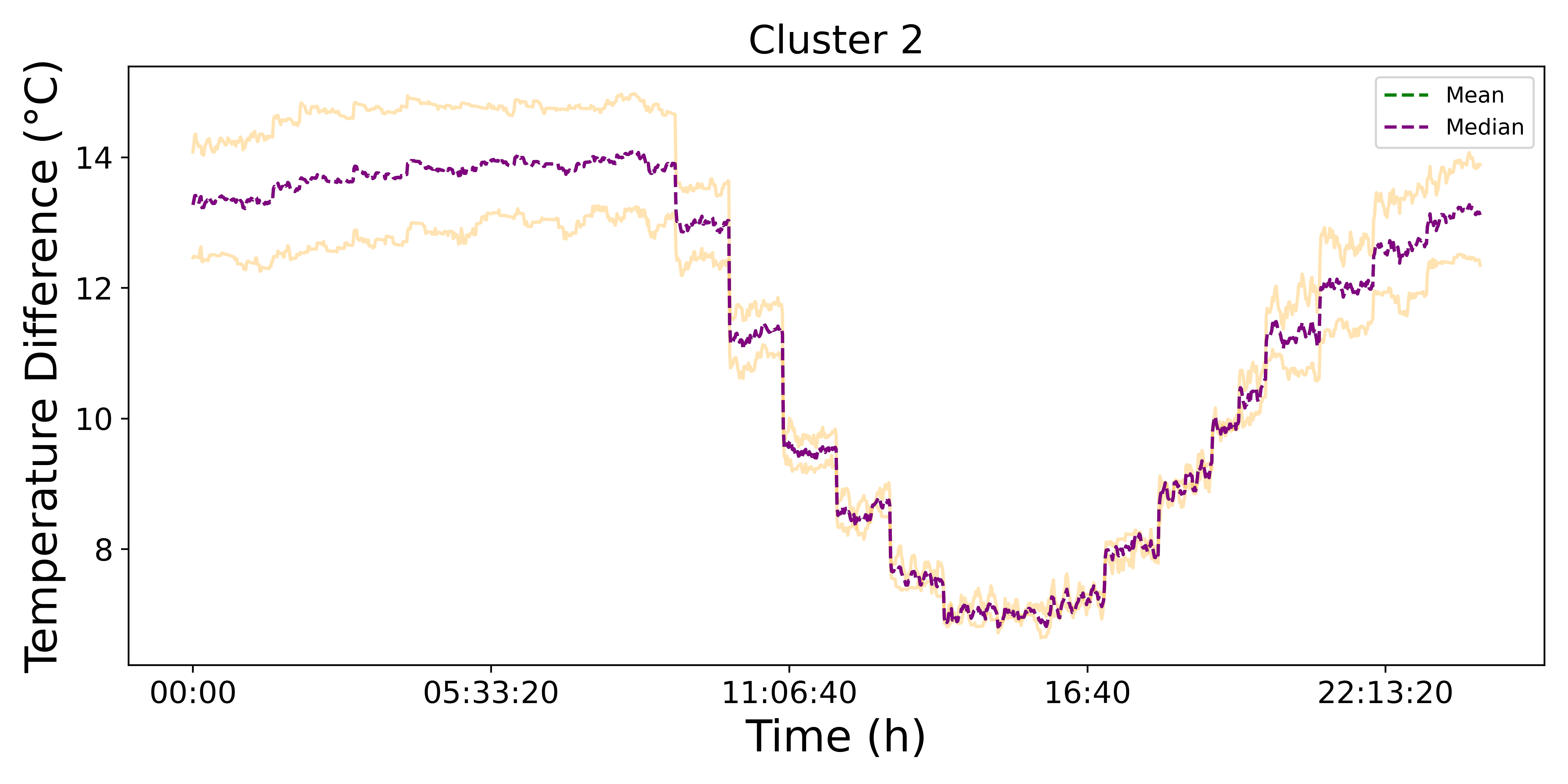}
    \caption{Cluster 2}
    \label{fig:cluster_2_building}
  \end{subfigure}
  \caption{Clusters based on K-means and DTW for the building dimension}
  \label{fig:clusters_building}
\end{figure}

\subsubsection{Scenario V: User Dimension}

Our examination delved into discerning the target temperature preferences set by each user for their respective households throughout the day. This approach underscored the significant influence of individual preferences on the alignment temperature trends and thus offer insights into user preferences regarding comfort. Figure \ref{fig:user} portrays the broad spectrum of target temperatures established by various households, highlighting the nuances in preferences at each hour of the day.

\begin{figure*}[h!] 
\centering
 \makebox[\textwidth]{\includegraphics[width=.8\paperwidth]{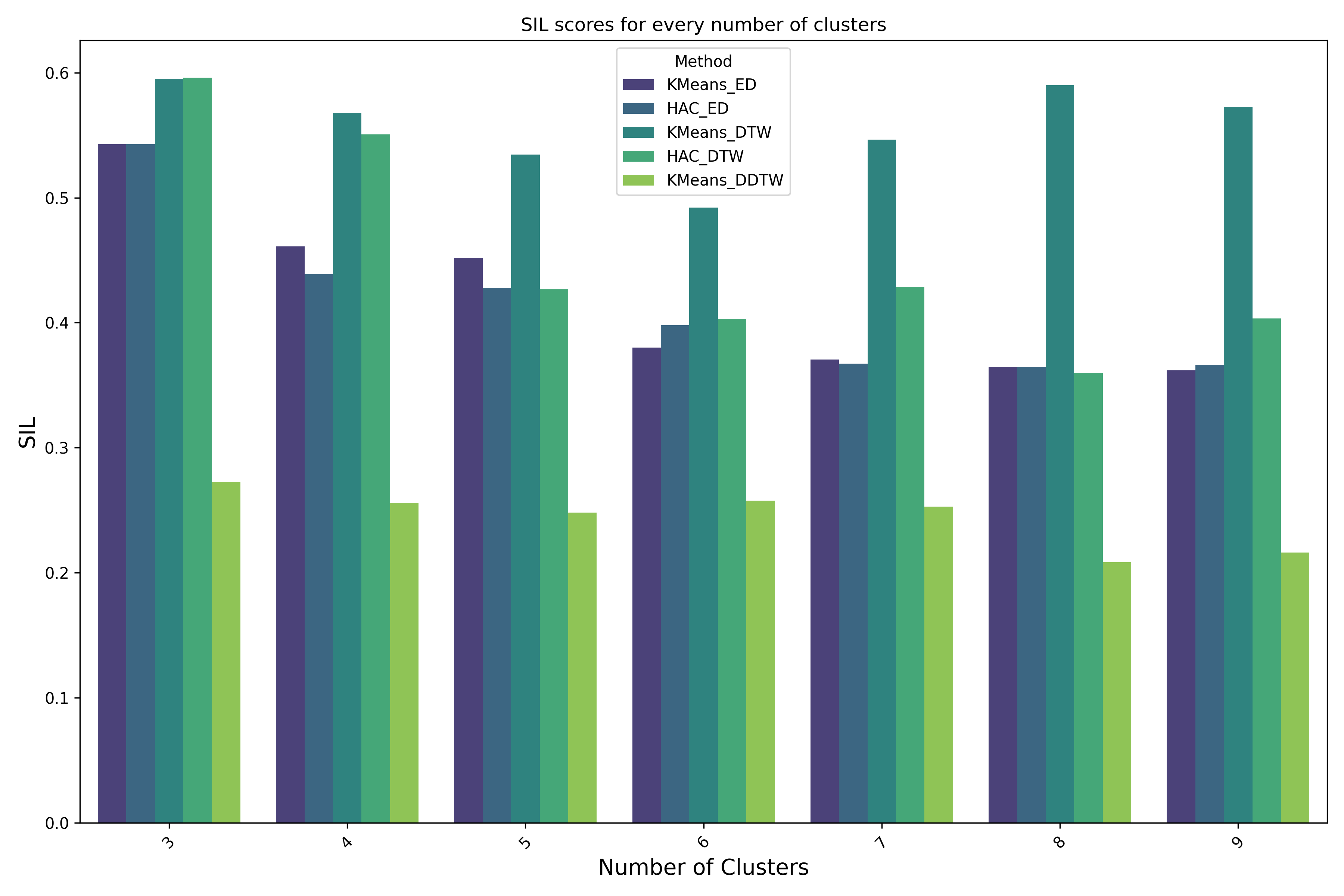}}
\caption{IL score for every number of clusters and algorithm for the three distance metrics in the user dimension}
\label{fig:user_sil}
\end{figure*}

In the present context, both the HAC and K-means algorithms yielded identical labels for the ED, while for DTW, only one label differed, resulting in nearly indistinguishable outcomes. Figure \ref{fig:user_sil} displays the SIL scores for various clustering algorithms and distance metrics. Across all algorithms and distance metrics, the highest SIL is again consistently observed for three clusters. An intriguing observation is evident in the Figure, where SIL peaks after the sixth cluster, nearly matching the performance of three clusters by the eighth. Furthermore, Table \ref{tab:user_dimension_metrics} offers a comprehensive overview of evaluation metric scores for each scenario, with three clusters being the selected count. Noteworthy is the superior performance of HAC with DTW across most metrics, except for CHI, where K-means with DTW outperformed it in a close call. Consequently, we proceed with our analysis utilizing the labels generated by HAC with DTW.

\begin{table}[h!]
\centering
\caption{Cluster evaluation metrics for the user dimension}
\label{tab:user_dimension_metrics}
\begin{tabular}{|l|c|c|c|}
\hline
Algorithm & Silhouette & DBI & CHI \\ \hline
K-Means with ED & 0.542979 & 0.588931 & 91.201546 \\ \hline
HAC with ED & 0.542979 & 0.588931 & 91.201546 \\ \hline
K-Means with DTW & 0.595037 & 0.512780 & \textbf{100.773726} \\ \hline
\textbf{HAC with DTW} & \textbf{0.596143} & \textbf{0.510645} & 98.951570 \\ \hline
K-Means with DDTW & 0.272560 & 1.149919 & 19.108795 \\ \hline
\end{tabular}
\end{table}

At this juncture, Table \ref{tab:similarity_matrices_user} presents a comparison of labels generated ED, DTW and DDTW methods, revealing notable disparities in some cases. Subtable \ref{subtab:ed_vs_dtw_user} demonstrates that both ED and DTW methods yielded identical labels, despite their divergent scores in the preceding Table \ref{tab:user_dimension_metrics}. This discrepancy arises from the additional features incorporated in the ED method, which, although resulting in identical labels, introduce some complexity in assessing similarity. This similarity underscores the resemblance between the two methods within this dimension. Conversely, Subtables \ref{subtab:dtw_vs_ddtw_building} and \ref{subtab:ed_vs_ddtw_building} highlight differences between clustering labels produced by ED and DDTW, as well as between DTW and DDTW, even though less substantial than observed in other dimensions.

\begin{table}[h!]
\centering
\caption{Similarity matrices for the user dimension}
\label{tab:similarity_matrices_user}
\begin{subtable}{0.3\linewidth}
\centering
\caption{Euclidean vs. DTW Matrix}
\label{subtab:ed_vs_dtw_user}
\begin{tabular}{c|ccc}
\hline
& 0 & 1 & 2 \\ \hline
0 & 0 & 9 & 0 \\
1 & 0 & 0 & 10 \\
2 & 9 & 0 & 0 \\
\hline
\end{tabular}
\end{subtable}
\begin{subtable}{0.3\linewidth}
\centering
\caption{Euclidean vs. DDTW Matrix}
\label{subtab:ed_vs_ddtw_user}
\begin{tabular}{c|ccc}
\hline
& 0 & 1 & 2 \\ \hline
0 & 0 & 0 & 9 \\
1 & 3 & 6 & 1 \\
2 & 2 & 7 & 0 \\
\hline
\end{tabular}
\end{subtable}
\begin{subtable}{0.3\linewidth}
\centering
\caption{DTW vs. DDTW Matrix}
\label{subtab:dtw_vs_ddtw_user}
\begin{tabular}{c|ccc}
\hline
& 0 & 1 & 2 \\ \hline
0 & 2 & 7 & 0 \\
1 & 0 & 0 & 9 \\
2 & 3 & 6 & 1 \\
\hline
\end{tabular}
\end{subtable}
\end{table}

Continuing forward, Figure \ref{fig:user_pca} displays the results of PCA conducted on each set of labels generated by the clustering algorithms across various distance metrics for the user dimension. Notably, despite the fact that both ED and DTW yielded identical labels, they are depicted in different colors to differentiate between cluster names. This visual representation showcases the promising outcomes of the ED and DTW methods, as the clusters appear to be distinctly separated and well-ordered, indicating a strong level of similarity between them. Additionally, the clarity of cluster separation suggests that these algorithms effectively captured the underlying patterns in the data, further validating their potential for accurate clustering in this context. However, DDTW once more exhibited limited similarity in cluster separation, consistent with the corresponding evaluation metrics presented earlier in this analysis.

\begin{figure}[ht!]
  \centering
  \begin{subfigure}[b]{0.3\textwidth}
    \includegraphics[width=\textwidth]{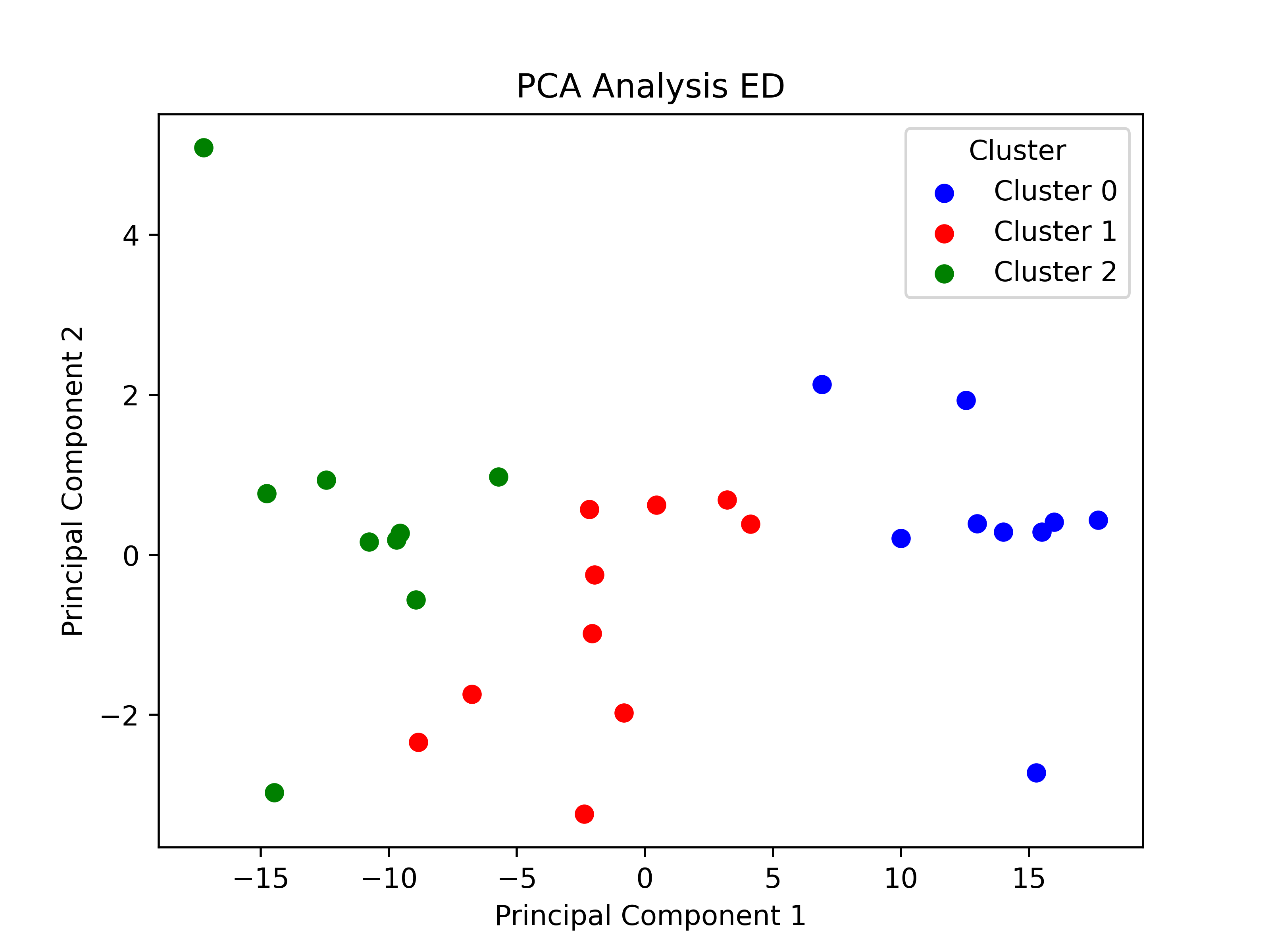}
    \caption{PCA Analysis - ED}
    \label{fig:pca_ed_user}
  \end{subfigure}
  \begin{subfigure}[b]{0.3\textwidth}
    \includegraphics[width=\textwidth]{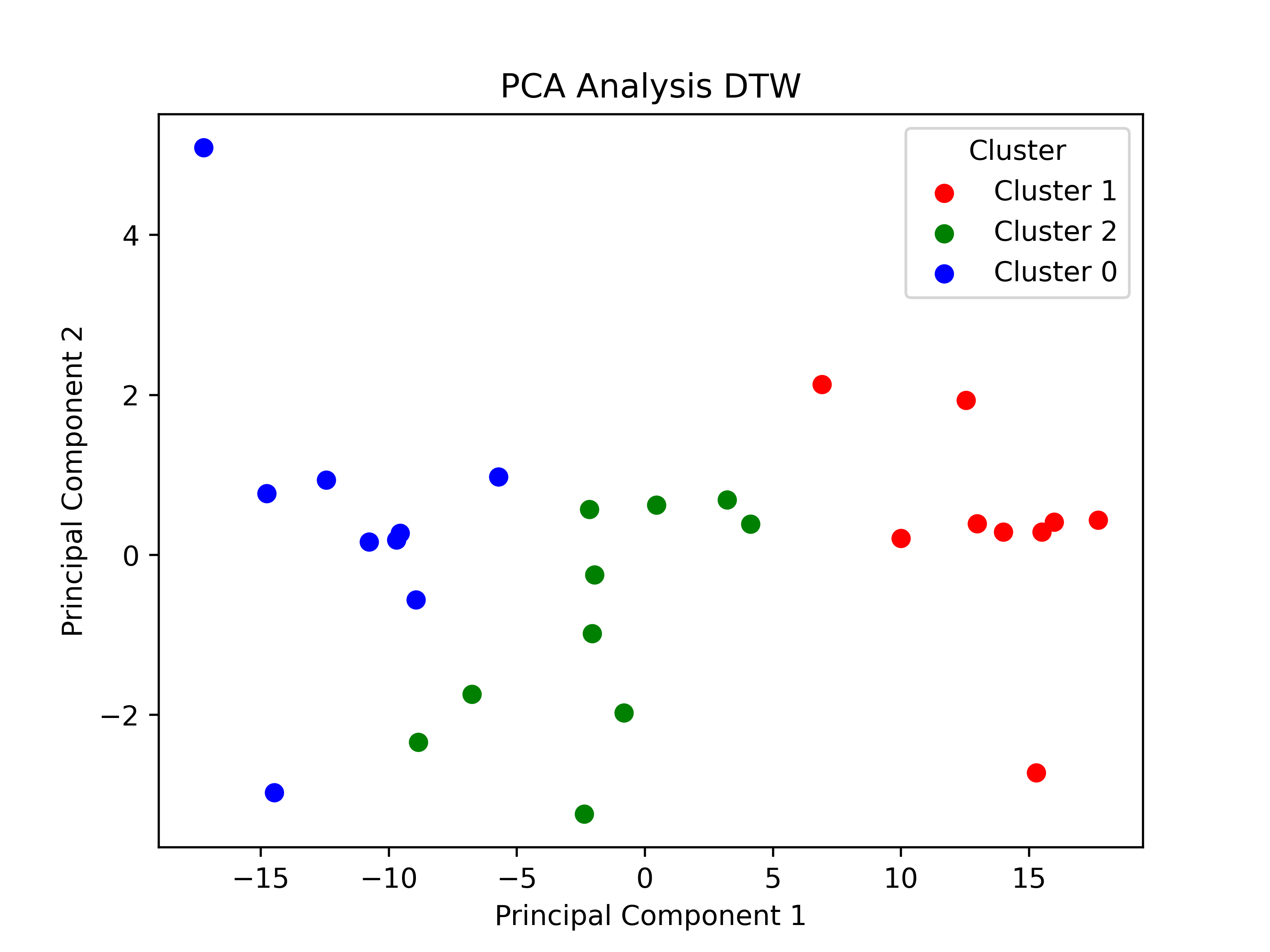}
    \caption{PCA Analysis - DTW}
    \label{fig:pca_dtw_user}
  \end{subfigure}
  \begin{subfigure}[b]{0.3\textwidth}
    \includegraphics[width=\textwidth]{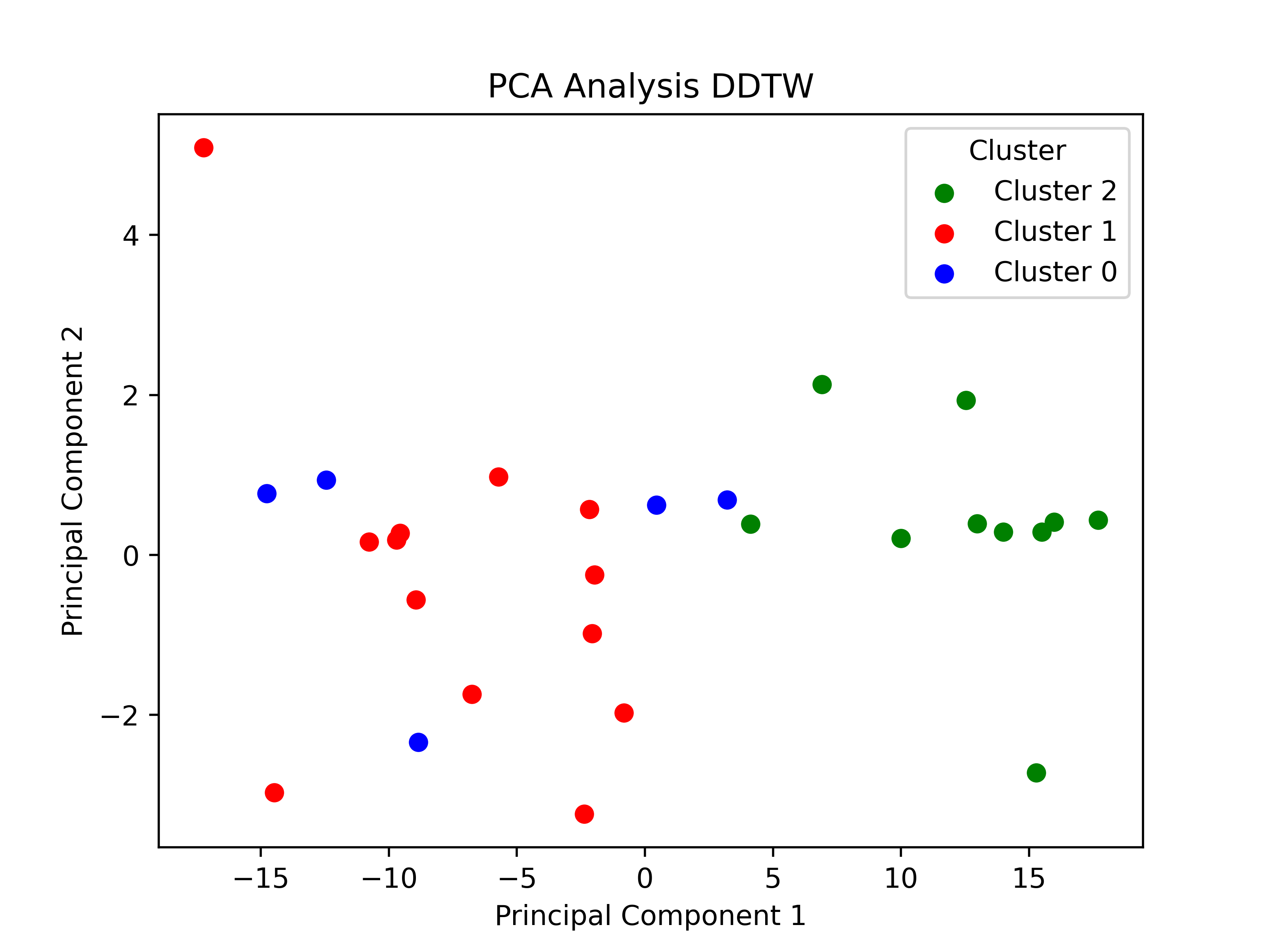}
    \caption{PCA Analysis - DDTW}
    \label{fig:pca_ddtw_user}
  \end{subfigure}
  \caption{PCA analysis in the user dimension}
  \label{fig:user_pca}
\end{figure}

In summary, when reviewing the findings detailed in Table \ref{tab:user_dimension_metrics} alongside the accompanying figures, the superiority of the  labels produced by DTW and ED in addressing this issue becomes apparent. Despite both ED and DTW producing identical labels, the visual representations in Figure \ref{fig:clusters_user} offer valuable insights into how data points are organized based on similarities in user characteristics, leveraging both HAC clustering and the DTW similarity measure. It is intriguing to note that the fluctuations impacting the clustering algorithms were mainly observed during the midday hours. This observation is particularly interesting considering that most users tend to set higher temperatures during the night or early morning, according to their specific behavioral patterns as can be seen in Figure \ref{fig:user}.

\begin{figure}[ht!]
  \centering
  \begin{subfigure}[b]{0.3\textwidth}
    \includegraphics[width=\textwidth]{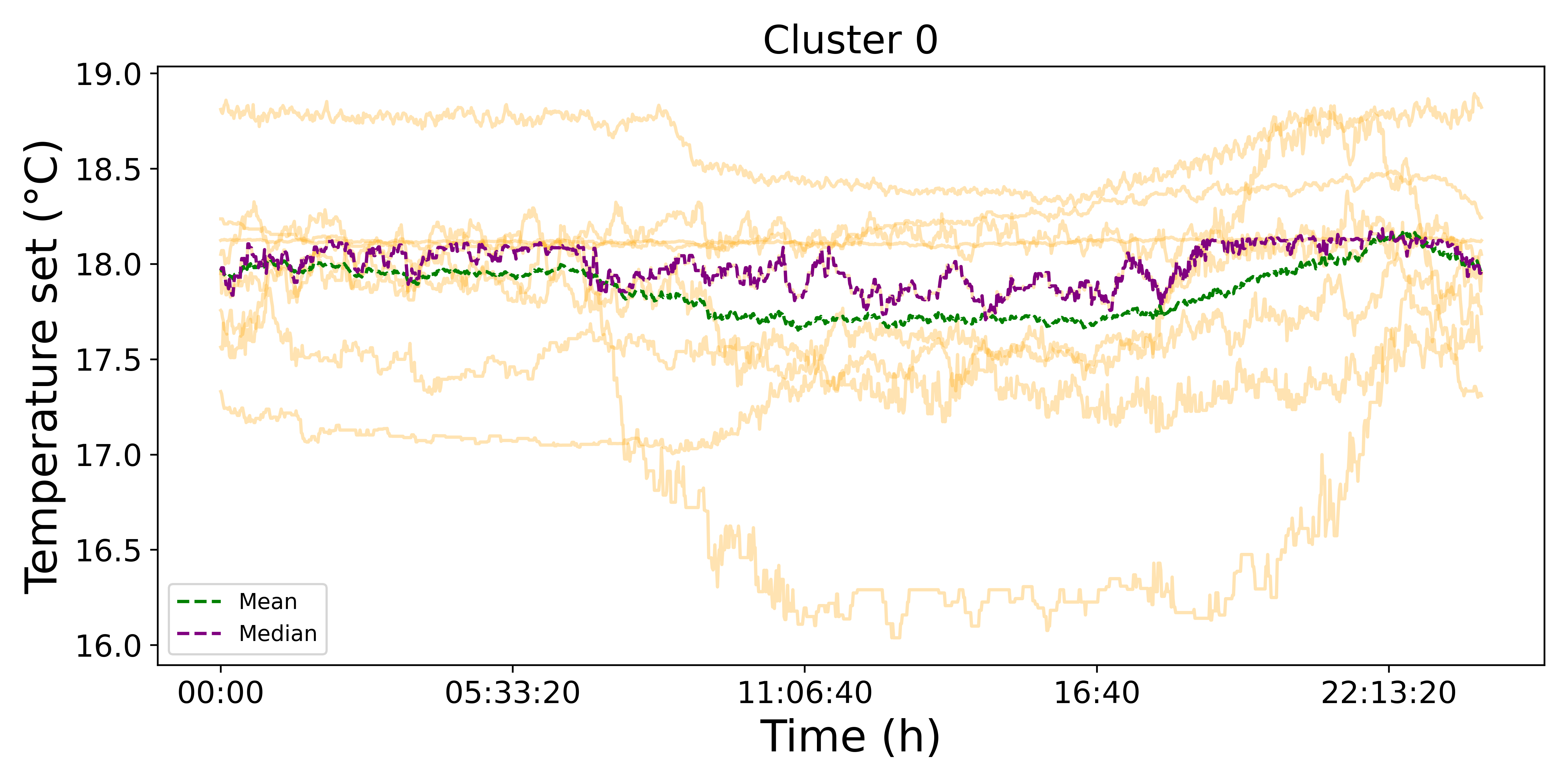}
    \caption{Cluster 0}
    \label{fig:cluster_0_user}
  \end{subfigure}
  \begin{subfigure}[b]{0.3\textwidth}
    \includegraphics[width=\textwidth]{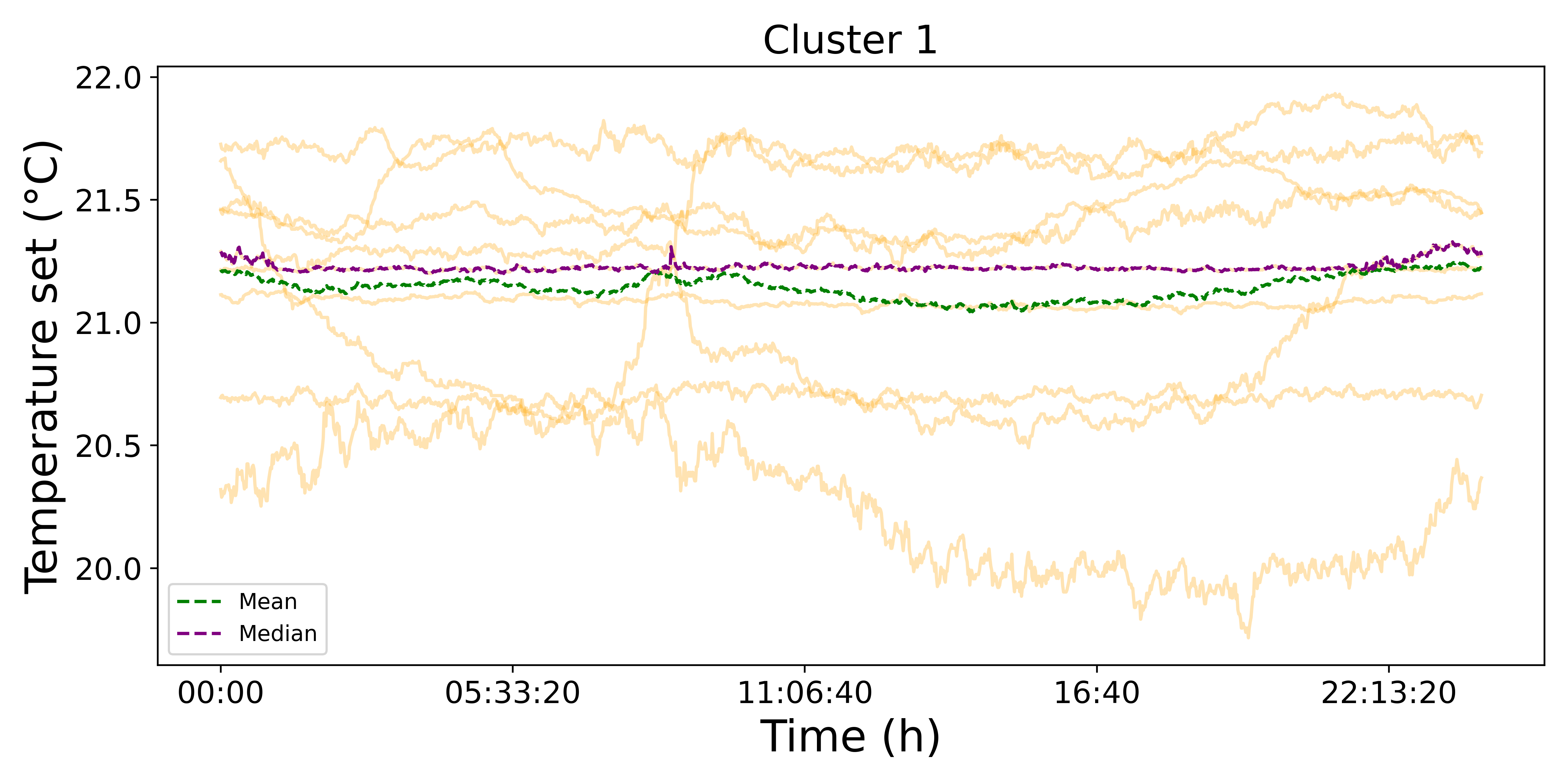}
    \caption{Cluster 1}
    \label{fig:cluster_1_user}
  \end{subfigure}
  \begin{subfigure}[b]{0.3\textwidth}
    \includegraphics[width=\textwidth]{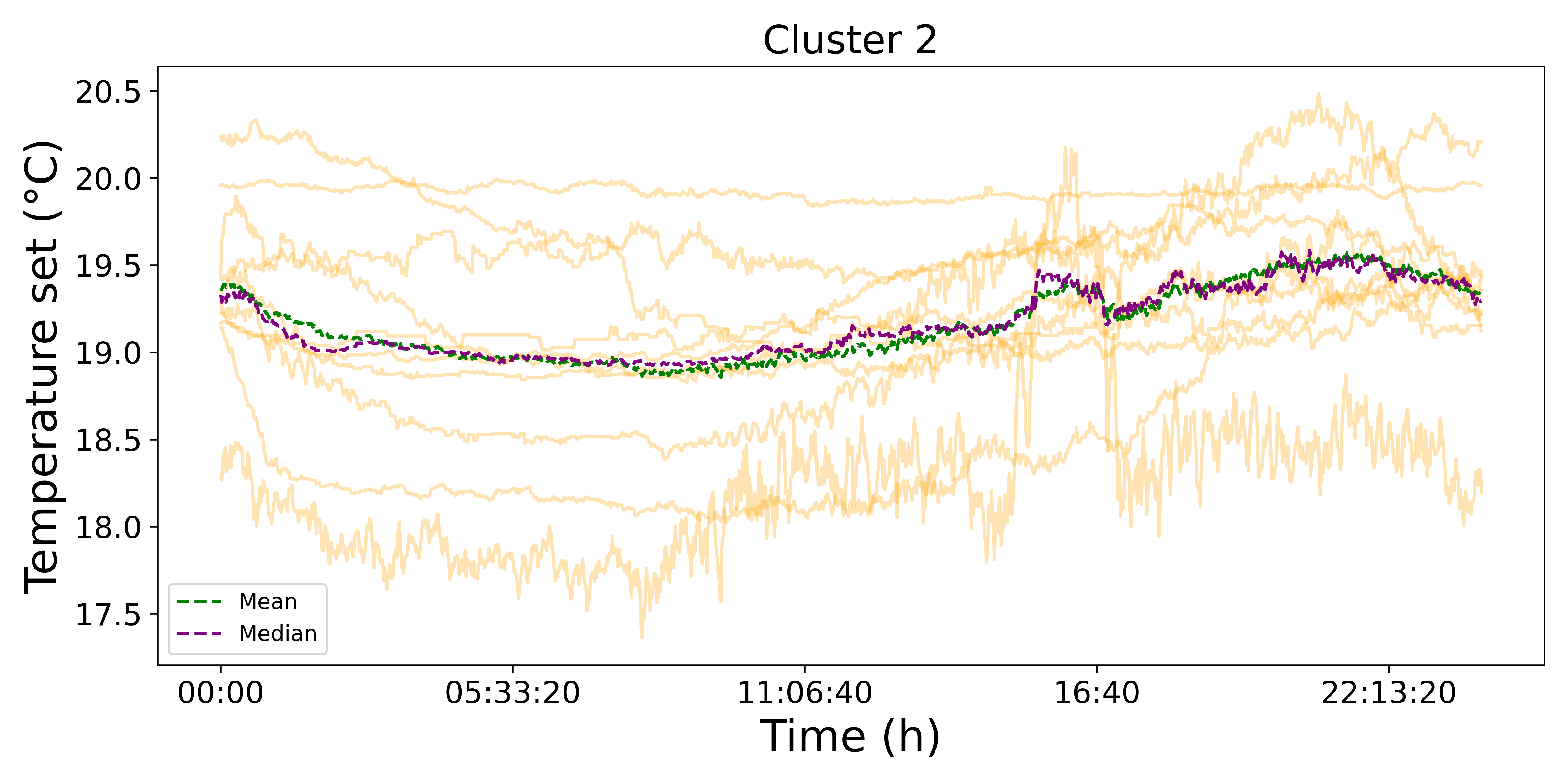}
    \caption{Cluster 2}
    \label{fig:cluster_2_user}
  \end{subfigure}
  \caption{Clusters based on K-means and DTW for the user dimension}
  \label{fig:clusters_user}
\end{figure}

\section{Discussion}
\label{sec:discusion}

In this study, we conducted an in-depth analysis of heating load profiles for the five distinct dimensions identified for this problem, presented as different scenarios in the previous section. Our exploration aimed to uncover underlying patterns within the data and evaluate the effectiveness of clustering algorithms and distance metrics in capturing these patterns. The five dimensions considered in our analysis encompassed heat demand, user, boiler, weather-related, and building-specific factors. 


The results from the previous section clearly indicate that DTW is the most suitable distance metric in many cases, closely followed by ED, based on the evaluation metrics used. Another key finding is that DDTW doesn't align well with the specifics of those dimensions, leading to clusters that are not quite similar in terms of time series values. However, in Figure \ref{fig:proportion}, we illustrate the proportion of matching labels between pairs of dimensions for each distance metric. Ideally, the main diagonal of all three heat maps should have a value of one, but for clarity, we've omitted it.

\begin{figure}[ht!]
  \centering
  \begin{subfigure}[b]{0.49\textwidth}
    \includegraphics[width=\textwidth]{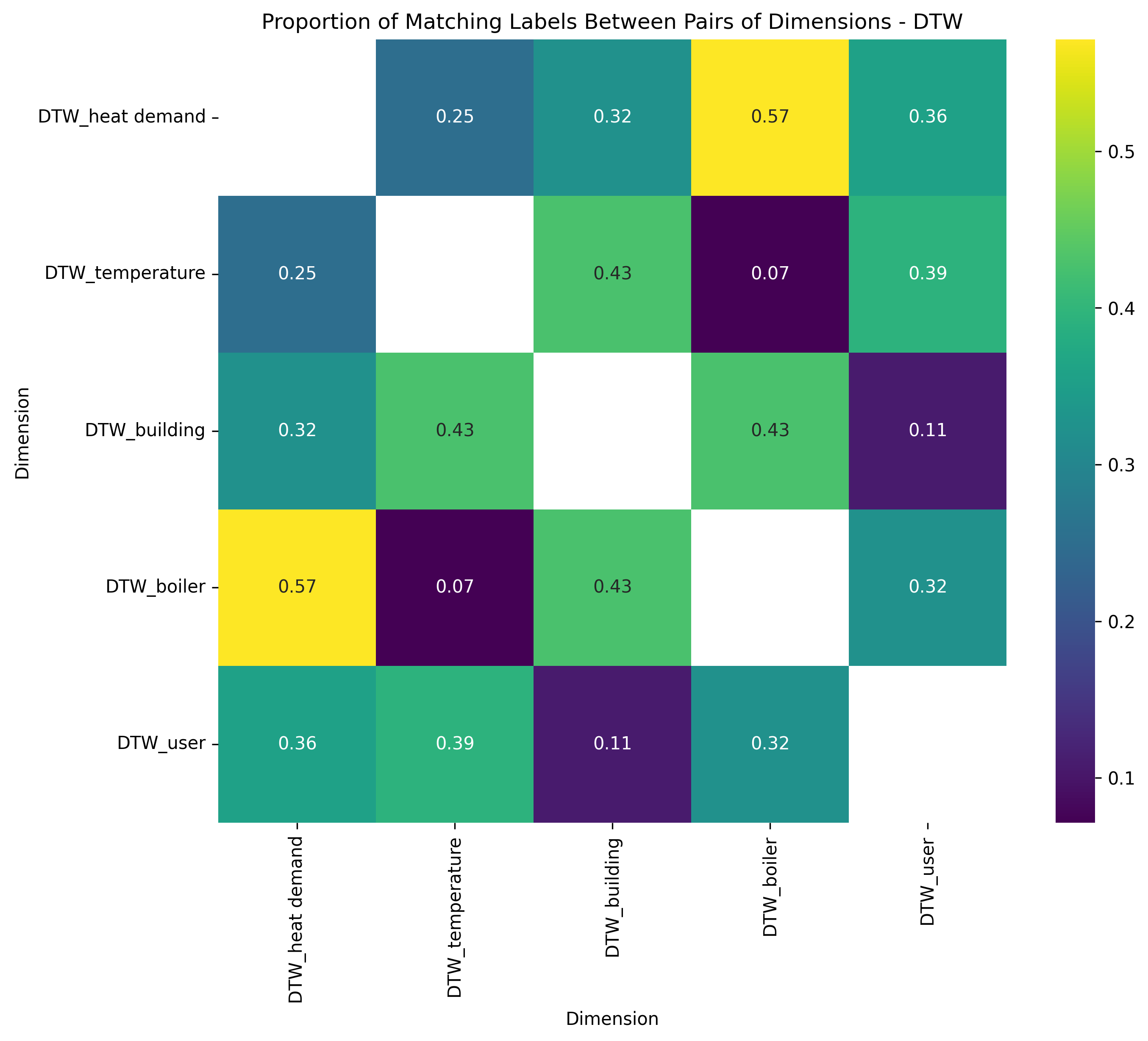}

    \label{fig:proportion_dtw}
  \end{subfigure}
  \hfill
  \begin{subfigure}[b]{0.49\textwidth}
    \includegraphics[width=\textwidth]{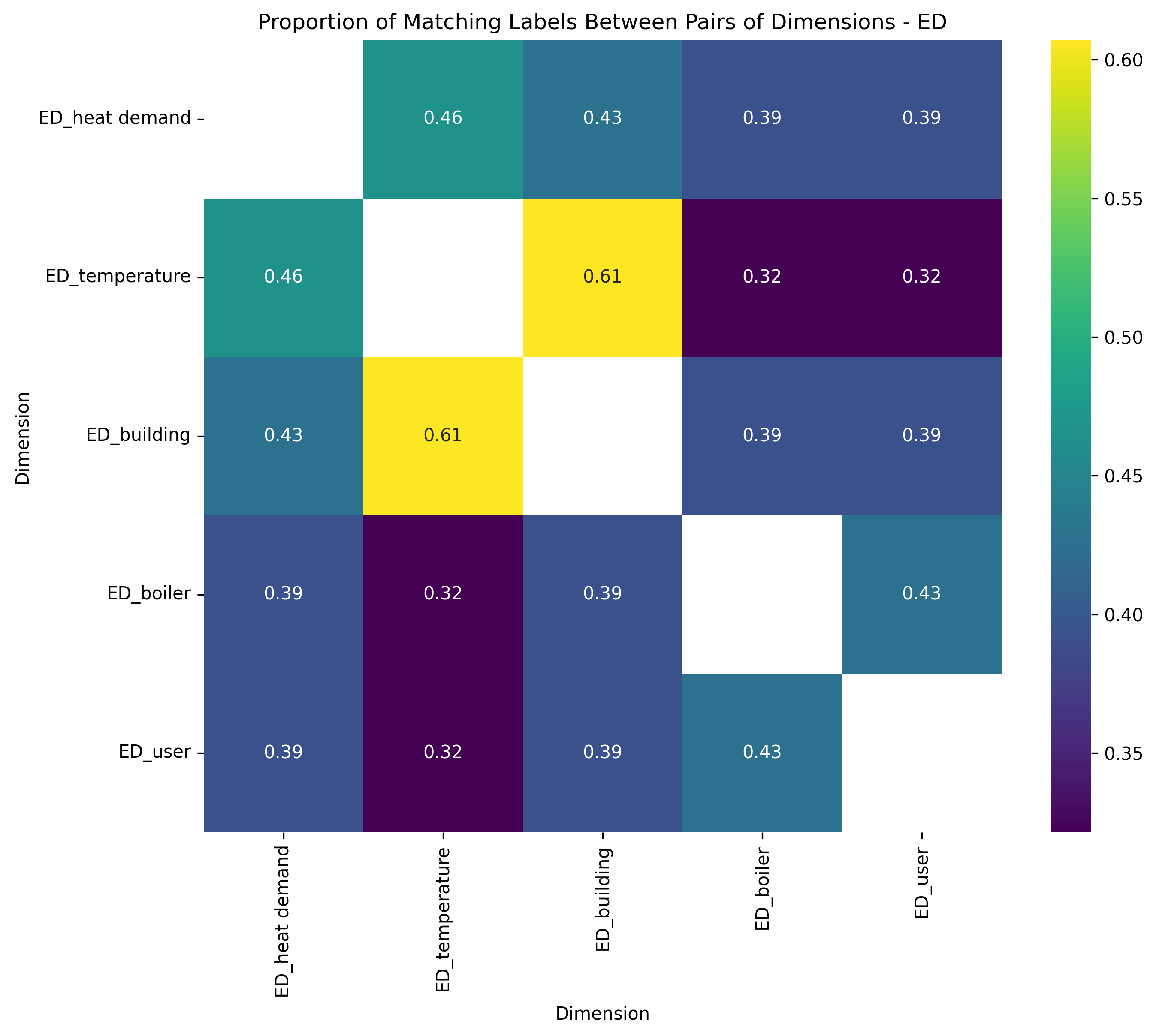}

    \label{fig:proportion_ed}
  \end{subfigure}
  
  \begin{subfigure}[b]{0.5\textwidth}
    \includegraphics[width=\textwidth]{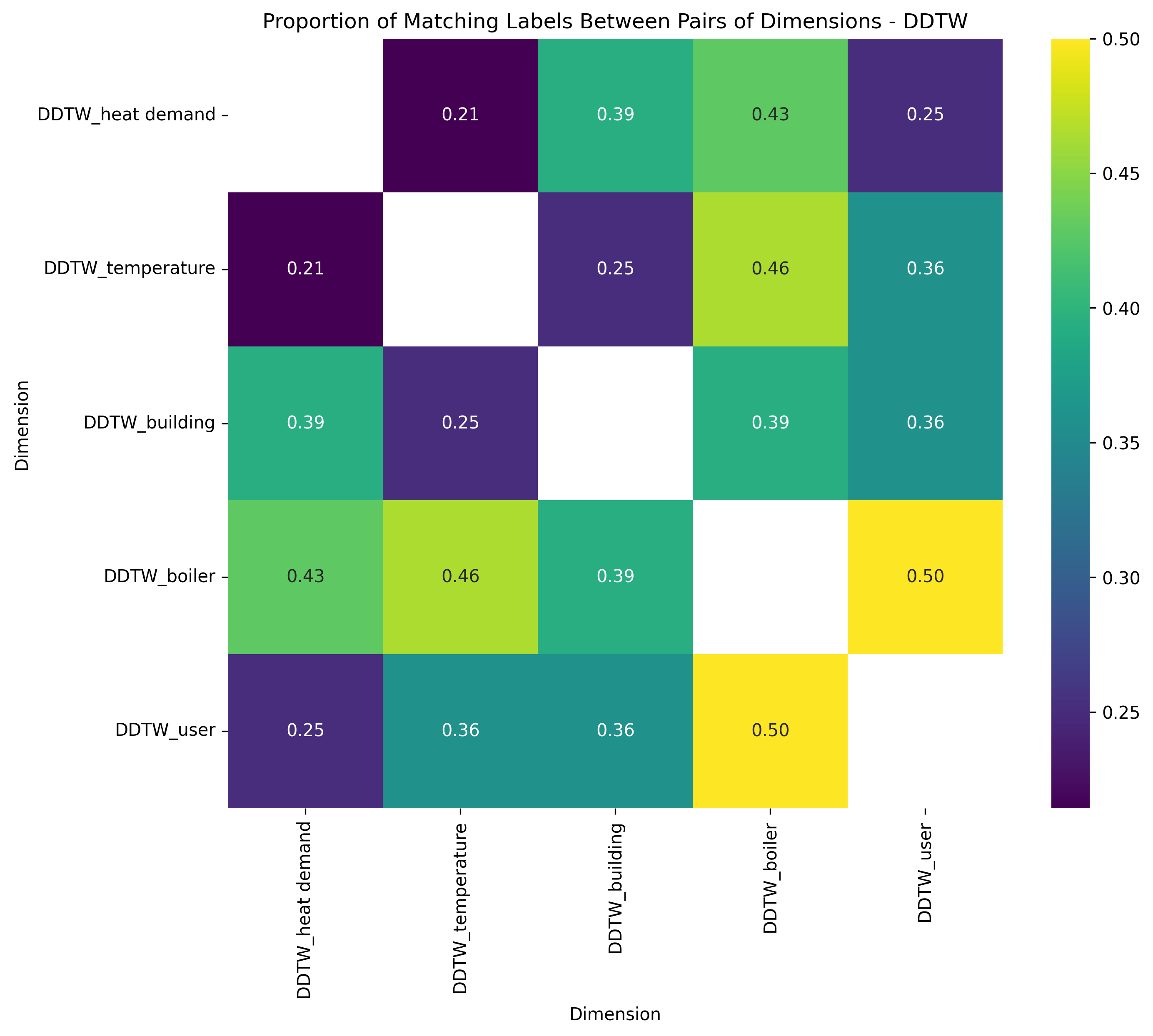}
    \label{fig:proportion_ddtw}
  \end{subfigure}
  \caption{Visual representation illustrating the degree of label agreement across paired dimensions for all distance metrics}
  \label{fig:proportion}
\end{figure}

Our analysis uncovered correlations between the dimensions examined, with each distance metric providing distinct insights. DTW aptly identified strong correlations between boiler usage and heat demand dimensions, underscoring the pivotal influence of boiler usage patterns and heat demand on heating load profiles. Additionally, DTW revealed correlations among boiler usage, temperature, and building-specific characteristics, as well as between user behavior and temperature, elucidating the collective impact of these dimensions on heating requirements. 
While ED also detected correlations between building and weather-related dimensions, it demonstrated a notable tendency to unveil robust associations across various dimensions, indicating a comprehensive interrelation among them. Notably, the correlations exhibited a degree of strength, particularly highlighting the connection between temperature and heat demand, with the strongest correlation observed between building characteristics and temperature. However, it's worth noting that while ED provided valuable insights, it might offer a broader perspective with less detailed granularity compared to DTW. Nevertheless, it still unveiled meaningful connections between the majority of dimensions. In contrast, DDTW showcased less pronounced links among clusters for each dimension, indicating weaker associations. Nevertheless, similar to DTW's findings, the principal correlation highlighted was between boiler usage patterns and user behavior. Overall, while DTW emerged as the most effective metric, ED provided valuable insights, while DDTW's performance was less optimal, suggesting its limitations in capturing the complexities of the dataset.

Given the thorough examination conducted in the paper, it is clear that the dimension most proficient in addressing the complexities of heating load profiles in buildings is the interaction between boiler usage patterns and heat demand. This dimension distinguishes itself through its consistent robust correlations and significant impact on shaping heating load behavior. However, the importance of temperature, evident in both the building's characteristics and the weather-related dimension, highlights its potential as an alternative when heating data are unavailable. Additionally, the strong correlation observed in the user dimension emphasizes the need for more data in these dimensions to further enhance  their refinement.

\section{Conclusions and Future Work}
\label{sec:conclusions}

Summing up, the current research endeavor highlights the critical importance of understanding heating load profiles and their intricacies for optimizing heating system operations, particularly in the context of energy savings and DR participation. Through the evaluation of clustering algorithms with different distance metrics across five dimensions, we have developed a methodology for identifying distinct patterns and correlations within each boiler dataset, providing valuable insights for stakeholders and interested parties. Furthermore, identifying the underlying patterns in heating load profiles allows building operators to
optimize energy usage, enhance comfort levels, and reduce operational costs through customized DR programs. 

More specifically, we experimented with two clustering algorithms, namely K-means and HAC and each algorithm demonstrated varying degrees of effectiveness in capturing the inherent structure of the data. Additionally, we evaluated  three different distance metrics, ED, DTW and DDTW, to quantify the similarity between heating load profiles. Our findings suggest that the choice of distance metric significantly influences clustering results, with certain metrics better suited for specific types of data distributions, but as a results in the majority of the cases DTW was the best choice. This underscores the importance of selecting appropriate algorithms and metrics tailored to the characteristics of the dataset.


However, in evaluating the proposed system, several limitations and practical considerations must be addressed to fully acknowledge its applicability and implications. First of all, every work related to data is naturally introduced with several limitations. Nevertheless, the data used in our analysis, even though spanning only one heating season, is sourced from different locations, household types and demographic groups. To address these limitations, we have implemented measures to ensure that our dataset is as comprehensive and up-to-date as possible. The system is designed to be adaptable, allowing for recalculation to account for variability in external conditions like weather patterns and occupant behavior and sensor or data errors. Future research could explore incorporating larger datasets that encompass a wider variety of building types, geographic regions, and climatic conditions to improve the applicability of the clustering models. Additionally, including more dimensions such as occupant behavior, socio-economic characteristics, and building occupancy could provide a more comprehensive understanding of energy consumption patterns. Another avenue for future work is investigating the temporal aspects of energy usage, such as seasonal variations and daily usage patterns, which could refine the clustering process and at the same time improve the accuracy of heating demand forecasting.

\section*{Acknowledgments}

The work presented is based on research conducted within the framework of the Horizon Europe European Commission project DEDALUS (Grant Agreement No. 101103998). The content of the paper is the sole responsibility of its authors and does not necessary reflect the views of the EC.

\bibliographystyle{unsrtnat}

\bibliography{annot}

\clearpage

\appendix
\setcounter{figure}{0}
\section{Appendix: Dimensional Distribution Figures}

This Section contains figures referenced in the main text, providing further detail on the dimensional data discussed. The figures are supplementary to those in the main sections and offer additional visualization of the key aspects that drive the clustering algorithms.

\begin{figure*}[ht!] 
\centering
 \makebox[\textwidth]{\includegraphics[width=.8\paperwidth]{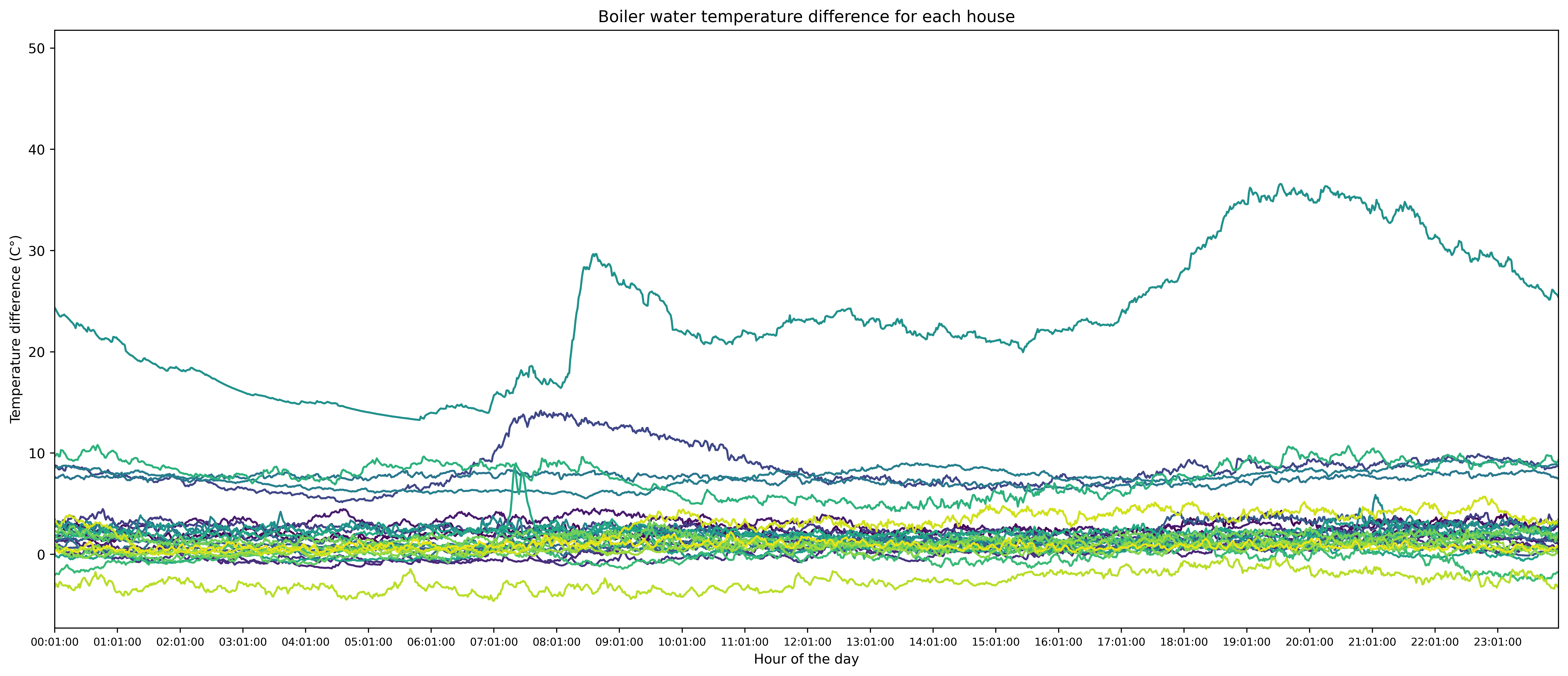}}
\caption{Boiler and return water temperature difference for each house throughout the day}
\label{fig:appendix1}
\end{figure*}

\begin{figure*}[h!] 
\centering
 \makebox[\textwidth]{\includegraphics[width=.8\paperwidth]{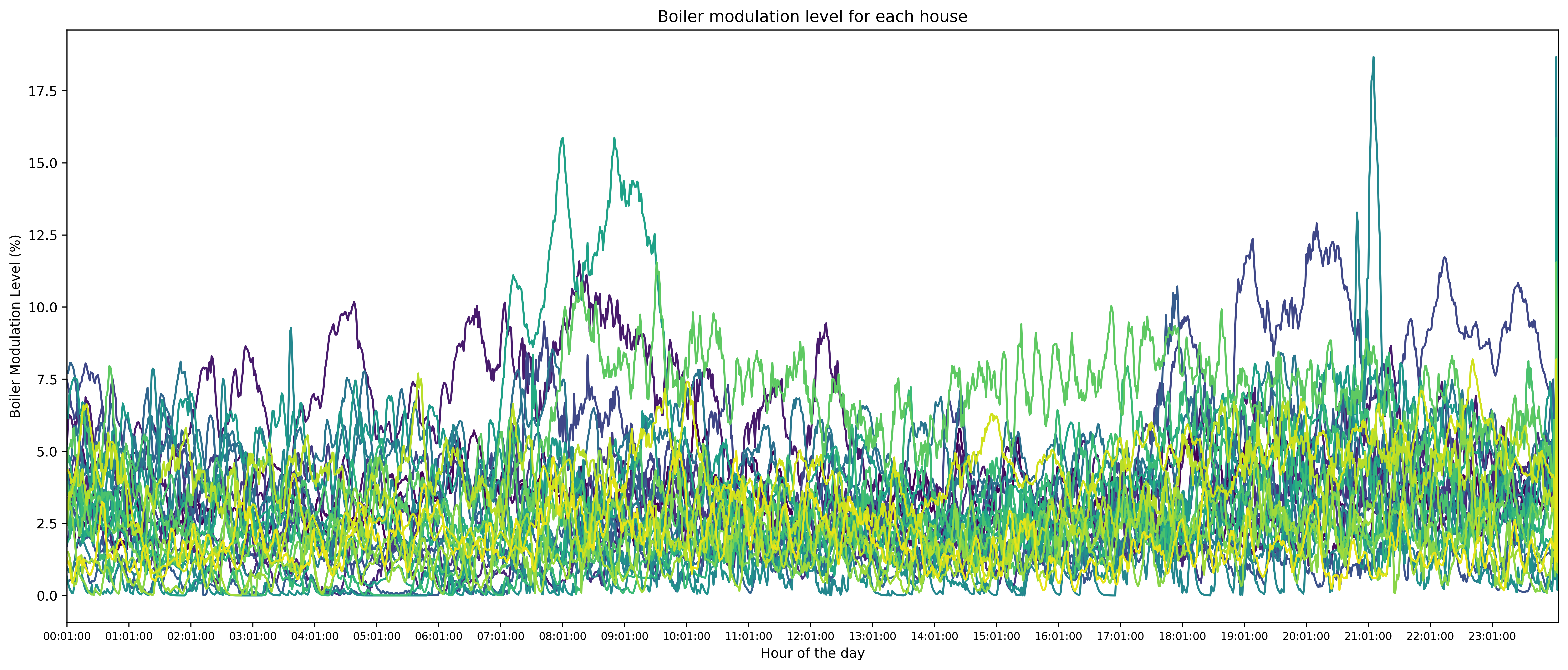}}
\caption{Boiler modulation level for each house throughout the day}
\label{fig:modulation_level}
\end{figure*}

\begin{figure*}[t!] 
\centering
 \makebox[\textwidth]{\includegraphics[width=.8\paperwidth]{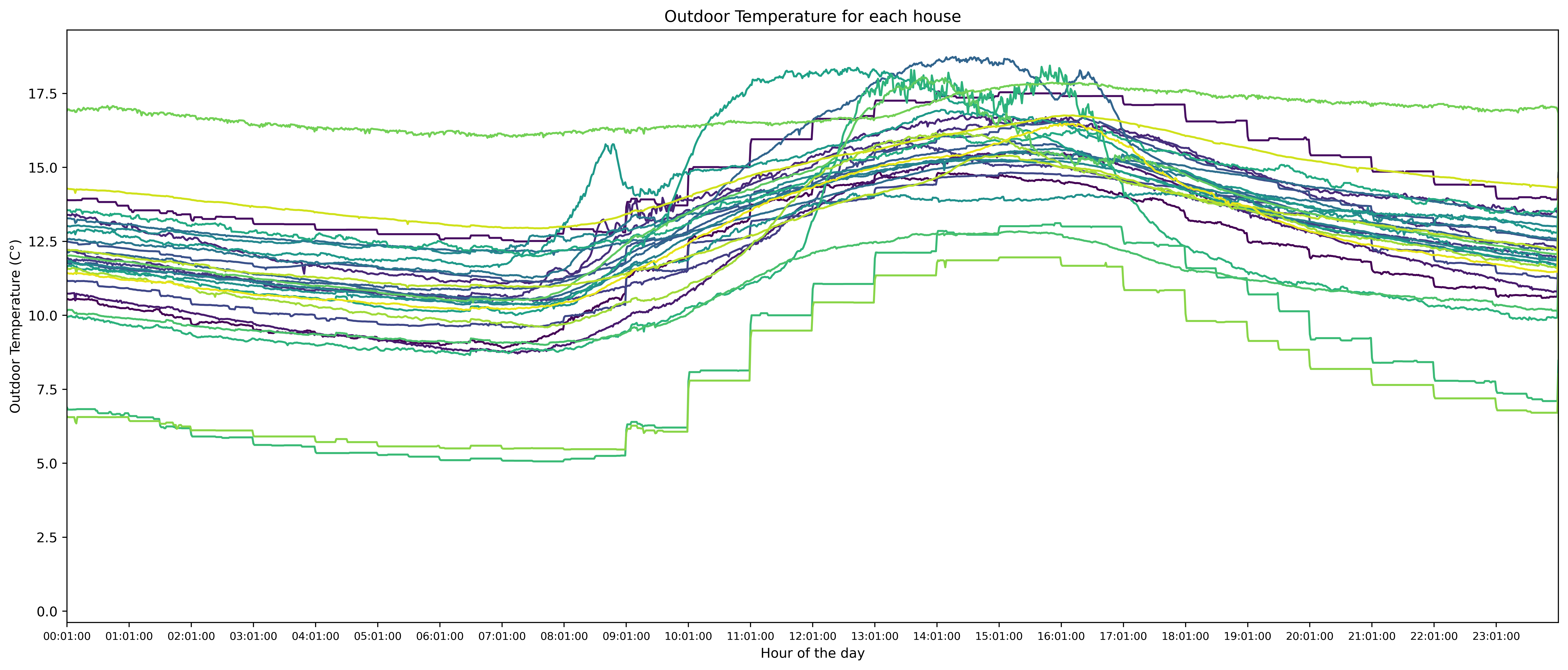}}
\caption{Outdoor temperature for each house throughout the day}
\label{fig:outdoor_temperature}
\end{figure*}

\begin{figure*}[h!] 
\centering
 \makebox[\textwidth]{\includegraphics[width=.8\paperwidth]{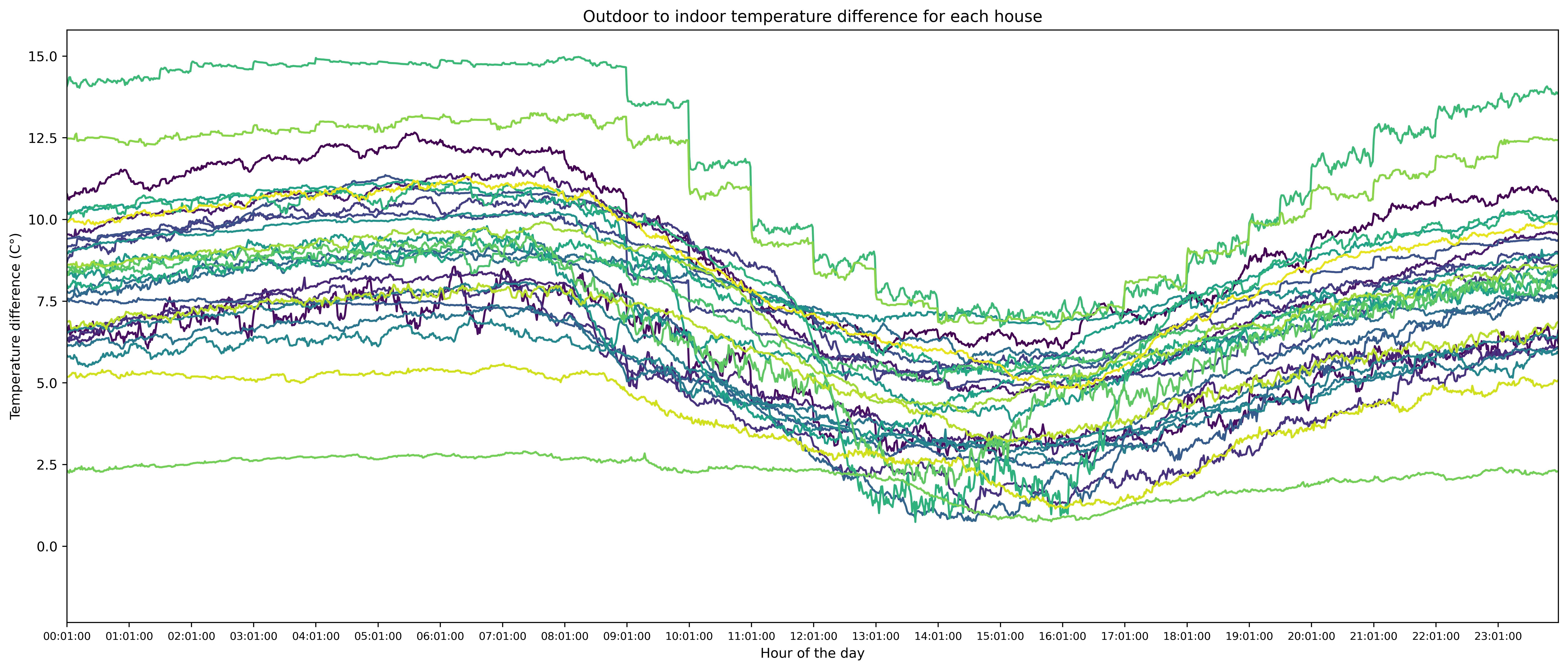}}
\caption{Variation in the disparity between indoor and outdoor temperatures for each house throughout the day}
\label{fig:building}
\end{figure*}

\begin{figure*}[h!] 
\centering
 \makebox[\textwidth]{\includegraphics[width=.8\paperwidth]{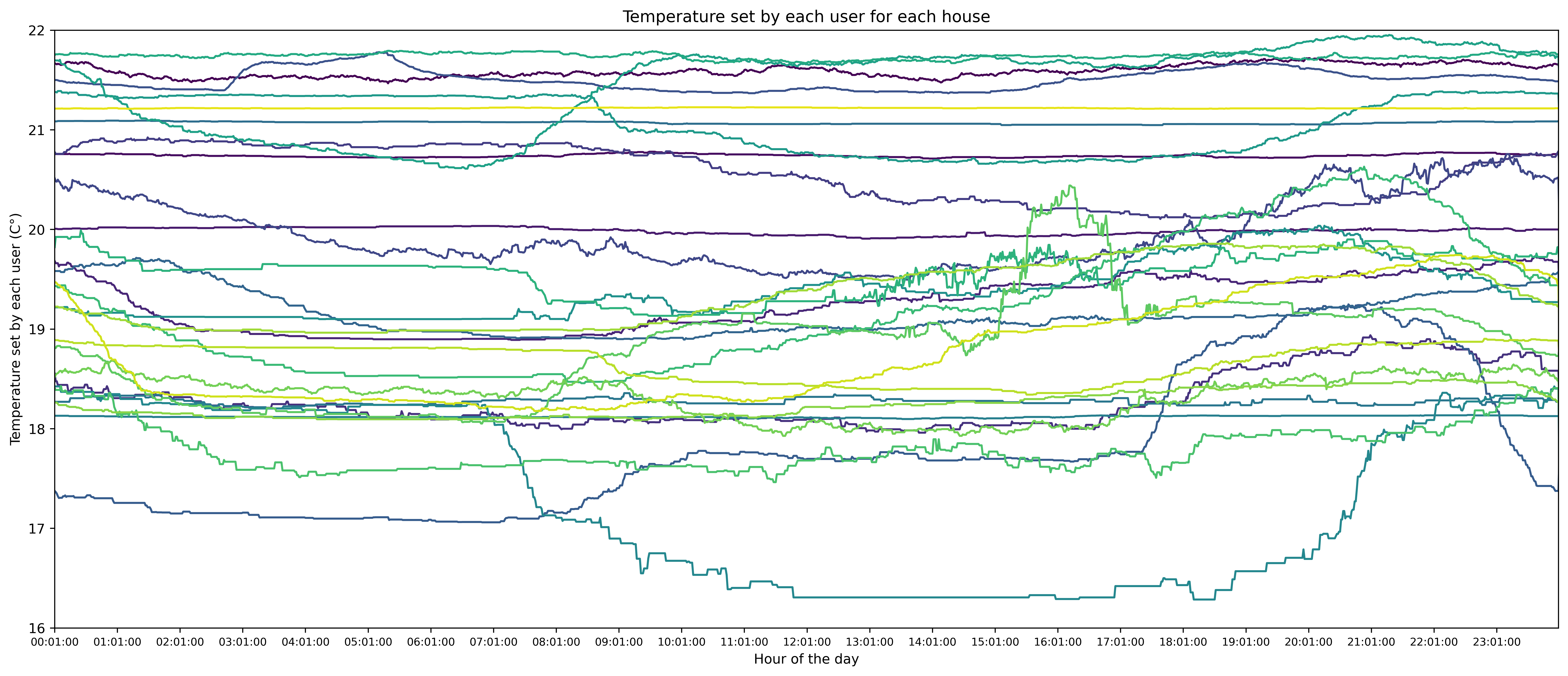}}
\caption{Target temperature set by each user for each house}
\label{fig:user}
\end{figure*}

\end{document}